\documentclass[11pt]{article}
\pdfoutput=1
\usepackage{jcappub}
\usepackage{booktabs}
\usepackage[english]{babel}
\usepackage{amsmath, amssymb, amsfonts, amsbsy, amstext, amsthm}
\usepackage{graphicx}
\usepackage{tabularx}
\usepackage[toc,page]{appendix}
\usepackage{comment}
\usepackage{exscale}
\usepackage[makeroom]{cancel}
\usepackage{soul}
\usepackage{siunitx}
\usepackage{mathtools}
\usepackage{multirow}
\usepackage[usenames,dvipsnames,table,xcdraw]{xcolor}
\usepackage{aas_macros}
\usepackage{xspace}
\usepackage{ulem}
\usepackage{mathrsfs}

\newcommand{\burst}{\textsc{burst}\xspace}
\newcommand{\neff}{\ensuremath{N_{\rm eff}}\xspace}
\newcommand{\cnub}{C$\nu$B\xspace}
\newcommand{\smnu}{CSM$\nu$\xspace}
\newcommand{\boldell}{\ensuremath{\boldsymbol{\ell}}\xspace}

\newcommand{\edit}[1]{{{#1}}}

\newcommand{\diffvis}{\ensuremath{\mathscr{V}}}

\newcommand{\nequil}{\ensuremath{n^{\rm (eq)}}}
\newcommand{\tcm}{\ensuremath{T_{\rm cm}}\xspace}

\newcommand\epsmax{\ensuremath{\epsilon_{\rm max}}\xspace}

\allowdisplaybreaks[1]
\numberwithin{equation}{section}
\def\beq{\begin{equation}}
\def\eeq{\end{equation}}

\def\d{{\rm d}}

\def\Nf{N_{\rm eff}}
\def\Tcm{T_{\rm cm}}

\DeclareRobustCommand{\SkipTocEntry}[4]{}

\newcommand{\smu}{Department of Physics,
Southern Methodist University, Dallas, TX 75275, USA}

\begin{document}

\pagenumbering{roman}
\begin{titlepage}
\baselineskip=15.5pt \thispagestyle{empty}

\null\hfill N3AS-24-011

%\bigskip\

%\vspace{1cm}
\vspace{0.5cm}
\begin{center}

\fontsize{19}{18}\selectfont \sffamily \bfseries Cosmic Neutrino Decoupling and its Observable Imprints: Insights from Entropic-Dual Transport

\end{center}
\vspace{0.1cm}
\begin{center}
{\fontsize{12}{30}\selectfont J.~Richard~Bond,$^{a}$ George~M.~Fuller,$^{b}$ Evan~Grohs,$^{c}$ Joel~Meyers,$^{d}$ and Matthew~James~Wilson$^{e}$}
\end{center}

\begin{center}

\small

\textsl{$^a$ Canadian Institute for Theoretical Astrophysics, Toronto, Ontario M5S 3H8, Canada}
\vskip 4pt

\textsl{$^b$ Department of Physics, University of California, San Diego, La Jolla, California 92093, USA}
\vskip 4pt

\textsl{$^c$ Department of Physics, North Carolina State University, Raleigh, NC 27695, USA}
\vskip 4pt

\textsl{$^d$ \smu}
\vskip 4pt

%\textsl{$^f$ Department of Physics, University of Toronto, Toronto, Ontario M5S 1A7, Canada}
%\vskip 7pt

\textsl{$^e$ Institute for Astroparticle Physics, Karlsruhe Institute of Technology, Karlsruhe, 76131, Germany}

\normalsize

\end{center}

\vspace{0.3cm}
\hrule \vspace{0.3cm}
\noindent {\sffamily \bfseries Abstract} \\[0.1cm]
Very different processes characterize the decoupling of neutrinos to form the cosmic neutrino background (\cnub) and the much later decoupling of photons from thermal equilibrium to form the cosmic microwave background (CMB).  The \cnub emerges from the fuzzy, energy-dependent neutrinosphere and  encodes the physics operating in the early universe in the temperature range $T\sim 10\,{\rm MeV}$ to $T\sim10\,{\rm keV}$.   This is the epoch where beyond Standard Model (BSM) physics, especially in the neutrino sector, may be influential in setting the light element abundances, the necessarily distorted fossil neutrino energy spectra, and other light particle energy density contributions. Here we use techniques honed in extensive CMB studies to analyze the \cnub as calculated in detailed neutrino energy transport and nuclear reaction simulations of the protracted weak decoupling and primordial nucleosynthesis epochs.
Our moment method, relative entropy, and differential visibility approach can leverage future high precision CMB and light element primordial abundance measurements to provide new insights into the \cnub and any BSM physics it encodes.
We demonstrate that the evolution of the energy spectrum of the \cnub throughout the weak decoupling epoch is accurately captured in the Standard Model by only three parameters per species, a non-trivial conclusion given the deviation from thermal equilibrium and the impact of the decrease of electron-positron pairs.
Furthermore, we can interpret each of the three parameters as physical characteristics of a non-equilibrium system.
\edit{Though the treatment presented here makes some simplifying assumptions including ignoring neutrino flavor oscillations, t}he success of our compact description within the Standard Model motivates its use also in BSM scenarios.
We further demonstrate how observations of primordial light element abundances can be used to place constraints on the \cnub energy spectrum, deriving response functions that can be applied for general deviations from a thermal spectrum.  
Combined with the description of those deviations that we develop here, our methods provide a convenient and powerful framework to constrain the impact of BSM physics on the \cnub.

\vspace{0.3cm}
\hrule
%\vskip 10pt

%\vspace{0.6cm}
\end{titlepage}

\thispagestyle{empty}
\setcounter{page}{2}
\tableofcontents

\clearpage
\pagenumbering{arabic}
\setcounter{page}{1}

%%%%%%%%%%%%%%%%%%%%%%%%%%%%%%%%%%%%%%%%%

%%%%%%%%%%%%%%%%%%
\section{Introduction}
\label{sec:Intro}
%%%%%%%%%%%%%%%%

Our focus in this paper is to provide a new way of analyzing the cosmic neutrino background (\cnub). The \lq\lq fingerprints\rq\rq\ of the \cnub can be found on the Cosmic Microwave Background (CMB) and large scale structure as well as on the primordial light element abundances which emerge from the epoch in the early universe when the weak interactions involving neutrinos and the nuclear reactions freeze out. Revealing these fingerprints may leverage future experimental and observational results into new insights of the early universe and the physics operating therein. There is urgency about this effort, as observation and experiment are poised to provide higher precision data in cosmology and on physics in the neutrino sector~\cite{Amin:2022soj,Gerbino:2022nvz,Chang:2022tzj,Dvorkin:2022jyg,DUNE:2022aul,Hyper-Kamiokande:2022smq,JUNO:2015zny}, while the origin of neutrino mass is a frontier issue in elementary particle physics~\cite{2016ARNPS..66..197D}. 

Next generation ground-based cosmic microwave background (CMB) experiments such as CMB-S4~\cite{CMB-S4:2016ple,Abazajian:2019eic} promise to measure \neff (a measure of the energy density in particles with relativistic kinematics around the epoch of photon decoupling) with a precision of about one percent, \edit{place} tight constraints on the primordial helium abundance, and \edit{provide} a significant measurement of the \lq\lq sum of the light neutrino masses\rq\rq\ (a measure of the suppression of matter clustering associated with massive cosmic neutrinos~\cite{Dolgov:2002wy,Lesgourgues:2006nd,Wong:2011ip,Lesgourgues:2012uu,Dvorkin:2019jgs,Green:2021xzn}). In parallel, laboratory- and accelerator-based experiments promise to elucidate key issues in neutrino physics~\cite{Gerbino:2022nvz}. Within five years, long baseline neutrino oscillation experiments will give us the neutrino mass hierarchy to fair confidence~\cite{Patterson:2015xja}. Likewise, we can expect important neutrino rest mass constraints from front line and future tritium endpoint experiments (e.g., KATRIN~\cite{KATRIN:2022ayy}, Project-8~\cite{Project8:2022wqh}), and tonne-scale neutrino-less double beta decay experiments (e.g., nEXO~\cite{nEXO:2021ujk}) that may pin down the character (Majorana or Dirac) of neutrinos and provide neutrino mass and hierarchy data complementary to that derived from cosmology. A known neutrino mass hierarchy not only extends the utility of the CMB- and large scale structure-derived constraints on, or measurements of the sum of the light neutrino masses, but it also feeds back on the results of the laboratory neutrino mass probes~\cite{Gerbino:2022nvz}.

The existence of the \cnub is a central prediction of hot big bang cosmology.  Within the Standard Model of particle physics, assuming a standard thermal history of our universe, we can make very definite predictions about the properties of cosmic neutrinos.  Up to some small corrections, the Standard Model predicts that cosmic neutrinos and antineutrinos of all flavors were well described by a relativistic, nearly zero degeneracy parameter Fermi-Dirac (FD) distribution.
In this simple and approximate description, these distributions share a common temperature with photons when the temperature is high.  At lower temperatures when the electron-positron pairs have disappeared, co-moving entropy-conservation yields a neutrino-to-photon-temperature ratio of $(4/11)^{1/3}$.
Subsequent to decoupling, neutrinos free-fall and their momenta redshift with the expanding universe, implying that some may have non-relativistic kinematics at the current epoch~\cite{Dolgov:2002wy}.
In what follows, we will use the results of detailed neutrino energy transport calculations to characterize deviations from this simplistic picture.

Cosmic neutrinos made up a sizable fraction of the energy budget of the early universe.  The gravitational influence of cosmic neutrinos had a significant impact on the expansion of the universe and the evolution of structure during the radiation-dominated era.   Observations of primordial light element abundances, the angular power spectra of the CMB, and the power spectrum of large-scale structure are therefore sensitive to the presence of cosmic neutrinos.  These observations are most sensitive to the total energy density of the cosmic neutrino background, parameterized through $\Nf$, defined such that the total radiation energy density of the universe is
\begin{equation}
    \rho_r = \rho_\gamma \left(1+\frac{7}{8}\left(\frac{4}{11}\right)^{4/3}\Nf \right) \, ,
    \label{eq:Neff_def}
\end{equation}
where $\rho_\gamma$ is the energy density of photons.
The Standard Model prediction is $\Nf=3.044(1)$~\cite{Akita:2020szl,Froustey:2020mcq,Bennett:2020zkv}.  Current observational constraints on $\Nf$ are consistent with the predictions of the Standard Model; data from the Planck satellite provides a measurement $\Nf=2.99\pm0.17~(68\%~\mathrm{CL})$~\cite{Planck:2018vyg} \edit{in $\Lambda$CDM+$\Nf$ cosmology. In the same model, measurements of the primordial helium and deuterium abundances along with BBN calculations suggest a constraint $\Nf=2.89\pm0.23$~\cite{Yeh:2022heq}, and CMB combined with helium and deuterium measurements gives $\Nf = 2.90\pm0.14$~\cite{Yeh:2022heq}.}

Upcoming CMB experiments will greatly improve the precision with which $\Nf$ can be measured~\cite{CMB-S4:2016ple,SimonsObservatory:2018koc,NASAPICO:2019thw,Abazajian:2019eic,Sehgal:2019ewc,Chang:2022tzj}.  On the other hand, future CMB observations by themselves are unlikely to give much insight into more detailed properties of cosmic neutrinos such as neutrino flavor content or the energy spectrum.
However, when coupled with Big Bang Nucleosynthesis (BBN) and primordial light-element abundance determinations~\cite{Cyburt:2015mya}, we can gain insights into these issues.
BBN depends not just on the gravitational influence of cosmic neutrinos but also on their weak interactions -- with exquisite sensitivity \cite{2016NuPhB.911..955G}.  Primordial abundances therefore in principle provide a window into measuring properties of cosmic neutrinos which are inaccessible through other observations.

Physics Beyond the Standard Model (BSM) can cause the cosmic neutrinos to have an energy distribution function which differs significantly from a relativistic FD distribution \cite{2011arXiv1110.6479F,Rasmussen:2021kbf}.  For example, the decay or annihilation of some new particles could inject additional neutrinos compared to the standard thermal history, or oscillation into new sterile states could alter the neutrino distribution \cite{2005PhRvD..72f3004A}.  If these processes occur at sufficiently early times, the neutrino distribution will tend to relax into thermal equilibrium.  However, late changes to the neutrino distribution may become frozen-in leading to an altered spectrum at late times, perhaps accompanied by alterations in $N_{\rm eff}$ and light element abundances relative to the Standard Model.
New light degrees of freedom may also contribute to $\Nf$ and impact associated observables~\cite{Dvorkin:2022jyg}.

Large-scale numerical simulations have been employed to follow this evolution of the neutrino component and the associated Standard Model baryonic physics like nuclear freeze-out and BBN~\cite{Dicus:1982bz, Cambier:1982pc, Rana:1991xk, Dodelson:1992km, Hannestad:1995rs, Dolgov:1997mb, Gnedin:1997vn, Dolgov:1998sf,2014PhRvD..89b3008B,  EscuderoAbenza:2020cmq,Akita:2020szl,Froustey:2020mcq,Bennett:2020zkv,Grohs:2015tfy,2024EPJC...84...86B}. However, these coupled neutrino transport and baryonic physics problems are not simple to model with Boltzmann calculations, and will be even more difficult to model with general treatments of neutrino flavor quantum coherence. The heart of this paper is the introduction of a technique for capturing and physically interpreting the salient features in these large-scale numerical simulations. \edit{Our goal for the present work is not to directly replace these simulations, but rather to introduce a framework that can be used to describe the relevant physics and which may be further developed to compactly treat more general BSM scenarios.  As such, we leave out some effects, such as neutrino flavor oscillations, that are included in other treatments of C$\nu$B transport that are aimed at precision calculation of $N_\mathrm{eff}$ in the Standard Model, for example. }

We demonstrate that we can capture the essential aspects of neutrino transport in the early universe through use of generalized statistical techniques. The generalized entropy approach that we develop allows us to faithfully represent the evolution of the cosmic neutrino background, as derived from detailed Boltzmann transport, using just three parameters \edit{per species}.  We show how the distribution function of each neutrino species can be described at high precision as $n(t) = \left[\exp(\alpha(t))+1\right]^{-1}$, with the function $\alpha = \alpha_0 + \alpha_1 \epsilon + \alpha_2 \epsilon^2$, where $\epsilon$ is related to the energy of the neutrinos, to be defined more precisely below.
The two parameters $\alpha_0$ and $\alpha_1$ are familiar in the context of the Grand Canonical Ensemble (GCE) from statistical mechanics, and relate to average number and average energy.  On the other hand, the third parameter is a new aspect of the neutrino distributions.  It is conjugate to the average of the energy-squared of the distribution, and by extension is related to the variance in energy.

In general the total energy-dependent quantity $\alpha$ is conjugate to the Wigner distribution function matrix of Boltzmann transport theory, and fully encodes the transport information, as the response in entropy due to a change in the distribution of number. This duality is general for all transport networks.
For example, in our nuclear reaction network, each nuclide $J$ is described by a chemical potential $\mu_J$, with a common temperature $T$.  The abundance is conjugate to a chemical potential, which would give a Maxwell-Boltzmann distribution of energies, and hence entropy, when used in tandem with the common temperature.  The nuclear reaction network we use in this work follows abundances, but it is useful to consider how it maps into the $\{\alpha_J\}$ dual description:  the $\alpha_J$ give the response of the entropy to changes in the number of $J$ nuclei. If the nuclear reactions which connect a subset of the nuclide network are rapid, equilibrium is locally attained and the total chemical potentials of the products and reactants in the subset are in detailed balance, with entropy differences related to the latent heat, i.e., the nuclear binding energies. At temperatures well above $1\,{\rm MeV}$, the system is in chemical and weak equilibrium, so all nuclide abundances are describable in terms of a single chemical potential related to the baryon density. As the temperature decreases, more chemical potentials ({\it i.e.}, $\alpha_J$) are needed, first in  quasi-nuclear statistical equilibrium among a subset of nuclides.  The partition becomes finer with decreasing reaction rates until eventually every nuclide requires its own $\alpha_J$ in an out-of-equilibrium configuration. As in the neutrino network case of distribution function numerical solutions, our nucleosynthesis network is solved numerically in abundances, but the dual description gives insight into such chemical equilibrium relations. The total response in the nuclear entropy for a change in baryon number is $\sum_J \langle \alpha_J \rangle Y_J$, where $\langle\alpha_J\rangle$ is the energy-averaged value of $\alpha_J$ and $Y_J$ is the abundance. Similarly, the total response in the neutrino entropy is a weighted sum over the individual responses for each energy and flavor.  If oscillations are ignored, the $\nu_\mu - \nu_\tau$ neutrino flavor blocks are related by the same microscopic rates and have the same responses (likewise for the $\overline{\nu}_\mu - \overline{\nu}_\tau$ blocks). The distribution function progression we describe below as the temperature drops is akin to a full neutrino equilibrium needing one $\alpha_1$, {\it i.e.}, one neutrino temperature, then a $\alpha_1$ for temperature and $\alpha_0$ for number, then an additional $\alpha_2$.   To complete the network description in $\alpha$ terms, in the photon-electron-positron plasma, the equilibrium relation $\mu_{e^-} + \mu_{e^+} = 2\mu_\gamma=0$ is obeyed, implying $\mu_{e^+}=-\mu_{e^-}$: the $\gamma$-$e^-$-$e^+$ plasma reduces to a full statistical equilibrium description, with the asymmetry in  $\alpha_{e^\pm}$ encoding the residual electron number per baryon once the $e^+ e^- \rightarrow \gamma \gamma $ phase transition is complete.  
 
This technique applied to the neutrino network is a simple-to-use parameterization which captures the most relevant physics and can offer cogent physical interpretations of key features emerging from the numerical simulations. It allows us to understand the origin of the deviations from the thermal spectrum in the Standard Model and BSM scenarios. That, in turn, helps us to quantify the degree to which observations could be sensitive to the energy distribution of cosmic neutrinos.  Using our method, we can describe a `last scattering surface' for cosmic neutrinos in close analogy with the CMB, finding a broad peak with energy-dependent width around $z \sim 10^{10}$ (corresponding to a temperature $T\sim1\,{\rm MeV}$).  We show how primordial abundances respond to changes in the cosmic neutrino energy spectrum, and thereby show how abundance measurements and CMB parameters can be used to place constraints on physics beyond the Standard Model.

In Section~\ref{sec:WeakDecoupling}, we give an overview of the physics of weak decoupling and weak freeze-out.  Section~\ref{sec:TransportWigner} describes the general framework of neutrino transport in the language of Wigner functions and their conjugates.  We compare aspects of neutrino decoupling to CMB transport in Section~\ref{sec:CMBTransport}.  Section~\ref{sec:CollisionRate} discusses the weak interactions relevant to neutrino decoupling.  We describe the numerical implementation of neutrino transport and primordial nucleosynthesis in Section~\ref{sec:BURST}.  Section~\ref{sec:SMnu} describes neutrino transport in the Standard Model and develops a compact treatment of how the occupation probability of neutrinos deviate from equilibrium.  In Section~\ref{sec:NuDiffVis}, we introduce the neutrino differential visibility and productivity as a novel means to visualize the neutrino decoupling process.    We discuss how weak nucleon interactions affect primordial abundance yields in Section~\ref{sec:WeakInteraction}.  Section~\ref{ssec:abunds_resps} introduces the linear response of light element abundances to distortions of the cosmic neutrino spectrum.
We conclude in Section~\ref{sec:Conclusions}.
In what follows, we use natural units where $c=\hbar=k_B=1$.

%%%%%%%%%%%%%%
\section{Overview of Weak Decoupling and Weak Freeze-Out}
\label{sec:WeakDecoupling}
%%%%%%%%%%%%%%

\begin{figure}[t!]
    \begin{center}
    \includegraphics[width = \columnwidth]{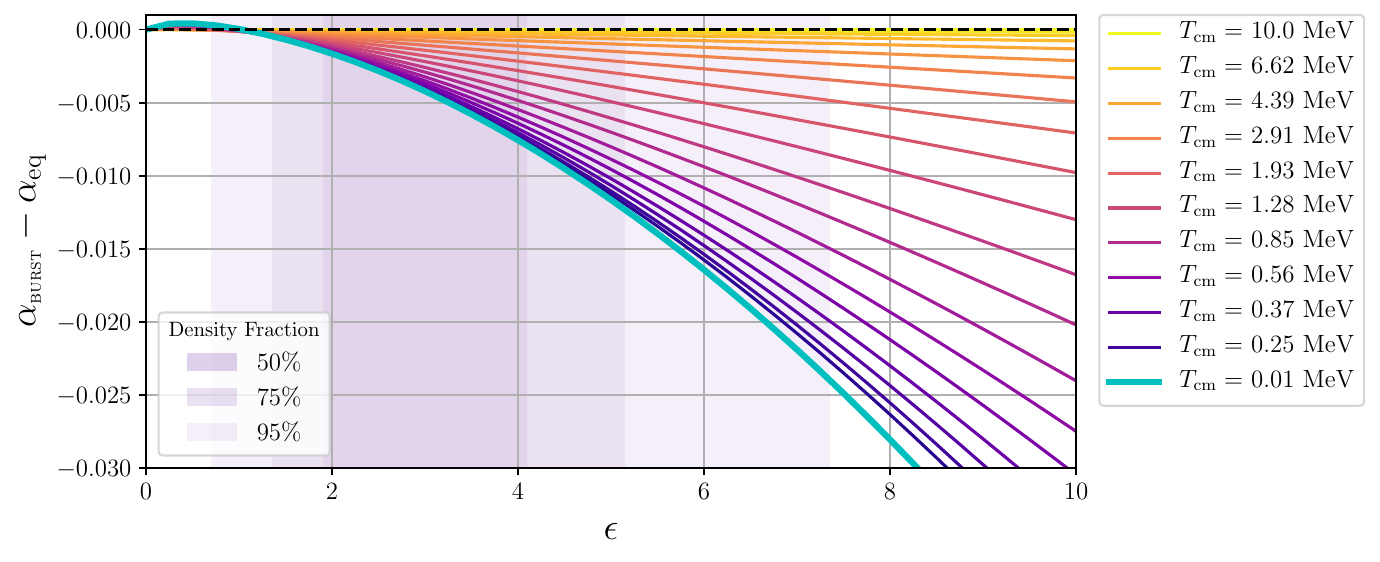}
    \caption{
    Difference between neutrino distribution functions and a relativistic Fermi-Dirac distribution with vanishing chemical potential plotted against $\epsilon=E/\tcm$ at epochs labeled by \tcm.  $E$ is the neutrino energy and \tcm is the comoving temperature quantity.
    Here, the neutrino distribution function for flavor $f$ is parameterized as $n_{\nu_{f}} = (\exp[{\alpha^{(f)}(\epsilon,T_{\text{cm}})}]+1)^{-1}$.  For this figure, $f=e$.
    A relativistic Fermi-Dirac distribution with vanishing chemical potential is described by $\alpha_\mathrm{eq} =  \epsilon$, and represented by the dashed black line.
    Here we show how the evolution of the $\nu_e$ distribution function resulting from a numerical transport calculation, described by $\alpha_{\textsc{\scriptsize burst}}$, differs from the  Fermi-Dirac distribution as neutrinos go out of equilibrium.
    Note that in this and similar figures below, we label only every other line with its corresponding value of $\Tcm{}$; the unlabeled lines should be understood to have $\Tcm$ given by the geometric mean of that of the neighboring lines.
    The Density Fraction color contours indicate the neutrino occupation fraction as a function of $\epsilon$.
    }
    \label{fig:alpha_evolution}
    \end{center}
\end{figure}

In this section we review the process of weak decoupling and weak freeze-out in the Standard Model, assuming a standard thermal history, following Ref.~\cite{Grohs:2015tfy}.  We will describe how out-of-equilibrium processes cause small distortions in the cosmic neutrino spectrum even in the standard case.
In the next section we introduce a full parameterization of these distortions in terms of a thermodynamic potential conjugate to the general neutrino distribution functions.
We will show how this distribution function can be represented in an expansion in terms of the coefficients $\{\alpha^{(f)}\}$, with $f$ denoting a flavor label. Surprisingly, with very few of these coefficients, our procedure can reproduce the results of detailed transport calculations for \cnub decoupling.
This conjugate description can illuminate when and how the \cnub in the Cosmological Standard Model, the \smnu, is altered by BSM physics.
Figure \ref{fig:alpha_evolution} shows how the transport-derived distribution functions evolve away from Fermi-Dirac character as the universe expands and cools.
We will show the subsequent sections how we can compactly describe the evolution of the \smnu, accurately accounting for this deviation.

%%%%%%%%%%%%%%%%% Alpha evolution %%%%%%%%%%%%%%%
%%%%%%%%%%%%%%%%%%%%%%%%%%%%%%

%%%%%%%%%%%%%%%%%%%%%%%%%%%
\subsection{Neutrino Last Scattering}
\label{sec:LastScattering}
%%%%%%%%%%%%%%
In this section we focus on the transition between the times of frequent neutrino scattering and free expansion, defining an energy-dependent region of last scattering for cosmic neutrinos.  As described in the previous section, weak decoupling is a gradual process, and we will therefore show that the neutrino last scattering surface is much broader than its CMB analog.

\begin{figure}[t!]
    \begin{center}
    \includegraphics[width = 0.85\columnwidth]{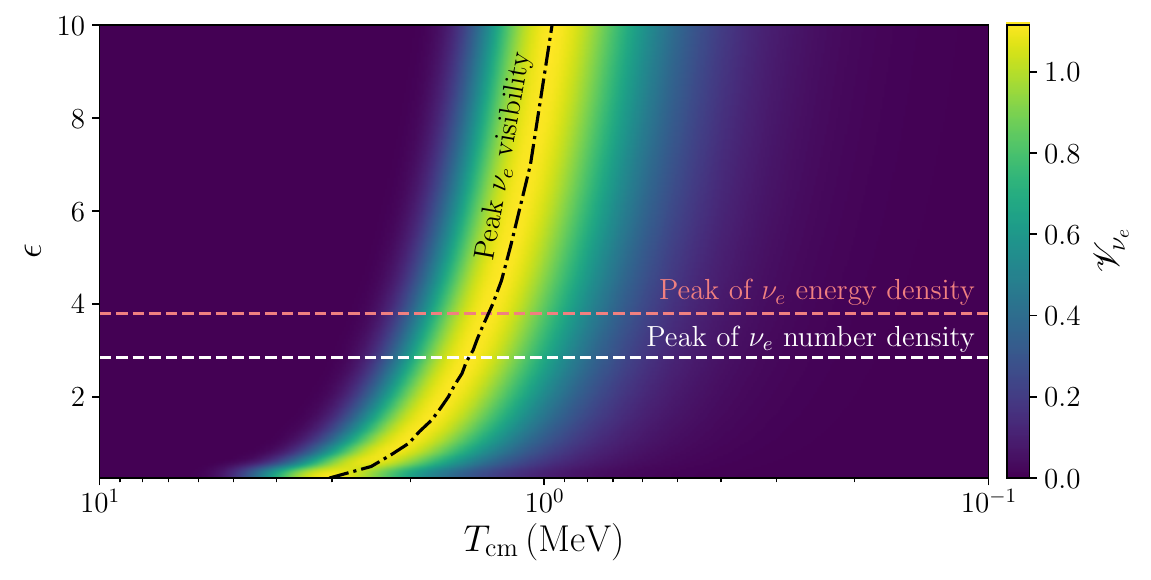}
    \caption{
    The broad energy-dependent last scattering surface of the cosmic neutrino background can be seen from this plot of the differential visibility function for electron-type neutrinos, namely, $\diffvis_{\nu_e} = d(e^{-\tau_{\nu_e}})/d\ln a$ for scale factor $a$ and optical depth $\tau_{\nu_e}$ (see Sec.\ \ref{sec:NuDiffVis} for more details).
    Contours of $\diffvis_{\nu_e}$ are shown in the $\epsilon$ versus \tcm space.  The black dot-dash line gives the peak value of $\diffvis_{\nu_e}$, and the dashed pink (white) give the peak values of the energy (number) density for massless fermions.
    This should be contrasted with the last scattering surface of the cosmic microwave background, which is more localized in temperature and is essentially independent of the photon frequency.
    }
    \label{fig:diff_vis}
    \end{center}
\end{figure}

In the early universe during the weak decoupling and BBN epochs, SM neutrinos interact only via weak interactions.  At tree-level, the scattering cross sections are proportional to the square of the Fermi constant multiplied by the square of the energy
\begin{align}
  \sigma_{\rm weak}&\propto G_F^2E^2 %\\
  %&
  =G_F^2\epsilon^2\tcm^2\label{eq:sigma_weak},
\end{align}
where $G_F\simeq1.166\times10^{-11}\,{\rm MeV}^{-2}$ and $E$ is an energy-scale typical of the weakly-interacting particles.
We have used Eq.\ \eqref{eq:sigma_weak} to define two quantities which we will return to frequently throughout this work.  \tcm is the comoving temperature quantity and is an energy scale which redshifts with increasing scale factor.  It replaces the conventional neutrino temperature scale which is no longer applicable in the presence of spectral distortions.  Using \tcm, we define $\epsilon\equiv E/\tcm$ as an alternative to $E$.  For massless particles, $\epsilon$ is a comoving invariant as $E$ and \tcm redshift in the same manner.
The cross section in Eq.\ \eqref{eq:sigma_weak} increases with increasing energy, implying that more energetic interactions will maintain equilibrium kinetics to lower temperatures.  Furthermore, we can approximate the scattering rates of the neutrinos as the weak cross section multiplied by the neutrino number density
\begin{equation}
  \Gamma_{\rm weak}\sim\sigma_{\rm weak}n_\nu\sim G_F^2T^5 \, ,
\end{equation}
where we have used the ultra-relativistic kinematics of $n_\nu\propto T^3$ and taken the typical energy scale also to be proportional to $T$.  When compared to the Hubble expansion rate for radiation-dominated conditions ($H\propto T^2$), we find
\begin{equation}\label{eq:gamma_to_H}
  \Gamma_{\rm weak}/H\propto T^3 \, .
\end{equation}
Note that we will always approximate neutrinos as ultra-relativistic during weak decoupling, an approximation that is well-justified by current upper bounds on the sum of neutrino masses \edit{from beta decay endpoint measurements by KATRIN~\cite{KATRIN:2021uub} and also from cosmological constraints~\cite{Planck:2018vyg}.}  Alternatively, if neutrinos were massive, and these masses were comparable to or smaller than the temperature, their number densities would include a Boltzmann factor $e^{-m/T}$ -- a much more sensitive function of temperature than the relationship in Eq.\ \eqref{eq:gamma_to_H}.  Combined with the fact that the cross section is energy dependent, neutrinos with different energies decouple at different times.  Neutrino decoupling is a protracted process and the surface of last scattering is broad; see Fig.~\ref{fig:diff_vis}.

The process of CMB photon decoupling differs significantly from cosmic neutrino decoupling.  Photons remain in thermal equilibrium with the plasma through Thomson scattering with free electrons, since the typical photon energy during the relevant period ($\lesssim1\,{\rm eV}$) is small compared to the electron rest mass, $m_e$.  The number density of free electrons drops precipitously when the formation of a neutral gas becomes energetically favorable (which occurs when the number density of photons with energy exceeding the atomic binding energy becomes small compared to the number density of nuclei).  The ionization fraction, $x_e$, is governed during the early stages of recombination by the Saha equation
\begin{equation}
    \frac{1-x_e}{x_e^2}\propto \eta \left(\frac{T}{m_e}\right)^{3/2} \exp(B_\mathrm{H}/T) \, ,
    \label{eq:Saha}
\end{equation}
where $\eta$ is the baryon-to-photon ratio and $B_\mathrm{H}=13.6\,{\rm eV}$ is the binding energy of hydrogen. A more careful treatment of recombination shows a somewhat slower decline in the later stages of recombination and a larger residual ionization fraction than predicted by the Saha equation~\cite{Peebles:1968ja,Zeldovich:1968qba,Seager:1999bc,Seager:1999km,Ali-Haimoud:2010hou,Chluba:2010ca}.
Small deviations from a purely thermal blackbody spectrum are expected due to dissipation of acoustic waves and photon interactions with adiabatically cooling electrons and baryons, though the fractional change to the intensity is expected to be $\mathcal{O}(10^{-8})$ in standard cosmology~\cite{Chluba:2011hw,Chluba:2016bvg}.  Additionally, the frequency-dependent Rayleigh scattering of CMB photons on neutral species leads to spectral distortion of CMB anisotropies, but does not lead to a change in the average frequency spectrum of the CMB~\cite{1991PThPh..86.1021T,Yu:2001gw,Lewis:2013yia,Alipour:2014dza,Beringue:2020wxw}.
The end result is well-described by a sharp and energy-independent surface of last scattering and a nearly perfect blackbody spectrum for cosmic photons~\cite{Fixsen:1996nj}. This is in stark contrast to the protracted and energy-dependent neutrino decoupling process.

%%%%%%%%%%%%%% Differential Visibility %%%%%%%%%%%%%%%%%%
%%%%%%%%%%%%%%%%%%%%%%%%%%

%%%%%%%%%%%%%%%%%%%%%%%%
\subsection{Neutron-to-proton interconversion}
\label{sec:np_conversion}
%%%%%%%%%%%%%%

In our use of the term neutrino/weak decoupling, we describe the freeze-out of the weak interactions between neutrinos and charged leptons, and also between neutrinos and themselves.  Very importantly, we do not group the baryon-isospin-changing weak-interaction processes into neutrino decoupling.  Although the baryons and neutrinos interact via both charged-current and neutral-current reactions, the baryon density is so small that neutrino-baryon reactions have a negligible effect on the dynamics of neutrino decoupling.  Conversely, the neutrino fluences are so large that they can maintain weak-chemical equilibrium among the baryons until late times.  For the specific process of baryons going out of weak-equilibrium, we use the term weak freeze-out to describe the breakdown of chemical equilibrium.  Although weak freeze-out is not simultaneous with the process of neutrino decoupling, they are both protracted processes that occur concurrently.

The charged-current, isospin-changing, weak interactions that we consider are
\begin{align}
  \nu_{e} + n \leftrightarrow p + e^{-},\label{eq:nuc_reac1} \\
  e^{+} + n \leftrightarrow p + \bar{\nu}_{e} \label{eq:nuc_reac2},\\
  n \leftrightarrow p + e^{-} + \bar{\nu}_{e}\label{eq:nuc_reac3}.
\end{align}
The three forward reactions yield neutron destruction, or conversely proton production, and vice-versa for the reverse reactions.  For brevity, we call this set of 6 isospin-changing reactions the neutron-to-proton interconversion reactions.

Similar to the scattering reactions between neutrinos and charged leptons, the interconversion rates are fast at high temperature and are able to maintain a steady-state equilibrium between the number densities of neutrons and protons.
The cross section for these interconversion processes scales roughly like that of Eq.\ \eqref{eq:sigma_weak}, but with an important caveat.  Neutrons are heavier than protons.  Consequently, forward reactions of Eqs.~\eqref{eq:nuc_reac1} -- \eqref{eq:nuc_reac3}, the neutron destruction channels, have no energy threshold.  Correspondingly, there is an energy threshold for the reverse reactions that create neutrons from protons.
The upshot is that as the universe expands and cools (absent electron neutrino degeneracy), the competition between the forward and reverse of Eqs.~\eqref{eq:nuc_reac1} -- \eqref{eq:nuc_reac3} leads to an excess of protons over neutrons.
Phase space considerations imply that the cross sections for the reactions in Eqs.~\eqref{eq:nuc_reac1} -- \eqref{eq:nuc_reac2} are larger than those of neutrino-electron scattering for the energy scales characteristic of weak decoupling and BBN.
The energy scale for neutron-to-proton interconversion is a function of the neutron-proton mass difference, and so the cross section is much larger than the neutrino scattering rates at BBN energy scales.  The implications are two-fold.  First, the larger cross section implies larger rates and later epochs for freeze-out.  Second, the scaling relation between $\Gamma$ and $H$ is an even more gradual function in temperature than the analog in Eq.\ \eqref{eq:gamma_to_H}.  The end result is another protracted period of decoupling: the weak freeze-out epoch.

\subsection{Weak Decoupling, Weak Freeze-Out, and Precision Observables}
\label{sec:Decupling_Freezeout_Observables}

The processes of weak decoupling and weak freeze-out have implications for cosmological observables.  Weak decoupling determines the number density, energy density, and energy spectrum of cosmic neutrinos.  Gravitational influence of the cosmic neutrinos impacts the expansion history and the evolution of density perturbations (especially during the radiation-dominated era when neutrinos make up a significant fraction of the total energy density) and also the formation of cosmological structure (particularly at late times when neutrinos are non-relativistic).  Weak freeze-out is intimately related to the process of BBN and thereby determines the primordial abundance of light elements.  Precise prediction of these observables thus requires a precise treatment of these events.  Furthermore, weak decoupling and weak freeze-out overlap in time, and so it is not possible to treat them as separate, isolated processes.  Instead, one requires self-consistent modeling of neutrino transport throughout the whole epoch.

Detailed calculation of neutrino transport in the early universe is a non-trivial task even within the Standard Model where all relevant interactions are well-understood in principle~\cite{Mangano:2001iu,Mangano:2005cc, Grohs:2015tfy,Akita:2020szl,Froustey:2020mcq,Bennett:2020zkv}.  Neutrinos interact not only among themselves but can also exchange energy with electrons and positrons in the plasma.  Interaction with the plasma is complicated by the fact that there exist charged-current channels for electron-type neutrinos that are not present for the $\nu_\mu$ and $\nu_\tau$.  Neutrino decoupling occurs at temperatures around $m_e$, such that electron-positron annihilation takes place while neutrinos are going out of equilibrium.  Flavor oscillations among the neutrinos adds yet more complexity.

Even in the Standard Model, the neutrino decoupling process with flavor evolution included is a vexing problem. Mean-field quantum kinetic approaches to neutrino flavor evolution, as well as many-body alterations to the mean field picture, are frontier problems in computational physics. There has been ongoing rapid development of numerical techniques for solving quantum kinetic equations (QKEs)~\cite{2013PhRvD..87k3010V,2013PhRvD..88j5009Z,2014PhRvD..89j5004V} and accompanying physical understanding of this problem~\cite{Mangano:2005cc,EscuderoAbenza:2020cmq,Akita:2020szl,Froustey:2020mcq,Bennett:2020zkv,2021arXiv211011889F}.
\edit{We will neglect flavor oscillations in the analysis presented here, though the framework we develop can be extended to accommodate this additional physics.}

The problem becomes even more challenging when one introduces physics beyond the Standard Model.  New ingredients can alter the process in a myriad of complicated ways.  For example, cosmological asymmetry in lepton number~\cite{1991ApJ...368....1S,2002PhRvD..66a3008A,2002NuPhB.632..363D}, flavor oscillations among active and sterile states~\cite{2001PhRvD..64b3501A,2002APh....16..339D,2005PhRvD..72f3004A,2006PhRvL..97n1301K,Drewes:2016upu,2019PrPNP.104....1B,2019PhRvD.100b3533J,2020PhRvL.124h1802D,2020JCAP...06..008G,2020JCAP...09..051G,2020PhLB..80035113G,2024arXiv240213878F,Gariazzo:2019gyi}, out-of-equilibrium decays of massive particles~\cite{Scherrer:1987rr,Cyburt:2002uv,2011arXiv1110.6479F,2015PhRvD..92j3509P,Kawasaki:2017bqm,Rasmussen:2021kbf,2023PhRvD.108l3036A}, neutrino self-interactions~\cite{Bell:2005dr,Cyr-Racine:2013jua,Proceedings:2019qno,2019PhRvD.100b3533J,2020PhRvL.124h1802D,2020JCAP...07..001G,2022JCAP...09..036C,2024arXiv240213878F}, or scattering of neutrinos with new species~\cite{Mangano:2006mp,Laha:2013xua,Olivares-DelCampo:2017feq,2020JCAP...12..049L} could alter early universe neutrino transport.  The various observational handles on the cosmic neutrino background provided by primordial light element abundances, the cosmic microwave background, and large-scale structure can therefore be used to explore and constrain this new physics, but doing so requires precise calculation of neutrino transport in the early universe.

In the next section, we introduce a formalism that is well-suited to handle the complexities of neutrino transport in the early universe. This formalism, a generalized entropy approach, is built on Wigner operator conjugate-control variables. As we will demonstrate, this approach can capture the salient features of the phase space evolution of the cosmic neutrinos.  In later sections, we show how this generalized entropy approach allows for a compact yet highly precise description of the rich physics of neutrino transport in the Standard Model.  The techniques we employ are flexible enough to accommodate a wide range of phenomena expected in models of physics beyond the Standard Model.

%%%%%%%%%%%%%%%%
\section{Transport in the Generalized Entropy Approach}
\label{sec:TransportWigner}
%%%%%%%%%%%%%%%%

\subsection{Motivation and Grand Canonical Ensemble}

This section describes a generalized entropy approach to neutrino transport in the early universe. We use the term \lq\lq neutrino transport\rq\rq\ to denote a time-evolution calculation of the six  neutrino energy-momentum distribution functions in the Standard Model ($\nu_e$, $\bar\nu_e$, $\nu_\mu$, $\bar\nu_\mu$, $\nu_\tau$, $\bar\nu_\tau$). In a full mean field quantum kinetic equation (QKE) treatment this is a daunting problem. Our goal here is to develop a complementary approach, one that is based on generalized statistical techniques and that has been eminently successful in modeling the physics at the  photon decoupling epoch. The generalized entropy approach we develop, and hone on Boltzmann neutrino transport calculations, gives a description of the dynamically evolving system in the early universe based on physical constraints. The success of this treatment and the physical insights into neutrino decoupling it yields, all of which we present in subsequent sections, suggest that it could be applied to more sophisticated fully quantum mechanical neutrino transport calculations in future work. In addition, we anticipate an extension of this treatment to beyond the Standard Model (BSM) scenarios. To begin, we outline a generalized entropy approach that is inspired by full quantum mechanical many-body system calculations, but is applied here in a classical statistical guise.

As described above in Sec.\ \ref{sec:WeakDecoupling}, the neutrino component at very high temperature in the early universe will be in a thermal, chemically equilibrated distribution of neutrino phase space (momentum and position) occupation probabilities: a Fermi-Dirac (FD) distribution. As the universe expands and cools and the energy-dependent weak interactions of neutrinos slow down relative to the expansion, the phase space occupation probability distributions will deviate from the FD form. Our objective is to follow these deviations with a procedure that can capture the important aspects of detailed Boltzmann transport calculations, but does so in an efficient manner with as few parameters as possible. 

We have found a surprising result: the essential physical properties and effects as derived from detailed Boltzmann transport can be captured with just three parameters per species.
Refs.~\cite{Dolgov:1997mb,Esposito:2000hi} discuss a mathematical fitting procedure for the evolution of neutrino distribution functions. By contrast, our work here presents a statistical approach to this problem, one that is rooted in quantum transport and that allows for ready physical interpretation of fitting parameters.
We start with an exposition of this procedure directed at a quite general system, later specializing to the neutrino decoupling case.

In describing this general procedure, it will prove useful to think first about how one uses the grand canonical ensemble in equilibrium statistical mechanics to derive the occupation numbers per unit volume of phase-space.
After establishing that as a basis for comparison, we will generalize to non-equilibrium systems.
There exist multiple derivations of the occupation numbers per unit volume of phase-space -- called the distribution function $n$ -- in equilibrium statistical mechanics.  For our purposes of exposition, the general $n$ is a function of kinematic variables (position and momentum, i.e., ${\bf x, p}$) and quantum numbers (e.g., spin, flavor, lepton number, etc.).

One approach to derive $n$ employs the use of Lagrange multipliers with generalized operator-charges assuming a form for the entropy distribution $S$ in terms of $n$.  For concreteness, we label the operator-charges as $Q_A$ with index $A$ and associate to each $Q_A$ a Lagrange multiplier $\alpha_A$.
In the example of a fermionic system in equilibrium subject to no external constraints, only two operator-charges are needed, namely, $Q_0=\sum n$ for total particle number and $Q_1=\sum En$ for total energy, where the summation is over all of the independent kinematic and quantum variables which describe $n$.  The form of the entropy for this system is $S=-\sum [n\ln n+(1-n)\ln(1-n)]$ derived from independent particle combinatorics.  In this specific fermionic system without external constraints, the only kinematic variable is the energy $E=\sqrt{|{\bf p}|^2+m^2}$ for particle mass $m$.

The equilibrium condition is equivalent to the extremization of the entropy.  By proceeding to extremize the entropy with the Lagrange multipliers on $Q_0$ and $Q_1$, one can deduce the form of $n$ as the standard Fermi-Dirac distribution
\begin{equation}\label{eq:FD1}
  n_{\{\ell\}}(E) = \dfrac{1}{1+e^{(E-\mu)/T}} \, ,
\end{equation}
where the subscript ${\{\ell\}}$ denotes the set of quantum numbers $\{\ell\}$.  
Identifying the values of the Lagrange multipliers as
\begin{equation}
  \alpha_0 = -\frac{\mu}{T}\, , \quad \alpha_1=\frac{1}{T}\, ,
\end{equation}
where $\mu$ and $T$ are the chemical potential and temperature, respectively, makes the connection to classical thermodynamics and the grand potential.

In writing Eq.\ \eqref{eq:FD1}, we have assumed no dependence of the quantum numbers $\{\ell\}$ on the distribution function $n_{\{\ell\}}$ -- the quantum numbers simply supply a label for the entropy summation.  This implicit assumption that the quantum numbers only act to give the degeneracy in a given phase-space volume need not hold for a fermionic system under external constraints.  Indeed, if the kinematics depend on quantum numbers, we would need to adopt different temperature and chemical potential variables for each set of quantum numbers.  In fact, this is the case for the fermionic system of neutrinos in the early universe with an asymmetry in lepton number.  Momentum-changing interactions between neutrinos tend to bring all species to the same temperature $T$.  On the other hand, momentum-preserving-interactions would tend to bring all neutrinos to the same chemical potential, $\mu$, while simultaneously bringing all anti-neutrinos to the same chemical potential opposite in sign to the neutrinos, $-\mu$.  In this case, the temperature is the same, but the chemical potential depends on the quantum lepton number.

As emphasized in Sec.\ \ref{sec:WeakDecoupling}, the parameters describing the distribution function for each neutrino species evolve slowly with time until the rates of neutrino scattering begin to fall below the expansion rate of the universe and the neutrino component as a whole can no longer be described adequately by an equilibrium approximation. The out-of-equilibrium scattering of neutrinos even in the SM, and similarly in BSM physics, can lead to spectral distortions away from the thermal FD blackbody 2-parameter solution of Eq.\ \eqref{eq:FD1}.
Capturing the effects of these spectral distortions will require more than the two parameters $\alpha_0$ and $\alpha_1$.

\subsection{Entropy-content operator}

Consequently, we pursue an alternative procedure that we term the generalized entropy approach.  We stress that this approach is general and can be used for other systems, although we will tailor it to capture moments of the early universe neutrino energy distributions.
Inspired by a quantum-mechanical approach, we define a density operator $\widehat{\rho}$
\begin{equation}\label{eq:rho1}
  \widehat{\rho} = \sum_{k,k^{\prime}} w_{k,k^{\prime}}|k\rangle\langle k^{\prime}|.
\end{equation}
In writing Eq.\ \eqref{eq:rho1}, the density operator refers to an ensemble of multi-particle states.  This ensemble may contain many different kinds of multi-particle states, i.e., fully-uncorrelated states, fully-entangled states, or a combination of the two.  Each state has a well defined number of particles, $\mathcal{N}$, and total energy, $\mathcal{E}$, but not all states have the same $\mathcal{N}$ and $\mathcal{E}$.  As a result, we need to consider elements of a Fock space to describe the ensemble.  In quantum mechanics, each $\mathcal{N}$-particle state would have an associated wavefunction, which we denote as the ket $|\psi\rangle$, and can be written using a basis of wavefunctions in a $3\mathcal{N}$ Hilbert space.  The elements of the Fock space which we need to describe the ensemble are denoted as $|k\rangle$ and consist of the union of bases from each $3\mathcal{N}$ Hilbert space.  We stay general and assume that $|k\rangle$ is not coincident with the eigenbasis of $\widehat{\rho}$. Hence, this is why we need the double summation in Eq.\ \eqref{eq:rho1}. Therefore, $w_{k,k^{\prime}}$ gives the density matrix element $\langle k|\widehat{\rho}|k^{\prime}\rangle$.
Very importantly: $\widehat{\rho}$ contains off-diagonal elements.
A member of the ensemble of $\widehat{\rho}$ need not be limited to a single multi-particle wavefunction, or even a linear superposition of multi-particle wavefunctions in a single $3\mathcal{N}$ Hilbert space, but instead could be a superposition of states in the general Fock space.  In this work, we consider the former possibility but neglect the later.  In other words, $\widehat{\rho}$ is a block-diagonal matrix in the Fock space, where each block corresponds to a $3\mathcal{N}$ Hilbert sub-space.
We borrow the notion of superposition in a Hilbert space to write the quantity $|\Psi\rangle$ as a superposition of ensemble members in the Fock space.  We will call this quantity a ``configuration'' to discriminate it from the quantum-mechanical concept of wavefunction, although we utilize the bra-ket notation of quantum mechanics.
For some configuration $|\Psi\rangle$, we can write the weights as $w_{k,k^{\prime}}=\langle k|\Psi\rangle\langle\Psi|k^{\prime}\rangle$ and normalize them such that the trace of $\widehat{\rho}$ is unity, i.e., ${\rm Tr}[\widehat{\rho}]\equiv\sum_{k}\langle k|\widehat{\rho}|k\rangle=1$.

Now that we have oriented ourselves in the statistical Fock space, the next step of the generalized entropy approach is to promote the operator-charges $Q_A$ to operators $\widehat{Q}_A$ in the Fock space, and associate to each operator a control parameter $\alpha_A$, analogous to the Lagrange multipliers in the equilibrium treatment.  To provide another example, $\widehat{Q}_N$ may be the number operator: a summation of tensor products of appropriate creation and annihilation operators $\widehat{a}_{\boldell}^{\dagger}\widehat{a}_{\boldell}$ with identity operators
\begin{equation}\label{eq:num_op}
  \widehat{Q}_N = \sum\limits_{\boldell}\left(\widehat{a}_{\boldell}^{\dagger}\widehat{a}_{\boldell} \otimes\prod\limits_{\boldell^{\prime}\ne \boldell}\widehat{I}_{\boldell^{\prime}}\right),
\end{equation}
which would give the number of particles in a many-body state $|k\rangle$.
In Eq.\ \eqref{eq:num_op}, we have used the shorthand notation $\boldell=\{\ell\},\mathbf{x}$ to denote a single particle state in the position basis (although we could use $\mathbf{p}$ instead to write the state in the momentum basis).
A member of the Fock space which is also a member of a $3\mathcal{N}$ Hilbert space is an eigenvector of $\widehat{Q}_N$, namely, $\widehat{Q}_N|\psi\rangle = \mathcal{N}|\psi\rangle$.  A configuration in the Fock space is in general not an eigenvector.  However, we can write
\begin{equation}
\langle\Psi|\widehat{Q}_N|\Psi\rangle\equiv\langle\widehat{Q}_N\rangle_{\Psi} = N,
\end{equation}
where we have defined the expectation value of $\widehat{Q}_N$.  In this case, for the configuration $|\Psi\rangle$, we would expect $N$ number of particles when selecting a multi-particle state from $|\Psi\rangle$.  In other words, $N$ is the average number of particles in the configuration.

In addition, note that the control parameters $\alpha_A$ are not Lagrange multipliers.  They will be functions of the various conditions of the environment under study, including time.

Using the operators $\widehat{Q}_A$, we define the entropy content operator which drives a system from an $(i)$nitial configuration, e.g., $|\Psi_i\rangle$ to a $(f)$inal one ($|\Psi_f\rangle$)
\begin{equation}\label{eq:eco}
  \widehat{s}(\{\alpha^{(fi)}_A\}) = \left[\sum_A   \alpha^{(fi)}_A  \widehat{Q}_A\right] - \mathcal{F}(\{\alpha_A^{(fi)}\})\widehat{I},
\end{equation}
under a specific realization of the control parameters $\{\alpha^{(fi)}_A\}$.  These control parameters are specific to the initial and final configurations, $|\Psi_i\rangle$ and $|\Psi_f\rangle$, or equivalently $\widehat{\rho}_i$ and $\widehat{\rho}_f$.
$\mathcal{F}$ is a complex number and a functional of the control parameters. We scale it with the Fock space identity operator and explain its role below after we have introduced other key concepts.

The transformation under $\{\alpha^{(fi)}_A\}$ will take an initial operator $\widehat{\rho}_i$ to the final operator $\widehat{\rho}_f$ in the following manner
\begin{equation}\label{eq:transform1}
  \widehat{\rho}_f = \exp[-\widehat{s}(\{\alpha^{(fi)}_A\})/2] \,\widehat{\rho}_i\, \exp[-\widehat{s}(\{\alpha^{(fi)}_A\})/2].
\end{equation}
We consider a set of transformations above such that the trace of the density matrix is preserved, i.e., ${\rm Tr}[\widehat{\rho}_f]={\rm Tr}[\widehat{\rho}_i]=1$ for our normalization.
Note we use the same operator in the argument of the exponentials, implying the transformation is not necessarily unitary for an arbitrary choice of $\widehat{s}$.  For the trace to be preserved, we must have the following condition
\begin{equation}
 {\rm Tr}[e^{-\widehat{s}}\widehat{\rho}_i]=1,\label{eq:trace1}
\end{equation}
where we have used the cyclic property of the trace, the commutativity of identical operators, and abbreviated the exponential operator for brevity in notation.  If we associate the configuration $|\Psi_i\rangle$ to the density operator $\widehat{\rho}_i$, we can show
\begin{align}
  1
  &=\langle\Psi_i|e^{-\widehat{s}}|\Psi_i\rangle \nonumber \\
  &\equiv \langle e^{-\widehat{s}}\rangle_i \, .
  \label{eq:ens_avg}
\end{align}
By Jensen's inequality
\begin{equation}
  \langle e^{-\widehat{s}}\rangle_i \ge e^{-\langle\widehat{s}\rangle_i},
\end{equation}
implying a thermodynamic-second-law analog for the entropy-content operator
\begin{equation}
  \langle \widehat{s}\rangle_i\ge 0.
\end{equation}

We return to the definition of $\widehat{s}$ in Eq.\ \eqref{eq:eco} to discuss the functional $\mathcal{F}$.  Our goal is to use control parameters $\{\alpha_A^{(fi)}\}$ to follow the deformation of the density operator.  However, the first term on the right-hand side of Eq.\ \eqref{eq:eco} does not preserve unitarity for general deformations.  As a result, we introduce a normalization factor, i.e., $\mathcal{F}\widehat{I}$, to ensure the unitarity of the transformation in Eq.\ \eqref{eq:transform1}.  The identity operator commutes with any/all Fock-space operators $\widehat{Q}_A$.  Using Eq.\ \eqref{eq:ens_avg}, we can show
\begin{align}
  e^{\mathcal{F}} &= \left\langle\exp\left[-\sum \alpha_A^{(fi)}\widehat{Q}_A\right]\right\rangle_i^{-1} \, ,\\
  \implies\mathcal{F} &= -\ln\left\{\left\langle\exp\left[-\sum \alpha_A^{(fi)}\widehat{Q}_A\right]\right\rangle_i\right\} \nonumber \\
  &= -\ln\left\{\sum\limits_{k,k^\prime}w_{k,k^{\prime}}\left\langle k\biggr|\exp\left[-\sum \alpha_A^{(fi)}\widehat{Q}_A\right]\biggr| k^{\prime}\right\rangle\right\}.\label{eq:fe1}
\end{align}
To explain the significance of Eq.\ \eqref{eq:fe1}, we return to equilibrium statistical mechanics.  One can write the Grand Canonical Partition function as the following
\begin{equation}\label{eq:gpf}
  \mathcal{Z} = \sum_i \left\langle \psi_i\biggr|\exp\left(-\frac{\widetilde{H}_i - \mu\widetilde{N}_i}{T}\right)\biggr|\psi_i\right\rangle
\end{equation}
for the ensemble of states indexed by $i$, with associated multi-particle wavefunctions $|\psi_i\rangle$, and quantum-mechanical Hamiltonian/Number operators $\widetilde{H}_i/\widetilde{N}_i$ for the $i^{\rm th}$ Hilbert space.  Upon examination of Eq.\ \eqref{eq:gpf} and the argument in the natural logarithm of \eqref{eq:fe1}, we see the connection between the generalized-entropy approach and the grand canonical ensemble.  In equilibrium statistical mechanics, the grand potential, $\Phi_G$, is a generalization of the Helmholtz free energy from the canonical ensemble.  $\Phi_G$ is related to the grand canonical partition function by the following expression
\begin{equation}
 \Phi_G = -T\ln\mathcal{Z}.
\end{equation}
Identifying the argument of the natural logarithm in Eq.\ \eqref{eq:fe1} as a non-equilibrium analog of $\mathcal{Z}$ from Eq.\ \eqref{eq:gpf}, we will call $\mathcal{F}$ the free-energy generating functional.

One may be tempted to continue the analogy and extend the control parameters $\alpha_A^{(fi)}$ to the thermodynamic variables $1/T$ and $-\mu/T$ if/once the system reaches an equilibrium state.  This would be incorrect in our formalism.  In the grand canonical ensemble, all states of the combined system and reservoir are equally likely, and physical processes act to exchange momentum, particle number, quantum numbers, etc., to change the system from one microstate to another.  In our formalism, the control parameters act to drive the initial configuration into a final one.  If the final density matrix is diagonal with coefficients which scale with the number of states the reservoir can obtain, then the system has reached a state of equilibrium.  If $\widehat{\rho}_f$ is in equilibrium, the system stays in this configuration in the absence of external influences\footnote{It is possible for there to be generalized external forces on a system.  In that case, however, the entropy content operator is not Hermitian and ${\rm Tr}[\widehat{\rho}_f]\ne{\rm Tr}[\widehat{\rho}_i]$.  We leave this possibility for exploration in future work.}.  As a result, the control parameters are all zero and trivially $\mathcal{F}=0$.
For the rest of this subsection, we will drop the superscript label $(fi)$ from the control parameters for brevity in notation, although stressing that the $\alpha_A$ always take an initial configuration into a final one.
Following the analogy of the GCE, we will take derivatives of Eq.\ \eqref{eq:fe1} with respect to the control parameters to determine various thermodynamic/statistical analogs.  However, the $\alpha_A$ are functions of time, and also position perhaps.  Therefore, $\mathcal{F}$ is a functional and we need a functional derivative.  As no covariant derivatives, i.e., $\partial_\nu\alpha_A^{\nu}$, appear in Eq.\ \eqref{eq:fe1}, the functional derivative mimics a partial derivative.

Functional derivatives of the free-energy generating functional with respect to the $\alpha_A$ parameters provide the connected correlation functions of the operators $\widehat{Q}_A$
\begin{equation}
  \left. \frac{\delta^M \mathcal{F}(\{\alpha_A^{(fi)}\})}{\delta\alpha_{m_1}...\delta\alpha_{m_M}} \right|_{\alpha_A = 0} = {(-1)^{M+1}} \left\langle \widehat{Q}_{(m_1}...\widehat{Q}_{m_M)}\right\rangle_{i,\mathrm{cc}} \, ,
  \label{eq:ConnectedCorrelators}
\end{equation}
where the subscript `cc' denotes connected component, and the parentheses in the subscripts indicates symmetrization over the indices.  The free-energy generating functional therefore plays the same role as the natural log of the partition function in quantum field theory and statistical mechanics, or the cumulant generating functional in statistics.  In Appendix~\ref{app:ConnectedComponents}, we provide justification for this statement and show explicit derivation of the low-order connected correlation functions from the free-energy generating functional.

\subsection{Connection to Wigner functions and the Mean field}

Our formalism for calculating the deformation of a systems' distribution function is completely general to this point. In principle, it could encompass the dynamics of systems in Fock space with inherent quantum coherence and entanglement. This generality may be necessary for following the energy distribution function evolution in the neutrino decoupling epoch if, for example, we include neutrino flavor transformation or various BSM scenarios involving macroscopic coherence. In a general approach the next step would be to re-configure the problem by using quantum correlations and translation symmetries to reduce two-point positions into an average position and momentum representation. This could be done via a generalized $N$-particle Wigner transformation. 
Although the equivalence of this $N$-particle approach to full quantum mechanics may seem obvious, it was controversial in its day, suggested by Weyl and by J. Groenewold~\cite{Groenewold:1946kp}.
Moyal and collaborators further developed this phase space formulation of quantum mechanics~\cite{Moyal:1949sk},
and more recently C. Zachos, D. Fairlie, and T. Curtright~\cite{Zachos:2005gri}
explored it in more detail.

However, for our example case of neutrino decoupling with no assumed quantum flavor coherence, our task is much simpler. We seek to find phase-space representations for the neutrino wavefunctions via a generalized $N$-particle Wigner transformation as alluded to above.  Our configurations in Fock space are for multiple particles and anti-symmeterized to comply with FD statistics.  As our goal at this point is only to find single-particle correlation functions, we will employ a mean field statistical approach to reduce the $N$-particle quasi-distribution to a single-particle correlation function.  These objects will be necessary for evaluating the functional derivatives and operators in \eqref{eq:ConnectedCorrelators}.

We give a sketch of how we construct the single-particle correlation function.  First, we start with a generic multi-particle state $|\psi\rangle$.  If we write the multi-particle wavefunction as a tensor product of single-particle states, we can arbitrarily pick an entry in the tensor product which we label as the $j^{\rm th}$ particle.  
We will take a generalized $N$-particle Wigner transform on the outer product $|\psi\rangle\langle\psi|$.
To calculate the matrix element of $|\psi\rangle\langle\psi|$ in the integrand of the Wigner transform, we write a tensor-product of position-quantum-number eigen-kets and bras, where the $j^{\rm th}$ entry would have $|\mathbf{X}_j+\delta\mathbf{x}_j, s_j, f_j\rangle$, and $\langle\mathbf{X}_j-\delta\mathbf{x}_j, s_j^{{\prime}}, f_j^{{\prime}}|$, for spins $s$ and flavors $f$.  
If we take the Fourier conjugate of $\delta {\bf x}_j $ to be the momentum ${\bf p}_j $, we can calculate the generalized $N$-particle Wigner transform as a function of 6$N$ kinematic variables and 4$N$ quantum variables.  We marginalize, i.e., integrate/sum over all of the other phase-space/quantum numbers, to obtain the single-particle Wigner function, which we denote as $W_{s_j^{\prime} f_j^{\prime}; s_j f_j} ({\bf X}_j , {\bf p}_j , t)$.  The Wigner transform and associated function yield a phase-space distribution from input single-particle quantum states.  
With the phase-space distribution and the free-energy generating functional $\mathcal{F}$, we can proceed to derive a form for the distribution in terms of the $\alpha_A$ control parameters.

\subsection{Free-energy generating Functional for Fermions}

If we start in an initial configuration $i$ we will have a set of occupation numbers and corresponding constrained expectation values $\langle \widehat{Q}_A \rangle$ of the generalized charges.
The entropy-content operator that will generate the change in the distribution functions in going from the initial configuration to a final \lq\lq deformed\rq\rq configuration $f$ is 
\begin{equation}
 \widehat{s}_{fi}\vert_\alpha \equiv \widehat{s}(q_i; \{\alpha_A^{(fi)}\}) = \sum\limits_A   \alpha_A^{(fi)}  \widehat{Q}_A (q_i) - {\cal F}(\{ \alpha_A^{(fi)} \})\widehat{I}  \, ,  
\end{equation}
where $q_i$ denote the set of parameters characterizing the initial configuration of the system (e.g., a set of single particle occupation probabilities for each neutrino type) and $\{\alpha_A^{(fi)}\}$ are a corresponding set of deformation parameters.
Hereafter, we again drop the superscript $(fi)$ on the $\alpha$ parameters for brevity.
For a given realization of a set of $\alpha$ parameters, $\{ \alpha\}$, the mean of the entropy content operator (hereafter the \lq\lq mean entropy\rq\rq) is 
\begin{equation}
 \langle \widehat{s}_{fi} \rangle \vert_\alpha = \sum   \alpha_A \langle  \widehat{Q}_A \rangle \vert_\alpha -{\cal F}(\{ \alpha_A \}) \, ,  
\end{equation}
and the entropy fluctuation from the mean is the difference
\begin{equation}
 \delta \widehat{s}_{fi}(q)\vert_\alpha = \widehat{s}_{fi}(q)\vert_\alpha - \langle \widehat{s}_{fi}\rangle \vert_\alpha\widehat{I} = \sum   \alpha_A \left( \widehat{Q}_A (q) -\langle  \widehat{Q}_A \rangle \vert_\alpha\widehat{I} \right) \, ,
\end{equation}
where again $\langle  \widehat{Q}_A \rangle \vert_\alpha$ denotes an expectation value for ${\widehat{Q}}_A$ for a given set $\{\alpha\}$.

The (non-equilibrium) evolution equation for the density operator $\widehat{\rho}_i$ is the familiar $N$-point quantum Liouville equation, or master equation when written in terms of matrix elements.
Here, for our problem of neutrino energy distribution evolution in the early universe, we will make the usual approximation of mean-field evolution of neutrinos so that neutrinos can be treated as quasiparticles.  For example, this is the usual approach in neutrino flavor transformation in the early universe and other astrophysical environments ~\cite{Mikheyev:1985abc,Wolfenstein:1978abc,Wolfenstein:1979abc}.

The initial density matrix, $\widehat{\rho}_i$, is in an equilibrium configuration, where the many-body wave functions are appropriately weighted in the GCE (i.e., both thermal and chemical equilibrium), our starting point for the transport and evolution of the neutrino distribution functions.  Full transport here is captured with a time-sequence of non-equilibrium density operators. We have used the transformation in Eq.\ \eqref{eq:transform1} and the symbol $\widehat{\rho}_f$ for the deformed density operator.  At this point, we will denote the time-sequence of deformed density operators as $\widehat{\rho}(t)$, with the initial condition $\widehat{\rho}(t=t_i)=\widehat{\rho}_i$. The code \burst{} has been used to evolve the neutrino distribution functions using appropriate collision kernels \cite{Grohs:2015tfy}.  Here, our intent is to capture that result in terms of the evolution of an appropriate expansion of mathematical objects derived from the density operator $\widehat{\rho}(t)$, namely Eq.\ \eqref{eq:ConnectedCorrelators}.
We cast the density operator in terms of a single-particle representation.  This representaion follows from an underlying independent quasiparticle state basis, where each pure state in this basis has been Wigner-transformed in terms of a phase space, i.e., ${\bf X}$, ${\bf p}$.
We can summarize this picture with the following form of the density operator 
\begin{equation}\label{eq:rho_sps}
    \widehat{\rho}(t)
    = e^{\cal F} \bigotimes\limits_{\mathbf{X}\mathbf{p}s^{\prime}f^{\prime}sf} e^{- \alpha \widehat{n} }\, ,\\
\end{equation}
where the spin and flavor labels of the control parameters $\alpha_{s^\prime f^\prime; sf}(t;\mathbf{X},\mathbf{p})$ and the phase-space number operator $\widehat{n}_{s^\prime f^\prime; sf}(\mathbf{X},\mathbf{p})$ as well as their dependence on time, $\mathbf{X}$ and $\mathbf{p}$ are implicit.
In writing Eq.\ \eqref{eq:rho_sps}, we have put in two conventions.  Firstly, we have placed time dependence in the control parameters.  Recall that as defined, the control parameters are used for deformations of the density operator, and are zero when $\widehat{\rho}_f$ remains in equilibrium.  In Eq.\ \eqref{eq:rho_sps}, the initial $\alpha(t=t_i)$ are used to specify the initial density operator (e.g., a FD equilibrium configuration) and $\alpha(t>t_i)$ give deviations from that initial configuration.
Secondly, we no longer have the subscript $A$ used to associate a control parameter with a Fock space operator $\widehat{Q}_A$. 
Now, we have used the quantum numbers and phase-space variables to specify a control parameter for each single-particle quantum state.

Using the unitarity of the trace, we can derive an expression for the mean value of $\langle \widehat{n}_{s^{\prime}f^{\prime};sf}(\mathbf{X},\mathbf{p})\rangle$.
\begin{align}
   1 = {\rm Tr}[\widehat{\rho}(t)] &= {\rm Tr}\left[ e^{\cal F} \bigotimes\limits_{\mathbf{X}\mathbf{p}s^{\prime}f^{\prime}sf} e^{- \alpha \widehat{n} }\right] \nonumber \\
   &= e^{\cal F}{\rm Tr}\left[  \bigotimes\limits_{\mathbf{X}\mathbf{p}s^{\prime}f^{\prime}sf} e^{- \alpha \widehat{n} }\right] \nonumber \\
   &= e^{\cal F}  \prod\limits_{\mathbf{X}\mathbf{p}s^{\prime}f^{\prime}sf} {\rm Tr}[e^{- \alpha \widehat{n} }]
\end{align}
For particles obeying Fermi-Dirac statistics, the number operator for each cell in phase space has two distinct eigenvalues: 0 and 1.  Thus, the trace over the exponentiated number operator is
\begin{equation}
   {\rm Tr}[e^{-\alpha \widehat{n}}] = 1 + e^{-\alpha},
\end{equation}
which implies the trace expression becomes
\begin{equation}
   1 = e^{\mathcal{F}}\prod\limits_{\mathbf{X}\mathbf{p}s^{\prime}f^{\prime}sf} [1+e^{- \alpha \widehat{n} }]
\end{equation}
Solving for $\mathcal{F}$ gives
\begin{align}
    {\cal F }(t)
    &= -  \sum_{\mathbf{X}\mathbf{p}s^{\prime}f^{\prime}sf} [ \ln (1+e^{-\alpha})] \, ,
\end{align}
which we can substitute back into Eq.\ \eqref{eq:rho_sps} to yield
\begin{align}
    \widehat{\rho}(t)
    &= \exp\left[-\sum\limits_{\mathbf{X}\mathbf{p}s^{\prime}f^{\prime}sf}\ln(1+e^{-\alpha})\right] \bigotimes\limits_{\mathbf{X}\mathbf{p}s^{\prime}f^{\prime}sf} e^{- \alpha \widehat{n} } \nonumber \\
    &= \prod\limits_{\mathbf{X}\mathbf{p}s^{\prime}f^{\prime}sf}\exp\left[-\ln(1+e^{-\alpha})\right] \bigotimes\limits_{\mathbf{X}\mathbf{p}s^{\prime}f^{\prime}sf} e^{- \alpha \widehat{n} } \nonumber \\
    &= \prod\limits_{\mathbf{X}\mathbf{p}s^{\prime}f^{\prime}sf}(1+e^{-\alpha})^{-1} \bigotimes\limits_{\mathbf{X}\mathbf{p}s^{\prime}f^{\prime}sf} e^{- \alpha \widehat{n} } \nonumber \\
    &=  \bigotimes\limits_{\mathbf{X}\mathbf{p}s^{\prime}f^{\prime}sf} (1+e^{-\alpha})^{-1}e^{- \alpha \widehat{n} } \, .
    \label{eq:rho_direct_product}
\end{align}
We can find the expectation value for a given phase-space cell $(\mathbf{X}_1\mathbf{p}_1s^{\prime}_1f^{\prime}_1s_1f_1)$ by the following
\begin{align}
  \langle \widehat{n}_{\mathbf{X}_1\mathbf{p}_1s^{\prime}_1f^{\prime}_1s_1f_1}\rangle &= {\rm Tr}[\widehat{n}\widehat{\rho}] \nonumber \\
  &= {\rm Tr}\left[ \left( \frac{\widehat{n}e^{- \alpha \widehat{n}}}{ 1+e^{-\alpha}}\right)_{\mathbf{X}_1\mathbf{p}_1s^{\prime}_1f^{\prime}_1s_1f_1} \otimes \bigotimes\limits_{\mathbf{X}\mathbf{p}s^{\prime}f^{\prime}sf \neq \mathbf{X}_1\mathbf{p}_1s^{\prime}_1f^{\prime}_1s_1f_1} \frac{e^{- \alpha \widehat{n}}}{1+e^{-\alpha}} \right] \nonumber \\
  &=\left(\frac{e^{-\alpha}}{1+e^{-\alpha}} \right)_{\mathbf{X}_1\mathbf{p}_1s^{\prime}_1f^{\prime}_1s_1f_1} \nonumber \\ 
  &=\left(\frac{1}{1+e^{\alpha}} \right)_{\mathbf{X}_1\mathbf{p}_1s^{\prime}_1f^{\prime}_1s_1f_1} \, .
  \label{eq:exp_val_n}
\end{align}
This result can be rearranged to give $\alpha_{\mathbf{X}_1\mathbf{p}_1s^{\prime}_1f^{\prime}_1s_1f_1}$ in terms of $\langle \widehat{n}_{\mathbf{X}_1\mathbf{p}_1s^{\prime}_1f^{\prime}_1s_1f_1} \rangle$ as
\begin{equation}
    \alpha_{\mathbf{X}_1\mathbf{p}_1s^{\prime}_1f^{\prime}_1s_1f_1} = \left[\ln(1-\langle \widehat{n} \rangle) -\ln \langle \widehat{n} \rangle\right]_{\mathbf{X}_1\mathbf{p}_1s^{\prime}_1f^{\prime}_1s_1f_1} \, .
\end{equation}
The density matrix then can be written as
\begin{align}
    \rho(t) 
     &= e^{\cal F} \bigotimes_{\mathbf{X}\mathbf{p}s^{\prime}f^{\prime}sf} \exp \left( \left[ \widehat{n} \ln \langle \widehat{n} \rangle - \widehat{n} \ln (1-\langle \widehat{n} \rangle ) \right]\right) \, ,
\end{align}
and the free energy can then be expressed as
\begin{align}
    {\cal F } (t)
    &=  \sum_{\mathbf{X}\mathbf{p}s^{\prime}f^{\prime}sf} [ \ln (1-\langle \widehat{n} \rangle )]  \, ,
 \end{align}
such that the density matrix is
\begin{align}
    \rho(t)&=  \bigotimes_{\mathbf{X}\mathbf{p}s^{\prime}f^{\prime}sf} \exp  \left[ \widehat{n} \ln \langle \widehat{n} \rangle + (1-\widehat{n}) \ln (1-\langle \widehat{n} \rangle ) \right] \, .
\end{align}
The mean Wigner matrix can be obtained as the derivative of the free energy with respect to $\alpha$
\begin{equation} 
    \langle \widehat{n} \rangle \vert_\alpha = \frac{\partial {\cal F}} {\partial \alpha} = (e^\alpha +1)^{-1} \, ,
\end{equation}
and the mean entropy is given by the usual expression
\begin{align}
 \langle \widehat{s}_{fi} \rangle \vert_\alpha (t)
 &=  \sum_{\mathbf{X}\mathbf{p}s^{\prime}f^{\prime}sf} \alpha \langle \widehat{n} \rangle \vert_\alpha -{\cal F}  
 = -\sum_{\mathbf{X}\mathbf{p}s^{\prime}f^{\prime}sf} \left[ \langle \widehat{n} \rangle \ln \langle \widehat{n} \rangle + (1-\langle \widehat{n} \rangle) \ln (1-\langle \widehat{n} \rangle ) \right] \, .
\end{align}

Within the independent particle approximation, we can express the full density matrix in terms of a set of density matrices deformed by $\alpha_{s^\prime f^\prime; sf}(t;\mathbf{X},\mathbf{p}) $. That is, the transport is fully encoded by these control parameters, but there are as many control parameters as there are dimensions of phase space.  This change of perspective therefore retains all of the information, but it also retains all of the complexity of describing the evolution of each of the individual particles in the system.

In many situations, however, it is possible to describe the relevant features of the system with a much smaller set of parameters.  When such a compression is possible, we can express the evolution of the system in terms of a smaller set of deformation parameters $\{\alpha_A\}$.
Therefore, at a given time $t$ we can explore $\rho(t;q,q^\prime) = \rho (q,q^\prime) \vert_{\alpha_A(t)} $ via its deformations by imposing various applied $\alpha_A$, where the  $\alpha_A$ would be a truncated set relative to the full $\alpha$ phase-space control-matrix.
This set of control parameters $\alpha_A$ specifies the constraints to which the system is subjected at any fixed time.
The interesting question is how close we can get to an accurate description of the full evolution of the system $\rho(t+\delta t)/\rho(t)$ by using a truncated set of evolving control parameters $\alpha_A (t+\delta t) $. If the number of variables $A = 0,1,\ldots$ is small then we have reduced the transport problem considerably. Of course this is familiar if we have the densities described by a sequence of equilibrium states, characterized at each time by a chemical potential  $\alpha_0 = -\beta \mu$ and an inverse temperature $\alpha_1 = \beta E$.  This equilibrium case provides an example of how the salient features of the evolution can be captured with just a few parameters even when the system contains many degrees of freedom.  We will now apply this formalism to the case of neutrino decoupling.

The energy dependence of the neutrino cross sections implies that low energy neutrinos are not as tightly coupled as high energy ones. This might be thought to motivate a simple near-equilibrium distribution at high energy with a modification at low energy. Instead we have found that the expansion of  $\alpha_{\mathbf{X}\mathbf{p}s^\prime f^\prime sf}(t)$ in powers of momenta to one order higher than the equilibrium case captures the essence of the Standard Model C$\nu$B. Allowing for control parameters which describe the number density current and energy-momentum current, along with one additional control parameter related to the current of the energy-momentum variance, the expansion takes the form
\begin{equation}
\alpha_{s^\prime f^\prime; sf}(\mathbf{X}\mathbf{p}) = \alpha_{(0) \mu \,  s^\prime f^\prime; sf } \frac{p^\mu}{E} +     \alpha_{(1) \mu \nu \, s^\prime f^\prime; sf } \frac{p^\mu p^\nu} {\tcm E}  + \alpha_{(2) \mu \nu \lambda \, s^\prime f^\prime; sf   } \frac{p^\mu p^\nu  p^\lambda}{\tcm^2 E}  + \ldots \ . 
\end{equation}
This expansion remains quite general and could, in principle, be applied to a wide array of problems, including neutrino transport in neutron stars and supernovae (in which case the factors of $\tcm$ appearing in this expression would be replaced by a characteristic energy scale relevant to the problem).
However, for describing the cosmic neutrino background, the homogeneous, isotropic, and hot conditions of the early universe lead to further simplifications of this expression.
Cosmic neutrinos are ultra-relativistic in the early universe such that $E \simeq |p|$, and we further focus on the case diagonal in spin and flavor, leaving the problem of describing neutrino flavor oscillations within this framework to future work.
As a result, this expression simplifies to the form
\begin{equation}
    \alpha = \alpha_{(0)} + \alpha_{(1)} \epsilon + \alpha_{(2)} \epsilon^2 + \cdots \, ,
\end{equation}
for each neutrino species.
A new ingredient is  $\alpha_{(2)}$, which is, in the Lagrange multiplier framework, indicating that fluctuations in the total energy $\langle (\Delta U)^2 \rangle$ are constrained as well as the number and energy densities. As our figures show, a description of the system using just $\alpha_{(0)}$ and $\alpha_{(1)}$ is in fact quite accurate over much of the neutrino decoupling epoch, but adding $\alpha_{(2)}$ considerably sharpens the agreement at low $\Tcm$.

How do we measure the accuracy of such a radical compression of full transport into just three variables per neutrino species?
A very general approach is based on the the relative entropy (also called the negative of the Kullback-Liebler divergence) between the two distributions, which for fermionic systems takes the form  
\begin{align}
D (a \Vert b)  &= 
      - \sum_{\mathbf{X}\mathbf{p}s^{\prime}f^{\prime}sf} \left[ \langle \widehat{n} \rangle \vert_{\alpha_a} \ln \frac{ \langle \widehat{n} \rangle \vert_{\alpha_a} }{ \langle \widehat{n} \rangle \vert_{\alpha_b} } +  (1-\langle \widehat{n} \rangle \vert_{\alpha_a}) \ln \frac{ (1-\langle \widehat{n} \rangle \vert_{\alpha_a} ) }{ (1-\langle \widehat{n} \rangle \vert_{\alpha_b}  ) } \right] \, .
\end{align}
This treatment can be used to assess the similarity of the deformed state that results from the full transport calculation $\alpha_a$ (in our case provided by the \burst{} code) to the state that is approximated by our three-parameter fit $\alpha_b$.  This prescription works even if the distributions being compared are far from equilibrium.  In practice, for the near-equilibrium evolution of the Standard Model thermal history, we use a goodness-of-fit parameter that measures the weighted mean square deviation of the distribution at each sampled point throughout the decoupling epoch; see Sec.~\ref{sec:SMnu}.

Before applying this formalism to the \cnub, we will briefly discuss an application of the same formalism to the (perhaps more familiar) case of CMB transport.

%%%%%%%%%%%%%%%
\section{Relation to CMB transport}
\label{sec:CMBTransport}
%%%%%%%%%%%%%%%

The closest analogue of the calculations we are doing here is to the development of spectral distortions of the CMB~\cite{Zeldovich:1969ff,Sunyaev:1970er,Burigana:1991eub,Hu:1992dc,Hu:1993gc,Chluba:2011hw}, and it is instructive to compare CMB transport with the C$\nu$B case.  This section will summarize the treatment of CMB transport described in Ref.~\cite{Bond:1993qp}. 

For the decoupling of photons from the plasma in the early universe, there are three identifiable epochs: 
\begin{enumerate}
        
    \item The blackbody epoch at $z>z_\mathrm{Pl} \approx 10^{6.9} \left({\Omega_b h^2 \over 0.01}\right)^{-0.39}$, above which Bremsstrahlung $ep\rightarrow ep+\gamma$ and Double Compton scattering $\gamma e \rightarrow \gamma e + \gamma$ are sufficiently fast that a Planck equilibrium distribution forms.  The redshift $z_\mathrm{Pl}$ therefore defines the cosmic photosphere.  The photon distribution in this epoch is characterized by just $\alpha^{(1)} = \beta E$.

    \item The chemical potential $\mu_\gamma$ epoch, at 
    $z_\mathrm{Pl} > z > z_\mathrm{BE} \approx 10^{5.6} \left(\frac{\Omega_b h^2}{0.01}\right)^{-1/2}$
    when Compton scattering dominates, photon number is conserved with injected energy redistributed fast enough to give a Bose-Einstein shape, and $\alpha$ is well characterized by $\alpha = \alpha^{(0)} + \beta E$, with  $\alpha^{(0)} = -\beta \mu_\gamma$.
    
    \item The Compton scattering-dominated epoch of thermal Sunyaev-Zeldovich $y$-distortions, 
    at $z <  z_{y} \approx 10^{5} \left({\Omega_b h^2 \over 0.01}\right)^{-1/2}$,
    when the energy redistribution is slow, resulting in the characteristic perturbative thermal Sunyaev-Zeldovich $y$-distortion shape: $\Delta \alpha = -y x \psi_\mathrm{K} (x)$ 
    with \edit{$y\equiv \tau (T_e-T)/m_e$, $\tau$ is the optical depth,} $x\equiv \omega/T$, and $\psi_\mathrm{K}(x) \equiv 2 - \frac{x}{2}\frac{e^x+1}{e^x-1}$.
    Below $z_{y}$, energy redistribution by Compton scattering is achieved through small positive energy kicks, $\langle \delta E/E \rangle \propto T_e /m_e$, and root mean square diffusion in energy $\langle (\delta E/E)^2 \rangle \propto T_e /m_e$, encoded in the Kompaneets diffusion equation for the photon distribution function.  
    
\end{enumerate}
There are intermediate transport regimes between the three epochs; for example, in between $z_\mathrm{Pl}$ and $z_\mathrm{BE}$ some photons lock into a Planck distribution and others do not~\cite{Zeldovich:1969ff,Sunyaev:1970er,Bond:1993qp,Chluba:2013pya}.  

To understand how these epochs arise, and to estimate magnitudes of the three characteristic redshifts, an analytic treatment
based on Ref.~\cite{Zeldovich:1969ff} is quite adequate.
The transport equation for the photon distribution function can be written as
\begin{equation}
    \label{eq:PhotonTransport}
    \frac{d\langle \widehat{n} \rangle}{dt} = {\cal C}_\mathrm{K} +{\cal C}_\mathrm{bremss}+{\cal C}_\mathrm{DC} \, ,
\end{equation}
where the right hand side gives the sum of the Kompaneets, bremsstrahlung, and double Compton source terms.  We can write the photon distribution function as
\begin{equation}
    \label{eq:PhotonDistribution}
    \langle \widehat{n} \rangle = \left(e^{\alpha}-1\right)^{-1} \, .
\end{equation}
We linearize the transport equation in $\Delta \alpha =\alpha - \epsilon$.  In the tight coupling regime, ${\cal C}_\mathrm{K} +{\cal C}_\mathrm{bremss}+{\cal C}_\mathrm{DC}$ approximately vanishes; 
this condition is satisfied for small $\epsilon_e =E_\gamma /T_e$, where $T_e$ is the electron temperature, if $\alpha = \alpha_0 (t) \exp (-\epsilon_e/\epsilon_0)$. Thus for low frequencies, $\epsilon_e < \epsilon_0$, bremsstrahlung and the double Compton process dump photons in fast enough to yield a Planck form, but for $\epsilon_e > \epsilon_0$ the Bose-Einstein form prevails.
Here $\epsilon_0 = (4\epsilon^3 (\Gamma_\mathrm{bremss} +\Gamma_\mathrm{DC})/\Gamma_\mathrm{K})^{1/2}$, where the `Kompaneets' rate is $\Gamma_\mathrm{K}\equiv 4n_e\sigma_T T_e/m_e$, in terms of electron density and Thomson scattering cross section. The approximate constancy of $\epsilon^3(\Gamma_\mathrm{bremss} +\Gamma_\mathrm{DC})$ is exploited to obtain this result.  If we assume an injection rate of ${\dot Y}_{\gamma}$ photons per
baryon with average energy $\bar{E}_{\gamma}$ at time $t$, adding to the $Y_{\gamma 0}$ photons per baryon already there, then the scaling parameter $\alpha_0$ evolves according to 
\begin{eqnarray}
     {{d\alpha_0}\over {dt}} &=& -{{\alpha_0}\over {\tau_\mathrm{D}}} + 
    \left( \frac{\bar{E}_{\gamma}}{3.6 T_{\gamma}} -1 \right) 
    1.87{{{\dot Y}_{\gamma} }\over 
    {Y_{\gamma 0} } } \ , \nonumber \\
    \tau_\mathrm{D} &=& 1.29 x_{0} [(\Gamma_\mathrm{bremss} +\Gamma_\mathrm{DC})x^3]^{-1} =1.29 
    (\Gamma_\mathrm{K} /4)^{-1} x_{0}^{-1} \, . \label{eq:dalphadt}
\end{eqnarray}
Thus there is a damping term with timescale $\tau_\mathrm{D}$ which pushes
$\alpha_0$ towards zero, i.e.~toward a Planck distribution.  The injection term works in the opposite direction, driving the distortion. When the damping time is shorter than the expansion rate of the universe, any injected energy input would be re-thermalized into a Planckian distribution in equilibrium with the electrons within one Hubble time. The transition between these epochs defines $z_\mathrm{Pl}$. When the Kompaneets rate is a few times the expansion rate, $x_0$ will be low but $\alpha_0$ will not be zero, and the Bose-Einstein form is appropriate. This defines $z_\mathrm{BE}$. However, it is not until the Kompaneets rate is a few times below the expansion rate that the perturbative $y$-distortion solution prevails. This defines
$z_y\approx z_\mathrm{BE}/4$. Naturally $z_\mathrm{BE}$ and $z_y$ scale in the way
defined by $\Gamma_\mathrm{K}/H$.

%\bondcomment{An example where $\alpha$ is not the best approach: dust emission.}

We can see from this example that following the evolution of $\alpha$ provides a convenient method for calculating spectral distortions during CMB decoupling.
We would like to see if there is any relatively simple form for $\alpha$ for the C$\nu$B as there is for the CMB. A key ingredient in the CMB case is that the photons that are shaken off in $ep \rightarrow ep$ and $\gamma e \rightarrow \gamma e$ are soft photons which must be redistributed over all energies by Compton scattering. For neutrino production, the energies are not soft, with highly energy-dependent reaction rates $\propto E$. Indeed one of the main highlights in this paper is identifying the extreme energy-dependence of the C$\nu$B  neutrino-sphere above which equilibrium holds. By contrast, the CMB photosphere is essentially energy-independent. 

The inhomogeneous transport of photons through CMB photon decoupling has a largely fixed spectral shape, and has usually been done in  $\Delta T_{s^\prime s} $ variables (see e.g.~\cite{Bond:1993qp}), but for accuracy we now recommend using 
$\alpha_{s^\prime s} ({\bf X}, {\bf p} ,t)$
or, ignoring the spectral distortions  given the dominance of the unperturbed Planck spectrum, the comoving inverse-temperature and its fluctuations 
$a^{-1}\beta_{\gamma; s^\prime s} ({\bf X}, {\bf p} ,t)$, since it does not have the nonlinear corrections that $\Delta T$ has. 

The inhomogeneous C$\nu$B transport through the time of CMB photon decoupling and to later times as the effects of neutrino masses become important is also best done with the comoving inverse temperature, 
$a^{-1}\beta_{\nu; s^\prime f^\prime sf} ({\bf X}, {\bf p}, t)$.
Unlike for photons, the dependence of the distribution on the magnitude of the comoving momenta,  $\epsilon \equiv p/\Tcm $, is important because neutrinos of different momenta experience the effects of neutrino masses at different cosmological epochs. 
While the formalism presented here allows for treatment of the full inhomogeneous neutrino transport, without a means to directly detect the C$\nu$B, cosmological observations are only sensitive to the homogeneous evolution.
Perturbations in the C$\nu$B density do, however, leave observational imprints on the CMB and large scale structure power spectra since they freely stream after C$\nu$B decoupling~\cite{Bashinsky:2003tk,Baumann:2015rya,Baumann:2017gkg}.
The shift in phase of the acoustic peaks due to the effects of neutrino free streaming has been observed in the CMB~\cite{Follin:2015hya} and in baryon acoustic oscillation data~\cite{Baumann:2019keh}.

%%%%%%%%%%%%%
\section{The Collision Rate Operator}
\label{sec:CollisionRate}
%%%%%%%%%%%%%

In the previous sections we described the general transport framework that we will utilize and described how the framework can be applied to the (perhaps more familiar) case of CMB transport.
We now turn to a description of the interactions relevant to neutrino transport. \edit{As mentioned above, we will neglect neutrino flavor oscillations in the treatment presented here.}

The Boltzmann equation for neutrino transport can often be thought of as consisting of a coarse-grain change in $n$ as a result of quasiparticle free-streaming evolution plus a fine-grain flow of information to the coarse-grain from  the net result of collisions: $D\langle \widehat{n}_{\nu_f} \rangle \vert_\alpha/dt =  C_{\nu_{f}}[n_{\nu}]$, where $C_{\nu_{f}}[n_{\nu}]dt$ encodes the net change in the mean Wigner function due to short scale collisions. In the absence of collisions, $\langle \widehat{n}_\nu \rangle \vert_\alpha$ is conserved along geodesic trajectories. If we force $\alpha $ to be tightly-coupled to $n$ through $\alpha = \ln (\langle \widehat{n}_\nu \rangle \vert_\alpha^{-1} -1)$, then $D\alpha/dt  = (d\langle \widehat{n}_\nu \rangle \vert_\alpha/d\alpha )^{-1}C = \langle \delta \widehat{n} \delta \widehat{n}\rangle^{-1} C$. Here $\langle \delta \widehat{n} \delta \widehat{n}\rangle = \langle \widehat{n}_\nu \rangle \vert_\alpha(1-\langle \widehat{n}_\nu \rangle \vert_\alpha)$. Generally the collision operator evaluated in the independent particle approximation is full of products of $\widehat{n}$ up to the fourth power, and even if $\alpha$ is transported, the collision rate operator will still use this product expansion in $\widehat{n}$. The collision term is derivable directly from the density matrix formalism as long as we go beyond the independent particle approximation to include the interaction Hamiltonian in $\rho_i$ and $\rho_f$, and expand to second order in the Fermi 4-point coupling $G_F$. This is equivalent to the usual derivation of $C$ where factors of $ \widehat{n} $ are assumed to be correlated only through the independent particle approximation in which correlators of higher order can be expressed as polynomial products of the first order correlators $\langle \widehat{n} \rangle \vert_\alpha$, $ \langle \prod \widehat{n}_f \rangle \vert_{\alpha,cc} = (-1)^N \partial^N {\cal F} /(\partial \alpha_1 \cdots \partial \alpha_N)$. Given that, there is a clean separation of Fermi Golden rule squared matrix elements and products of the statistical factors. For 2-body interactions, the only factors entering are  $\langle \widehat{n} \rangle$ and Fermi-blockers $1-\langle \widehat{n} \rangle$. 

Neutrinos ($\nu_{f}$) and anti-neutrinos ($\bar{\nu}_{f}$) of flavors $f = e, \mu, \tau$ collide with each other and with electrons ($e^{-}$) and positrons ($e^{+}$) through various two-body interaction channels, the latter resulting in entropy flow between the $e^{-}$/$e^{+}$/photon plasma and the neutrino sea, resulting in distortions of the mean neutrino occupation probabilities. Here and below, where the context avoids potential confusion, we will use the notation $n_{\nu_{f}}$ to refer to the expectation value of the occupation probability $\langle \widehat{n} \rangle_f$. The $n_{\nu_{f}}$ distributions are time-dependent functions of coarse-grained positions ${\bf X}$ and relative 3-momenta ${\bf p}$, and time, $t$. We assume that the neutrino sea is spatially homogeneous, {\it i.e.}, ${\bf X}$-independent, and independent of the 3-momentum direction ${\bf \widehat{p}}$, so only a function of its magnitude $p$ and time $t$. The coarse-grained evolution is $Dn/dt = \left[ \frac{\partial}{\partial t} - H(a) p \frac{\partial}{\partial p}\right]n$,
%n_{\nu_{i}}
if we assume the neutrino masses are small compared with the momenta of interest. In that case it is advantageous to transform from $p\approx E_p$ to a comoving  energy parameter $\epsilon = \frac{E_{\nu}}{T_{\text{cm}}}$, where $T_{\text{cm}}\propto a^{-1}$ is a measure of the scale factor expressed in comoving temperature units: 
\begin{equation}
    T_{\text{cm}} = T_{\text{in}}a_{\text{in}} \left[ \frac{1}{a(t)} \right] \, ,
    \label{eq:Tcm_def}
\end{equation}
where $T_{\text{in}}$ and $a_{\text{in}}$ are the initial values of temperature and scale factor of the plasma before the weak decoupling epoch.

In the flavor eigenbasis, in the Boltzmann equation for neutrinos of flavor $f$ as a function of time and comoving energy $\epsilon$, 
\begin{equation} 
    \frac{d}{dt}n_{\nu_{f}}\vert_\epsilon = C_{\nu_{f}}[n_{f^\prime }] = \sum_{r}C_{\nu_{f}}^{(r)}[n_{f^\prime}] \, ,
    \label{eq:boltzmann}
\end{equation}
the collision rate operator is decomposed into interaction channels $r$ in which flavor $f^\prime$ stimulates changes in flavor $f$.

\edit{Equation \eqref{eq:boltzmann} gives the equations of motion for the neutrino distribution functions indexed by $\epsilon$.  Section \ref{sec:SMnu} presents results from the code \burst which integrate Eq.\ \eqref{eq:boltzmann} forward in time along with fits using the dual-description $\alpha$ from Sec.\ \ref{sec:TransportWigner}.  The description of $n_{\nu_f}|_\epsilon$ is indeed dual to $\alpha_{\nu_f}$ as can be seen by transforming Eq.\ \eqref{eq:boltzmann} into the new dependent variable
\begin{equation}
  \frac{d}{dt}\alpha_{\nu_f} = C_{\nu_f}[n_{f^{\prime}}]e^{-\alpha_{\nu_f}}(e^{\alpha_{\nu_f}}+1)^2,\label{eq:alpha_boltz}
\end{equation}
where the collision term would need to be rewritten by substituting $1/[e^{\alpha_{f^{\prime}}(\epsilon)}+1]$ for $n_{f^{\prime}}|_\epsilon$.  Although we show fits using $\alpha_{\nu_f}$ later in Sec.\ \ref{sec:SMnu}, one could in principle discretize Eq.\ \eqref{eq:alpha_boltz} and integrate the set of ODEs forward in time.  We leave that program to future work (see Appendix \ref{app:EvolutionAlpha} for details).
}

\edit{Returning to the discussion of the collision term using occupation numbers $n_{\nu_f}|_\epsilon$}, the general form for a 2-body interaction $1 + 2 \leftrightarrow 3 + 4$ is given by Fermi's Golden Rule, 
\begin{equation} 
    \begin{split}
    C_{\nu_{1}}^{(r)}[n_{f}] = \frac{1}{2E_{1}}\int& \frac{d^{3}p_{2}}{(2 \pi)^{3}2E_{2}}\frac{d^{3}p_{3}}{(2 \pi)^{3}2E_{3}}\frac{d^{3}p_{4}}{(2 \pi)^{3}2E_{4}} (2 \pi)^{4}\delta^{(4)}(p_{1}+p_{2}-p_{3}-p_{4}) \\
    & \times S_{r} \langle |\mathcal{M}_{r}|^{2} \rangle F_{r}(p_{1},p_{2},p_{3},p_{4}) \, .
    \end{split}
    \label{eq:collint}
\end{equation}
$\langle |\mathcal{M}_{r}|^{2} \rangle$ is the squared amplitude of the weak interaction $r$, averaged over initial spin states and summed over final spin states. $S_{r}$ is the symmetrization factor for identical particles, where $S_{r} = \frac{1}{2}$ for identical particle interactions, and $S_{r} = 1$ for all other interactions. $\delta^{(4)}(p_{1}+p_{2}-p_{3}-p_{4})$ is the energy and momentum conserving delta-function. The function  $F_{r}(p_{1},p_{2},p_{3},p_{4})$ includes the independent particle Fermi-blocking factors $1-\langle \widehat{n} \rangle \vert_\alpha$ in the outgoing channels as well as the occupation number product in the incoming channels:  
\begin{equation} \label{F_eqt}
\begin{split}
F_{r}(p_{1},p_{2},p_{3},p_{4}) &= (1-n_{1})(1-n_{2})n_{3}n_{4}  - n_{1}n_{2}(1-n_{3})(1-n_{4}) \\
&= F_{r}^{(\text{in})} - F_{r}^{(\text{out})}.
\end{split}
\end{equation}
Here $F_{r}^{(\text{in})}$ is the statistical factor associated with the probability of scattering into a state, and $F_{r}^{(\text{out})}$ is the statistical factor associated with the probability of scattering out of a state. It is assumed that electrons and positrons maintain a thermal Fermi-Dirac distribution during this time, with the electrons at the plasma temperature $T_e \approx T_\gamma$. Thus we need only solve the Boltzmann equations for neutrinos and anti-neutrinos. Thus $p_{1}$ always refers to either a neutrino or anti-neutrino, while $p_{2}$, $p_{3}$, $p_{4}$ refer to a neutrino, anti-neutrino, or charged lepton. Table (\ref{colltab}) outlines all the weak interaction processes, along with their scattering matrix term $G_{F}^{-2}S_{r}\langle |\mathcal{M}_{r}|^{2} \rangle$, where $G_{F} = 1.166\times 10^{-11}~\text{MeV}^{-2}$ is the Fermi 4-point interaction constant.
For completeness we list the terms $r$ as derived for use in the \burst{} code, as given by an analogous table in \cite{Grohs:2015tfy}. 

\begin{table}[htbp]
\centering
\begin{tabularx}{\linewidth}{|c|c|X|}
\hline
\rowcolor[HTML]{ECF4FF} $r$ & Weak Interaction Process                                                & $G_{F}^{-2}S_{r}\langle |\mathcal{M}_{r}|^{2} \rangle$                                                                                                                                                     \\ \hline \hline

 1   & $\nu_{f} + \nu_{f} \leftrightarrow \nu_{f}  + \nu_{f}$                  & $2^{6}(P_{1}\cdot P_{2})(P_{3}\cdot P_{4})$                                                                                                                                                                \\ \hline
\rowcolor[HTML]{EFEFEF} 2   & $\nu_{f} + \nu_{f^\prime} \leftrightarrow \nu_{f}+ \nu_{f^\prime}$                   & $2^{5}(P_{1}\cdot P_{2})(P_{3}\cdot P_{4})$                                                                                                                                                                \\ \hline
 3   & $\nu_{f} + \bar{\nu}_{f} \leftrightarrow \nu_{f}+ \bar{\nu}_{f}$        & $2^{7}(P_{1}\cdot P_{4})(P_{2}\cdot P_{3})$                                                                                                                                                                \\ \hline
\rowcolor[HTML]{EFEFEF} 4   & $\nu_{f} + \bar{\nu}_{f^\prime} \leftrightarrow \nu_{f}+ \bar{\nu}_{f^\prime}$        & $2^{5}(P_{1}\cdot P_{4})(P_{2}\cdot P_{3})$                                                                                                                                                                \\ \hline
 5   & $\nu_{f} + \bar{\nu}_{f} \leftrightarrow \nu_{f^\prime}+ \bar{\nu}_{f^\prime}$        & $2^{5}(P_{1}\cdot P_{4})(P_{2}\cdot P_{3})$                                                                                                                                                                \\ \hline
\rowcolor[HTML]{EFEFEF} 6   & $\nu_{e} + e^{-} \leftrightarrow e^{-} + \nu_{e}$                       & $2^{5}[(2\sin^{2}\theta_{W}+1)^{2}(P_{1}\cdot Q_{2})(Q_{3}\cdot P_{4}) + 4\sin^{4}\theta_{W}(P_{1}\cdot Q_{3})(Q_{2}\cdot P_{4}) -2\sin^{2}\theta_{W}(2\sin^{2}\theta_{W} +1)m_{e}^{2}(P_{1}\cdot P_{4})]$ \\ \hline
 7   & $\nu_{\mu(\tau)} + e^{-} \leftrightarrow e^{-} + \nu_{\mu(\tau)}$       & $2^{5}[(2\sin^{2}\theta_{W}-1)^{2}(P_{1}\cdot Q_{2})(Q_{3}\cdot P_{4}) +4\sin^{4}\theta_{W}(P_{1}\cdot Q_{3})(Q_{2}\cdot P_{4})-2\sin^{2}\theta_{W}(2\sin^{2}\theta_{W} - 1)m_{e}^{2}(P_{1}\cdot P_{4})]$  \\ \hline
\rowcolor[HTML]{EFEFEF} 8   & $\nu_{e} + e^{+} \leftrightarrow e^{+} + \nu_{e}$                       & $2^{5}[(2\sin^{2}\theta_{W}+1)^{2}(P_{1}\cdot Q_{3})(Q_{2}\cdot P_{4}) +4\sin^{4}\theta_{W}(P_{1}\cdot Q_{2})(Q_{3}\cdot P_{4})-2\sin^{2}\theta_{W}(2\sin^{2}\theta_{W} + 1)m_{e}^{2}(P_{1}\cdot P_{4})]$  \\ \hline
 9   & $\nu_{\mu(\tau)} + e^{+} \leftrightarrow e^{+} + \nu_{\mu(\tau)}$       & $2^{5}[(2\sin^{2}\theta_{W}-1)^{2}(P_{1}\cdot Q_{3})(Q_{2}\cdot P_{4}) +4\sin^{4}\theta_{W}(P_{1}\cdot Q_{2})(Q_{3}\cdot P_{4}) -2\sin^{2}\theta_{W}(2\sin^{2}\theta_{W} - 1)m_{e}^{2}(P_{1}\cdot P_{4})]$ \\ \hline
\rowcolor[HTML]{EFEFEF} 10  & $\nu_{e} + \bar{\nu}_{e} \leftrightarrow e^{-} + e^{+}$                 & $2^{5}[(2\sin^{2}\theta_{W}+1)^{2}(P_{1}\cdot Q_{4})(P_{2}\cdot Q_{3}) +4\sin^{4}\theta_{W}(P_{1}\cdot Q_{3})(P_{2}\cdot Q_{4})-2\sin^{2}\theta_{W}(2\sin^{2}\theta_{W} +1)m_{e}^{2}(P_{1}\cdot P_{2})]$   \\ \hline
 11  & $\nu_{\mu(\tau)} + \bar{\nu}_{\mu(\tau)} \leftrightarrow e^{-} + e^{+}$ & $2^{5}[(2\sin^{2}\theta_{W}-1)^{2}(P_{1}\cdot Q_{4})(P_{2}\cdot Q_{3})+4\sin^{4}\theta_{W}(P_{1}\cdot Q_{3})(P_{2}\cdot Q_{4})-2\sin^{2}\theta_{W}(2\sin^{2}\theta_{W}-1)m_{e}^{2}(P_{1}\cdot P_{2})]$     \\ \hline
\end{tabularx}
\caption[List of the weak interactions of neutrinos along with the corresponding scattering matrix term.]{This table displays the weak interaction processes for neutrinos. $r = 5, 10, 11$ represent annihilation channels, while the other processes are scattering channels. $P_{i}$ represents the momenta of neutrinos or anti-neutrinos, while $Q_{i}$ represents momenta of charged leptons. Anti-neutrinos undergo the same set of interactions, except the matrix amplitudes are the parity-conjugate for processes $r=6,\ldots,9$.}
\label{colltab}
\end{table}

There is an analogous set of weak interactions for anti-neutrinos, except for processes $r=6,\ldots,9$, where the squared matrix term is the parity-conjugate of the respective process in Table~\ref{colltab}. We are operating at temperatures far below the muon and $\tau$ mass, yet far above the neutrino masses, so in terms of reaction rates there is no difference between the behavior of $\nu_{\mu}$ and $\nu_{\tau}$.\footnote{There is an ongoing effort to include neutrino oscillations into the dynamics of neutrinos during the weak decoupling epoch. In Refs.~\cite{Mangano:2005cc,Akita:2020szl,Froustey:2020mcq,Bennett:2020zkv}, neutrino oscillations (or mixing) are included in the evolution of the neutrino distribution functions leading to slight differences among the final distribution functions for each neutrino flavor.} Of course we are just in the regime in which the temperature is comparable to the electron mass, hence the amplitudes of $\nu_{e}$ interactions includes charged current interactions, modifying the $(2\sin^{2}\theta_{W}-1)$ to a $(2\sin^{2}\theta_{W}+1)$ factor, where $\theta_{W}$ is the weak mixing angle (or Weinberg angle), with $\sin^{2}\theta_{W} \approx 0.23$.

The computation of neutrino transport is iterative: collision rates are calculated for all neutrino flavor distribution functions at a given time step $t$ which kick the distributions into a new state at $t+dt$. Those newly calculated distributions are then used to calculate the collision rates at the next time step, and so on. The 9-dimensional integral in Eq.~\eqref{eq:collint} can be reduced to a 2-dimensional integral over $\epsilon$ variables. The reduction and subsequent formulation of the rates is spelled out in detail in Appendix B and C of Ref.~\cite{Grohs:2015tfy}.

%%%%%%%%%%%%%%%%%%%% 
\section{The Neutrino BURST Numerical Computations}
\label{sec:BURST}
%%%%%%%%%%%%%%%%%%%%

One application of the maximum-entropy distributions discussed above is to
evaluate the response of the light-element abundances to the cosmic neutrino distributions.
To accomplish this task, we need to run numerical experiments on BBN using a computation with sufficient energy and time resolution.
We use the code \burst{} \cite{Grohs:2015tfy} for these experiments.  \burst{} calculates neutrino-energy transport without resorting to any approximations of equilibrium.  Using a suitable numerical implementation of the collision integrals in Sec.\ \ref{sec:CollisionRate}, we can time-evolve the neutrino distributions through the weak decoupling epoch and couple those distributions to the neutron-to-proton interconversion rates.  The free nucleon abundances are then input into the nuclear reaction network to calculate the light-element abundances.  The progression of neutrino scattering, isospin-changing reactions, nucleosynthesis is all within the context of the standard cosmology.  By extending the standard-model \burst{} computation, we have the ability to conduct
numerical experiments with perturbed neutrino distributions.

We refer the reader to Ref.\ \cite{Grohs:2015tfy} for details of the numerical implementations in \burst{}.  We give a few pertinent details here for the specific experiments in this work.  We use 6 distributions (3 flavors of neutrinos and anti-neutrinos)  with 101 energy bins, linearly spaced from $\epsilon=0$ to $25$.  All 101 energy bins are used when calculating energy densities and the weak-isospin-changing rates.  For the neutrino scattering rates, we do not evolve the distribution for $\epsilon=0$.
We do not set the cosmological parameter \neff{} as an input for our nucleosynthesis calculations.  \neff{} is an output from the neutrino-energy-transport and other electromagnetic processes.
Finally, we only consider neutrino kinetics for the energy scales of BBN -- scales much larger than the neutrino rest masses.  We always treat neutrinos as ultra-relativistic and ignore any rest-mass contribution to the total energy density of the universe at the time of BBN.  Our results are agnostic in regard to the cosmological parameter $\sum m_\nu$.

We neglect the role of neutrino flavor oscillations in this work.  The computational problem of including oscillations with collisions has been done within the standard cosmology with a mean field, quantum kinetic equation (QKE) approach~\cite{Akita:2020szl,Froustey:2020mcq,Bennett:2020zkv} and in a non-zero lepton-number extension~\cite{2021arXiv211011889F}.  For the case of the standard cosmology, the occupation numbers for the neutrino energy distributions have the same qualitative features between the Boltzmann and QKE calculations throughout the weak decoupling epoch.  Importantly, the electron neutrino (and anti-neutrino) distributions have smaller deviations from FD equilibrium for the QKE case compared to the Boltzmann case.  This result derives from the fact that with oscillations present, an electron neutrino has a non-zero probability to oscillate to a $\mu$ or $\tau$ flavor neutrino.  As an example, our results for the linear response of the nuclide abundances to spectral distortions using Boltzmann transport could be applied to the distributions with oscillations included if we were to rescale those responses to smaller values.  In theory, neutrinos could follow a much more complicated evolution during and after weak decoupling if collisions were to damp any coherence in the neutrino wave functions at early times and the energy-dependent vacuum Hamiltonian introduced a non-zero flavor coherence at later times.  However, the results of Refs.~\cite{Akita:2020szl,Froustey:2020mcq,Bennett:2020zkv} indicate the neutrino density matrices and the vacuum Hamiltonian are coincident with one another, i.e., that neutrino wave functions are parallel to the mass eigenvectors instead of the flavor eigenvectors and are therefore immutable from the vacuum potential at late times.  This finding holds over the entirety of weak decoupling implying that the distribution functions follow the same trajectories in both the Boltzmann and QKE treatments, but with different amplitudes.  As a result, we expect that our results here are transferable to the more general QKE problem in the standard cosmology, but the abundance-responses would need to be scaled down.  For a BSM model, the results could be qualitatively different between Boltzmann and QKE treatments. Moreover, even in the Standard Model case, the mean field QKE approach may not be adequate to capture the neutrino flavor evolution. Quantum many body effects might be important~
\cite{2009PhRvD..79j5003S}, though the status of studies of this issue are mixed~
\cite{2003JHEP...10..043F}. Some show the development of entanglement entropy with time and with the number of neutrinos, a marker for the break down of the mean field approach. However, these studies~
\cite{2007JPhG...34...47B,2021PhRvD.104l3035P,2021PhRvD.104f3009H}
focus on lower entropy, compact object-like conditions and it is not clear how that will scale to the early universe. In any case, a potential target for our analysis will be to look for signatures of BSM or many-body effects in altering the light element and $N_{\rm eff}$ observables. 

The code begins evolving cosmological variables at a temperature $T = 30\, \text{MeV}$, where all of the neutrino species are in thermal equilibrium with the electrons and positrons. During this time, the comoving temperature is equal to the plasma temperature. Neutrino transport begins at a temperature of $T = 10\, \text{MeV}$, after which the comoving temperature begins to deviate from the plasma temperature. Neutrino transport terminates at a temperature of $T_{\text{cm}} = 0.015\, \text{MeV}$, which is after the annihilation of the thermal electrons and positrons.  \burst uses an adaptive time step based on convergence criteria from the previous time step.  The code takes on order $10^4$ time steps to progress through weak decoupling and BBN.

%%%%%%%%%%%%%%%%%%%%%
\section{The Evolution of Neutrino Distribution Functions and their Conjugates in Weak Decoupling}
\label{sec:SMnu}
%%%%%%%%%%%%%%%%%%%%%

\begin{figure}[!t]
    \centering
    \includegraphics[width=\columnwidth]{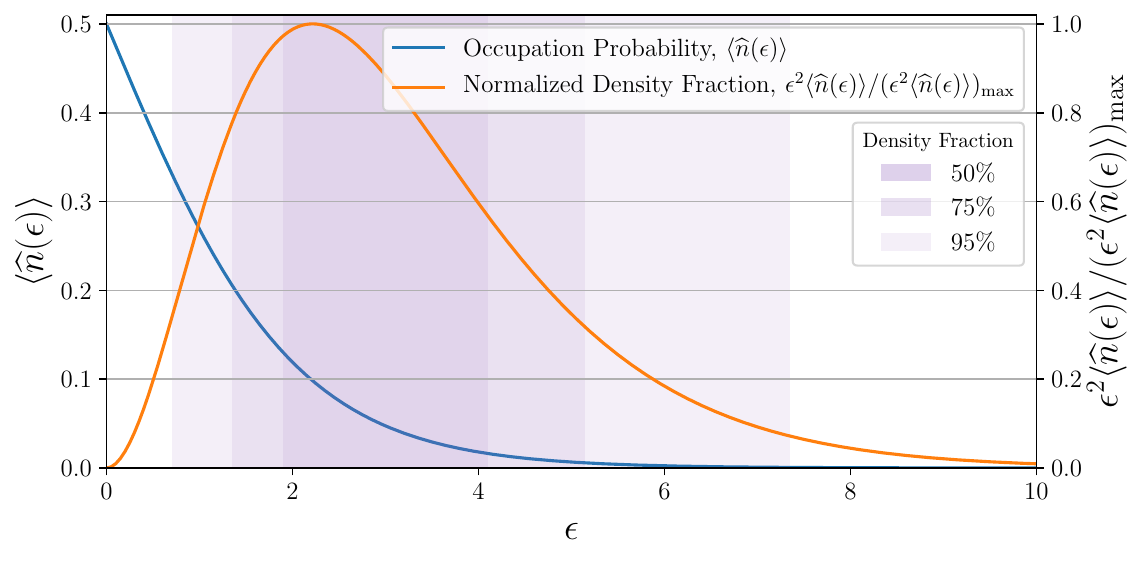}
    \caption[Equilibrium Fermi-Dirac distribution.]{
    The mean relativistic Fermi-Dirac occupation probability $\langle \widehat{n}(\epsilon) \rangle$ at temperature $T=\Tcm$ in terms of $\epsilon = p/\Tcm{}$ and the normalized density fraction $\epsilon^2 \langle \widehat{n}(\epsilon)\rangle/(\epsilon^2 \langle \widehat n(\epsilon) \rangle)_\mathrm{max}$ of a species described by this distribution.
    The equilibrium distribution function accurately describes cosmic neutrinos at high temperatures, before weak decoupling.
    The normalized density fraction is used below to weight parametric fits of the non-equilibrium distribution function.
    }
    \label{fig:feq}
\end{figure}

At high temperatures, cosmic neutrinos of each flavor $f$ were in thermal equilibrium with the plasma and were well described by a relativistic Fermi-Dirac distribution 
\begin{equation}
    \langle \widehat{n}_{\nu_{f}}\rangle = \frac{1}{e^{\epsilon}+1} \, ,
    \label{eq:FD_distribution}
\end{equation}
which is shown in Figure~\ref{fig:feq}.  As neutrinos decouple from the plasma, their distribution begins to deviate from the equilibrium Fermi-Dirac distribution.  We will characterize the evolution of the neutrino distributions by
\begin{equation} 
    \langle \widehat{n}_{\nu_{f}} \rangle \vert_\alpha = \frac{1}{e^{\alpha^{(f)}(\epsilon,T_{\text{cm}})}+1} \, ,
    \label{eq:distribution_alpha}
\end{equation}
where the quantity $\alpha(\epsilon,\Tcm{})$, introduced in Sec.~\ref{sec:TransportWigner}, describes the out-of-equilibrium neutrino transport.  In this notation, the equilibrium occupation probability is given by $\alpha(\epsilon,\Tcm{}) = \alpha_\mathrm{eq} = \epsilon$.

%%%%%%%%%%%%%%%%%% FD Distribution %%%%%%%%%%%%%%%%%%
%%%%%%%%%%%%%%%%%%%%%%%%%%%%%%%%%%%%

As described in Sec.~\ref{sec:TransportWigner}, the definition of the occupation probability in terms of $\alpha$ is motivated by its connection to the entropy. The entropy per unit phase space is given by
\begin{equation}
    \langle \widehat{s}_f \rangle = -\langle \widehat{n}_{\nu_{f}}\rangle \ln \langle \widehat{n}_{\nu_{f}}\rangle  - (1-\langle \widehat{n}_{\nu_{f}}\rangle )\ln(1-\langle \widehat{n}_{\nu_{f}}\rangle ).
    \label{eq:s_and_alpha}
\end{equation}
With the occupation probability given by Eq.~\eqref{eq:distribution_alpha}, the derivative of the entropy with respect to the distribution function is given by
\begin{equation}
    %\frac{\d \langle s_i \vert \alpha \rangle }{\d \langle n_{\nu_{i}}\vert \alpha \rangle } = \alpha^{(i)} \, .
    \frac{\d \langle \widehat{s}_f \rangle }{\d \langle \widehat{n}_{\nu_{f}} \rangle } = \alpha^{(f)} \, .
\end{equation}
This shows the direct link between the response in entropy to the stimulus of  distribution function change through $\alpha$, which acts here as a stimulus-response transport coefficient. 

As described below, we can capture the evolution of the non-equilibrium distribution function through the epoch of weak decoupling in the Standard Model very precisely by modeling the $\alpha^{(f)}(\epsilon,\Tcm{})$ as polynomials in $\epsilon$.  Furthermore, deviations from the equilibrium distribution are small in the Standard Model, and can be treated perturbatively in this case.

%%%%%%%%%%%%%%%%%%%%
\subsection{Linear Model for $\alpha$ Evolution}
\label{ssec:2ParamAlpha}
%%%%%%%%%%%%%%%%%%%%

First, we model the departure from equilibrium with a simple two-parameter model for the neutrino occupation probability, defined as
\begin{equation}
    \alpha^{(f)}_\mathrm{fit} = \alpha^{(f)}_0(\Tcm{}) + \alpha^{(f)}_1(\Tcm{}) \cdot \epsilon \, .
    \label{eq:2param_fit_def}
\end{equation}
In this model, the parameter $\alpha_0(\Tcm{})$ plays a role similar to the chemical potential divided by the temperature, but here the chemical potentials of the neutrinos and antineutrinos have the same sign.  A positive value of $\alpha_0(\Tcm{})$ represents a deficit of both neutrinos and antineutrinos compared to an equilibrium distribution with vanishing chemical potential.  Meanwhile the parameter $\alpha_1(\Tcm{})$ is related to the mean energy, and plays a role similar to the effective temperature of the neutrinos.

In order to determine the values of the parameters $\alpha^{(f)}_0(\Tcm{})$ and $\alpha^{(f)}_1(\Tcm{})$, we performed a density-weighted fit of the model shown in Eq.~\eqref{eq:2param_fit_def} to the output of the \burst code for each value of $\Tcm{}$ and for each neutrino flavor.  The weighting function is chosen  to be the normalized density fraction for the equilibrium Fermi-Dirac distribution  $\epsilon^2 \langle \widehat{n}(\epsilon)\rangle/(\epsilon^2 \langle \widehat n(\epsilon) \rangle)_\mathrm{max}$, plotted in Figure~\ref{fig:feq}.  Since we have not included flavor oscillations, the $\mu$ and $\tau$ neutrino distributions are identical to one another, though they are distinct from the electron neutrino distribution.

\begin{figure}[tbp!]
    \centering
    \includegraphics[width=1.0\textwidth]{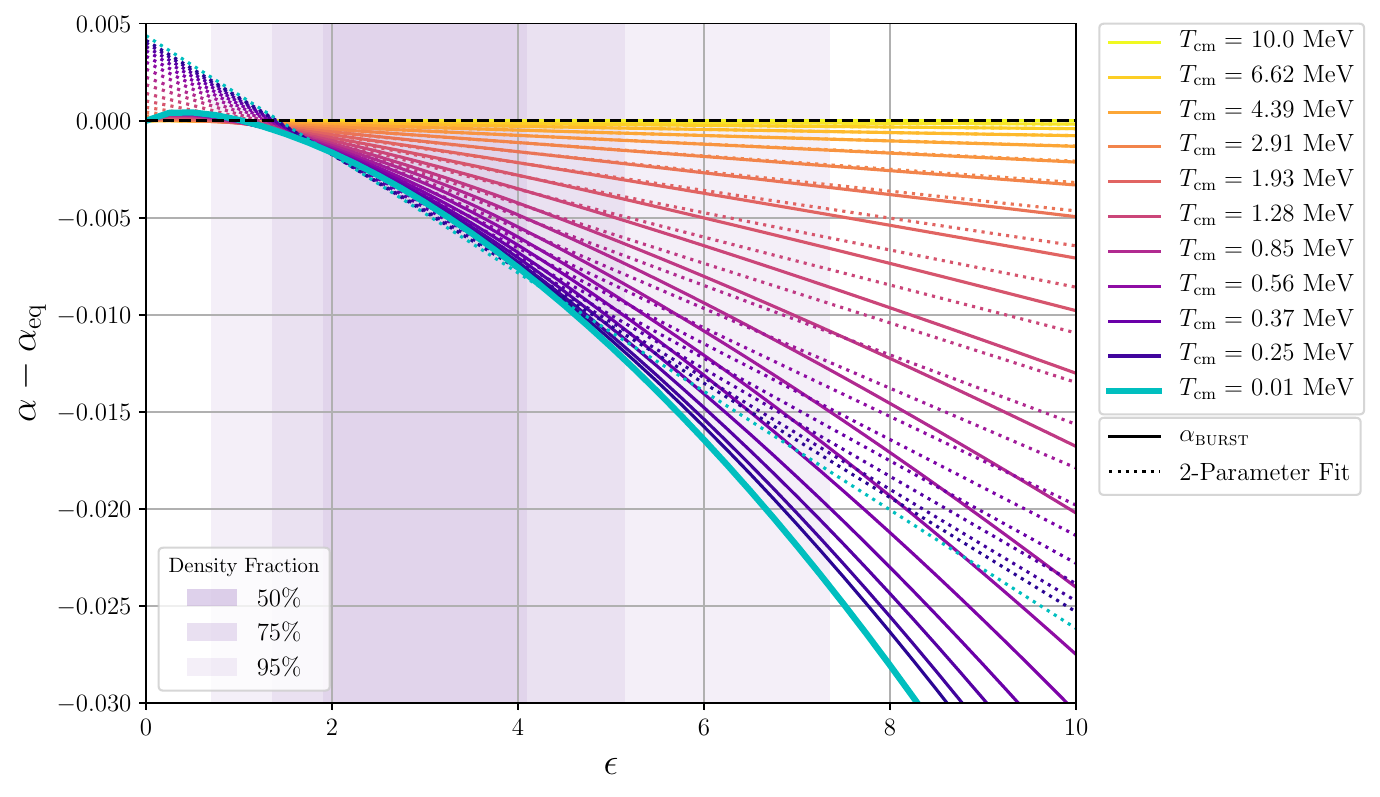}
    \caption[Evolution of $\alpha$ with linear fit]
    {
    We plot here the evolution of $\alpha(\epsilon,\Tcm{})$ for electron neutrinos through the weak decoupling epoch.
    The quantity $\alpha$ is related to the occupation probability through $\left\langle \widehat{n}_{\nu_{f}} \right\rangle = [\exp(\alpha)+1]^{-1}$.
    We show the difference of $\alpha(\epsilon,\Tcm{})$ calculated with \burst{}, labeled $\alpha_{\textsc{\scriptsize burst}}$ (solid lines), from the value that it would have in thermal equilibrium, $\alpha_\mathrm{eq} = \epsilon$.
    We also show $\alpha(\epsilon,\Tcm)$ obtained from weighted linear fit to the results of \burst{}, $\alpha_\mathrm{fit}(\epsilon,\Tcm) = \alpha_0(\Tcm{}) + \alpha_1(\Tcm{}) \epsilon$ (dotted lines).
    Shaded regions denote the fraction of neutrino density in a given range of $\epsilon = p/\Tcm$.
    Note that the 2-parameter model provides a good fit to $\alpha(\epsilon)$ at high temperatures during the initial phases of weak decoupling, but it begins to deviate at low temperatures during the later stages of weak decoupling.
    }
    \label{fig:alpha_evolution_linear_fit}
\end{figure}

\begin{figure}[!tbph]
    \centering
    \includegraphics[width=0.75\textwidth]{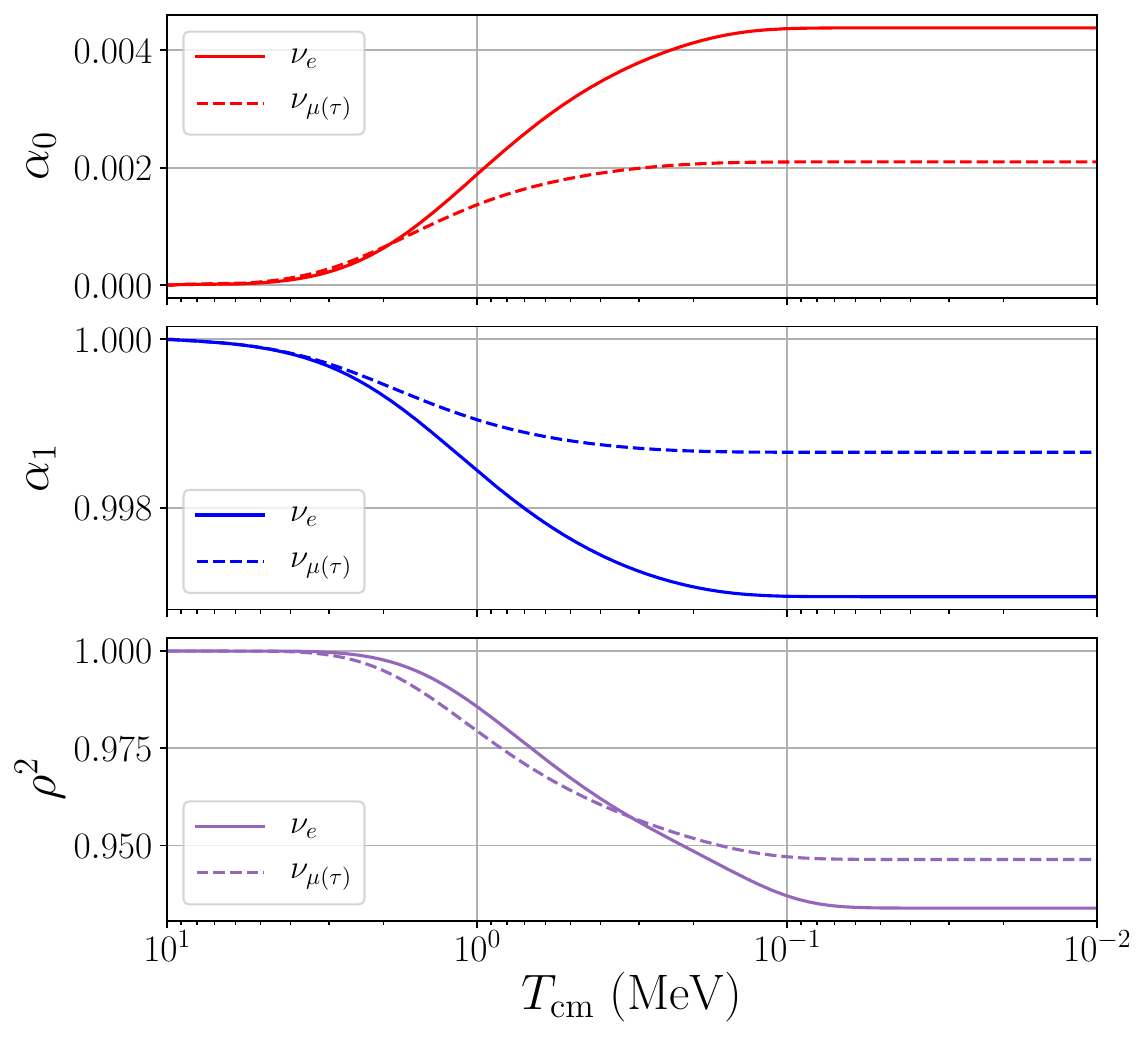}
    \caption[Linear fit coefficients]
    {
    The shape of $\alpha(\epsilon,\Tcm{})$ can be reasonably approximated with a 2-parameter fit early in the weak decoupling epoch.
    Here we show the best fit values for the coefficients in a 2-parameter model of $\alpha(\epsilon,\Tcm{})$ with the form $\alpha_\mathrm{fit}(\epsilon,\Tcm{}) = \alpha_{0}(T_{\text{cm}}) + \alpha_{1}(T_{\text{cm}})\cdot \epsilon$.
    These coefficients are related to the extensive variables of the neutrino ensemble: $\alpha_{0}$ is conjugate to the expected number of neutrinos in the system, $\alpha_{1}$ is conjugate to the expected energy.
    The goodness of fit, $\rho^2$ defined in Eq.~\eqref{eq:goodness_of_fit}, for this 2-parameter model is also shown as a function of $\Tcm$, showing that this model is an excellent fit early in the decoupling epoch, though less so as decoupling progresses.
    }
    \label{fig:alpha_evolve_linear}
\end{figure}

As shown in Figure~\ref{fig:alpha_evolution_linear_fit}, this model provides a reasonable fit to the full transport calculation early in the neutrino decoupling epoch, but it fails to accurately capture the occupation probability at the later stages of neutrino decoupling.  The evolution of the best-fit parameters for this linear model is shown in Figure~\ref{fig:alpha_evolve_linear}.
We also show the goodness of fit for this model, defined as
\begin{equation}
    \rho^2(\Tcm{}) \equiv 1 - \frac{ \int d\epsilon \, W(\epsilon) \left(\alpha_{\textsc{\scriptsize burst}}(\epsilon,\Tcm{})-\alpha_\mathrm{fit}(\epsilon,\Tcm{})\right)^2}{\int d\epsilon \, W(\epsilon) \left(\alpha_{\textsc{\scriptsize burst}}(\epsilon,\Tcm{})-\alpha_\mathrm{eq}(\epsilon,\Tcm{})\right)^2} \, ,
    \label{eq:goodness_of_fit}
\end{equation}
where $W(\epsilon)= \epsilon^2 \langle \widehat{n}(\epsilon)\rangle/(\epsilon^2 \langle \widehat n(\epsilon) \rangle)_\mathrm{max}$ is the weighting function, taken to be the normalized density fraction plotted in Fig.~\ref{fig:feq}, and the range of integration is taken to be $0<\epsilon<10$.  A perfect fit would have $\alpha_\mathrm{fit}=\alpha_{\textsc{\scriptsize burst}}$ and thus $\rho^2(\Tcm{})=1$.

%%%%%%%%%%%%%%%% Alpha evolution - linear fit %%%%%%%%%%%%%%%%%
%%%%%%%%%%%%%%%%%%%%%%%%%%%%%%%%%%%

%%%%%%% Linear Alpha Fit Coefficients %%%%%%%%%%%%
%%%%%%%%%%%%%%%%%%%%%%%%

One can see that this linear model provides a reasonable approximation to the evolution of the neutrino occupation probability at high temperature, but it begins to deviate from the numerical results as the temperature drops below about 1~MeV.  Next, we will describe a model which provides a better fit to the full transport calculation throughout the entire process of neutrino decoupling.

%%%%%%%%%%%%%
\subsection{Quadratic Model for $\alpha$ Evolution} 
\label{ssec:3ParamAlpha}
%%%%%%%%%%%%%%

Now we define a 3-parameter polynomial model of the form
\begin{equation}
    \alpha^{(f)}_\mathrm{fit} = \alpha^{(f)}_{0}(T_{\text{cm}}) + \alpha^{(f)}_{1}(T_{\text{cm}})\cdot \epsilon + \alpha^{(f)}_{2}(T_{\text{cm}})\cdot \epsilon^{2} \, .
    \label{eq:3param_fit_def}
\end{equation}
As in the 2-parameter model described above, the parameter $\alpha_0(\Tcm{})$ is conjugate to the mean number of neutrinos and $\alpha_1(\Tcm{})$ is conjugate to the mean energy.  The parameter $\alpha_2(\Tcm{})$ is conjugate to the average square of neutrino energy.

\begin{figure}[!htbp]
    \centering
    \includegraphics[width=1.0\textwidth]{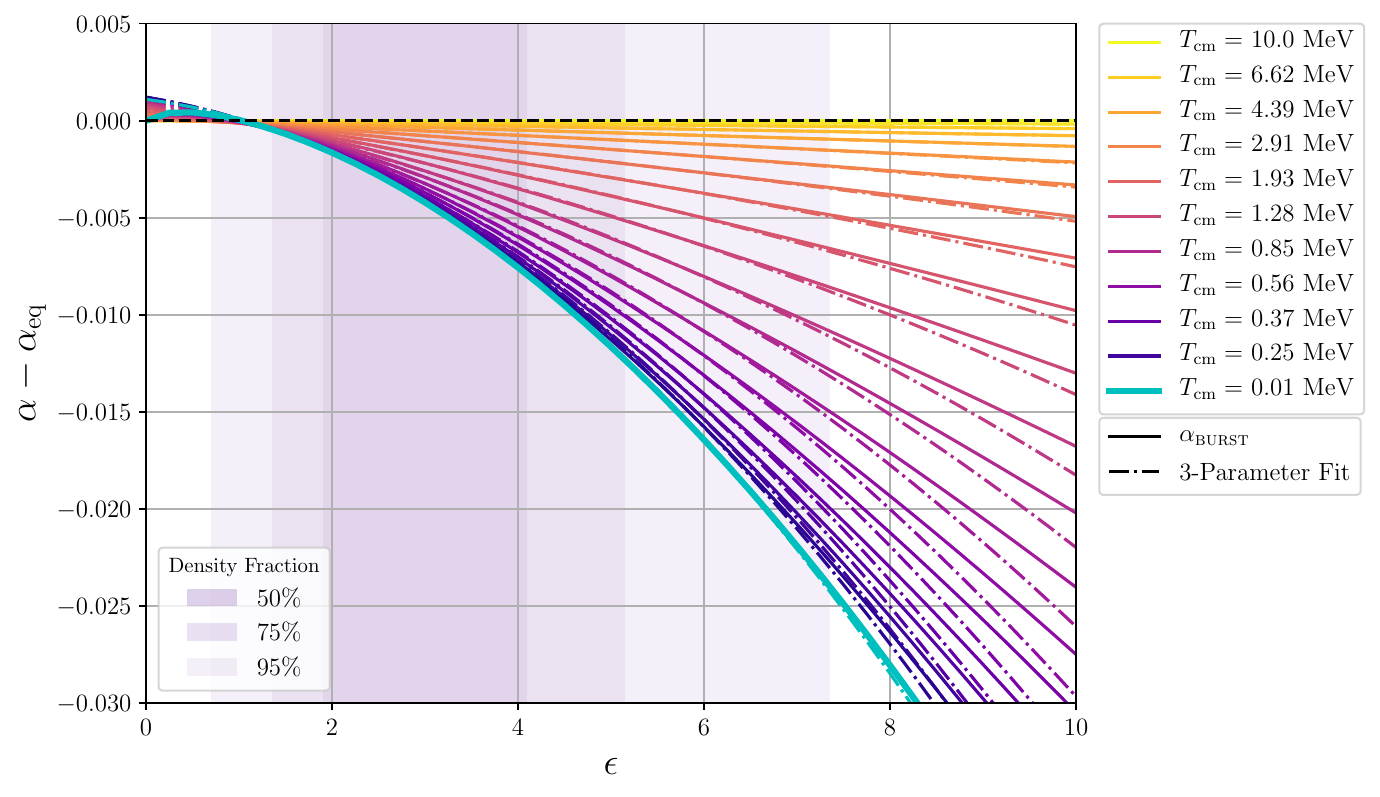}
    \caption[Evolution of $\alpha$ with quadratic fit]
    {
    Same as Figure~\ref{fig:alpha_evolution_linear_fit}, but for a 3-parameter fit of the form $\alpha(\epsilon,\Tcm{}) = \alpha_0(\Tcm{}) + \alpha_1(\Tcm{}) \epsilon + \alpha_2(\Tcm{}) \epsilon^2$.
    Note that the 3-parameter model provides a good fit throughout weak decoupling, especially in the region with the highest number density of neutrinos, and matches very well the asymptotic occupation probability at low temperatures.
    }
    \label{fig:alpha_evolution_quadratic_fit}
\end{figure}

\begin{figure}[!tbph]
    \centering
    \includegraphics[width=0.75\textwidth]{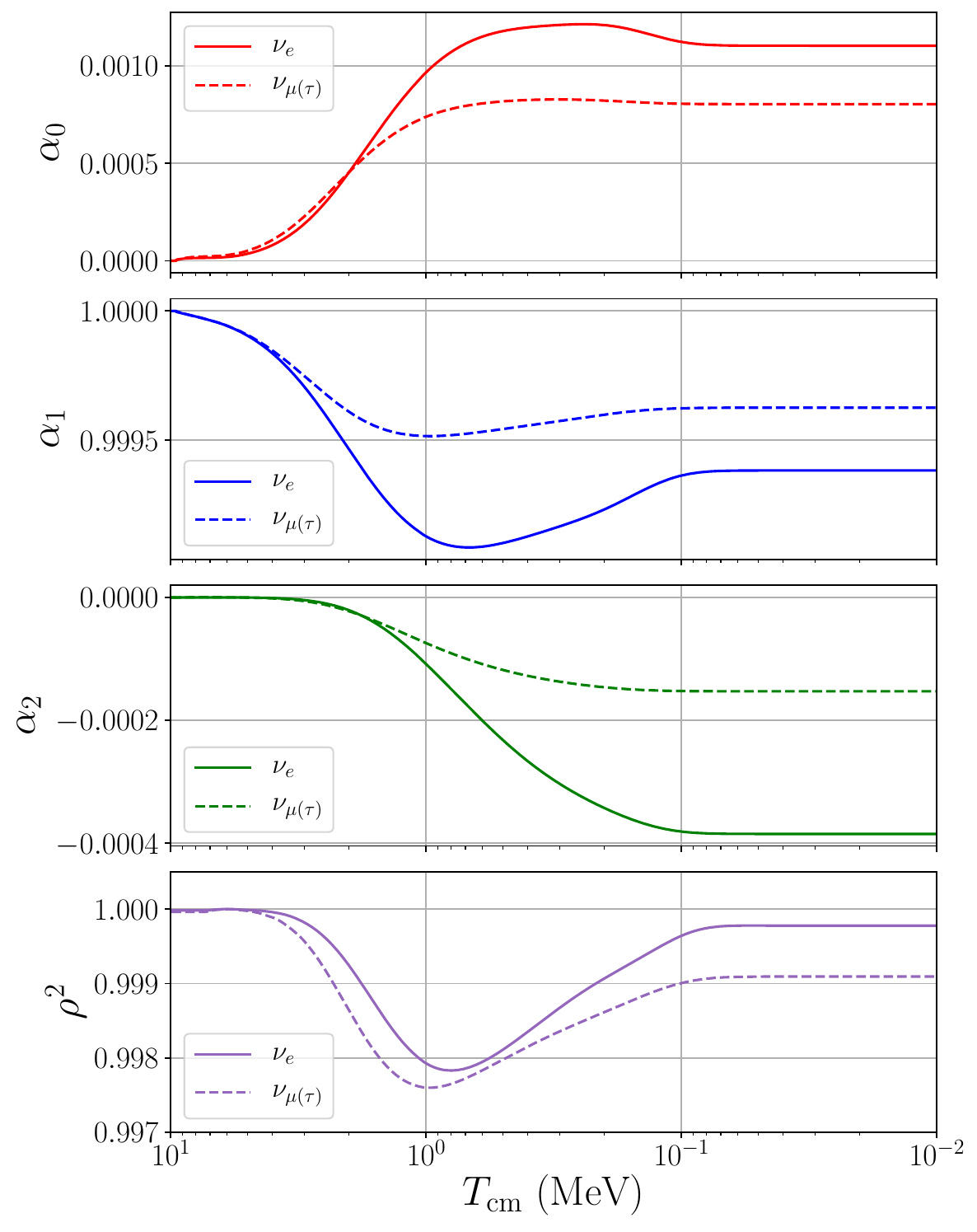}
    \caption[Quadratic fit coefficients]
    {
    Same as Figure~\ref{fig:alpha_evolve_linear} but for a 3-parameter model of the form $\alpha_\mathrm{fit}(\epsilon,\Tcm{}) = \alpha_{0}(T_{\text{cm}}) + \alpha_{1}(T_{\text{cm}})\cdot \epsilon + \alpha_{2}(T_{\text{cm}})\cdot \epsilon^2$.
    In this model, $\alpha_{0}$, $\alpha_{1}$, $\alpha_{2}$ are conjugate to the expected number, energy, and squared-energy, respectively, of the neutrino distributions.
    The goodness of fit $\rho^2$ for this model can be compared directly to that of the 2-parameter model shown in Figure~\ref{fig:alpha_evolve_linear}, demonstrating that the 3-parameter model provides an excellent approximation to the true evolution of $\alpha$ throughout the decoupling epoch.
    }
    \label{fig:alpha_evolve_quadratic}
\end{figure}

\begin{figure}[!tbph]
    \centering
    \includegraphics[width=1.0\textwidth]{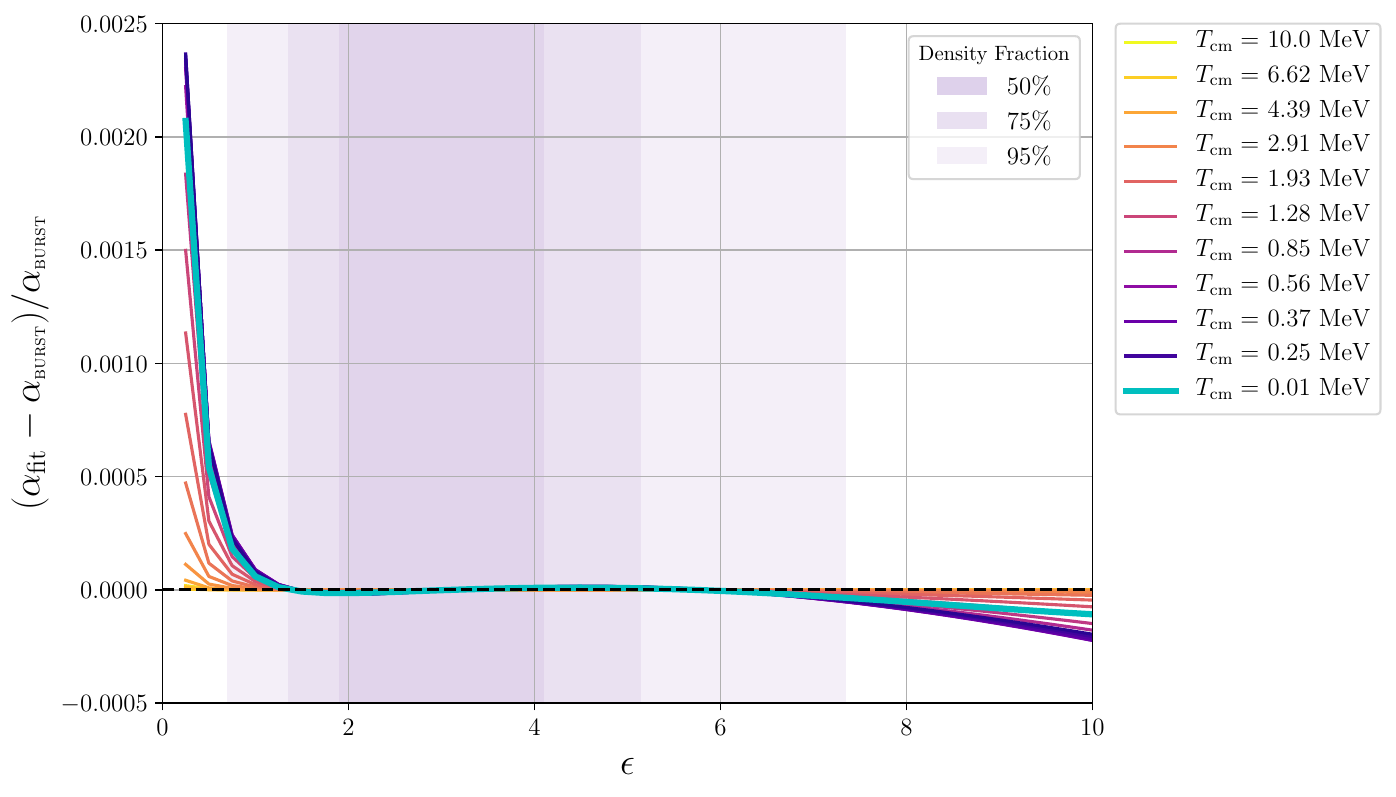}
    \caption[Percent difference between fitting model and output of $\alpha$.]
    {
    Fractional difference between the best fit 3-parameter model describing the occupation probability of electron neutrinos $\alpha_\mathrm{fit}(\epsilon,\Tcm{}) = \alpha_{0}(T_{\text{cm}}) + \alpha_{1}(T_{\text{cm}})\cdot \epsilon + \alpha_{2}(T_{\text{cm}})\cdot \epsilon^2$ and the value of $\alpha(\epsilon,\Tcm{})$ calculated with \burst{}, labeled $\alpha_{\textsc{\scriptsize burst}}$.  While the 3-parameter model is a remarkably good fit for all values of $\Tcm$ and $\epsilon$, the largest fractional deviations occur for low $\epsilon$ at low $\Tcm$ and high $\epsilon$ at intermediate $\Tcm$.
    }
    \label{fig:perdiff}
\end{figure}

\begin{figure}[!tbph]
    \centering
    \includegraphics[width=0.75\textwidth]{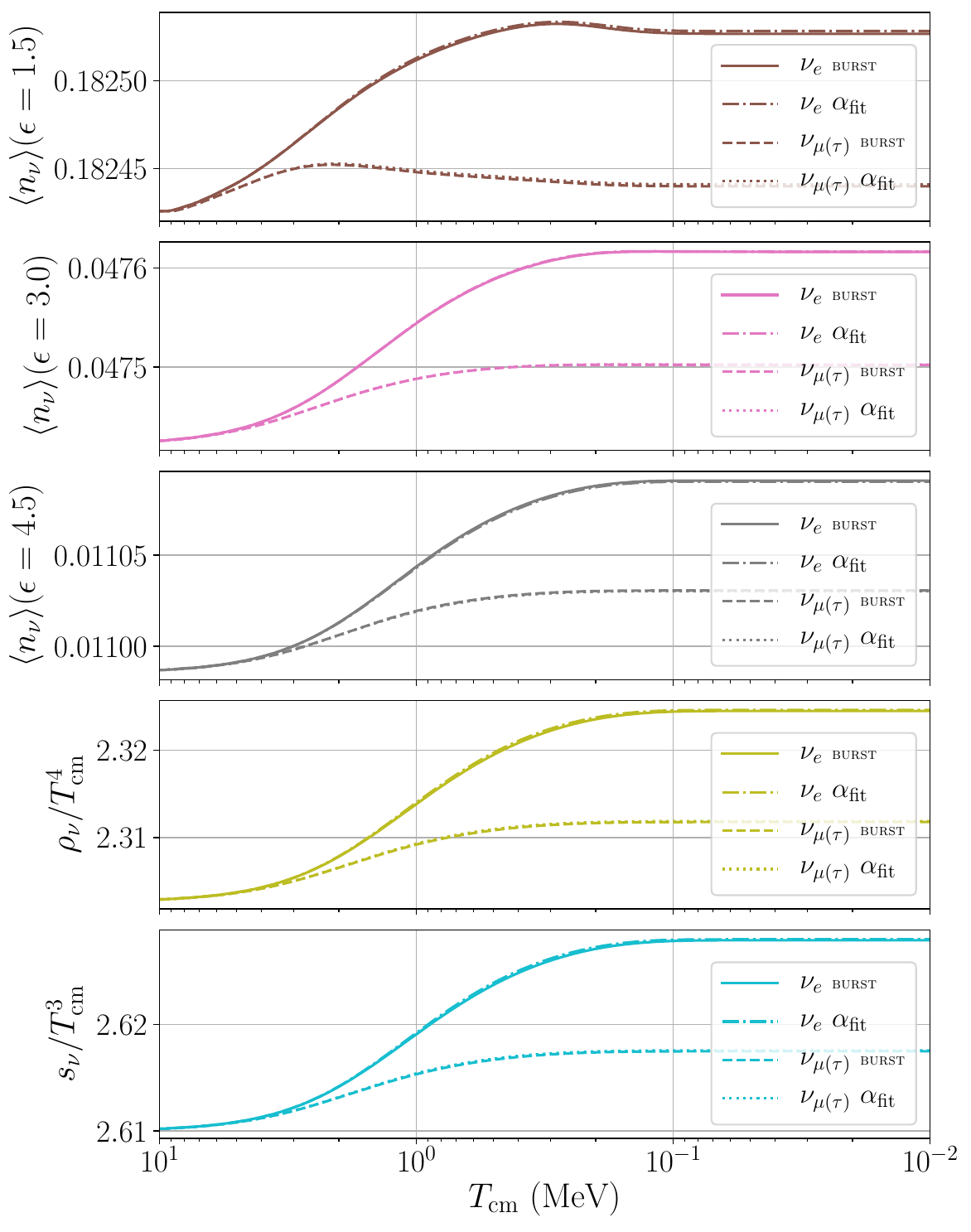}
    \caption[Comparison of occupation, energy density, and entropy]
    {
    \edit{Comparison of the distribution function at three values of $\epsilon$, the energy density, and the entropy density derived from \burst{} and from the quadratic $\alpha$ model.  It is evident from these comparisons that the quadratic model provides an excellent fit to the output of \burst{}.}
    }
    \label{fig:n_rho_s}
\end{figure}

Just as was done for the 2-parameter model, we performed a density-weighted fit of the model shown in Eq.~\eqref{eq:3param_fit_def} to the output of the \burst code, using the same weighting function.  Figure~\ref{fig:alpha_evolution_quadratic_fit} shows that this 3-parameter model provides an excellent fit to the neutrino occupation probability calculated from full transport calculation throughout the epoch of neutrino decoupling.  While the goodness of fit for the 2-parameter model was $1-\rho^2\approx 0.07$, the 3-parameter model fit achieves $1-\rho^2<0.0025$ throughout the whole epoch of interest, and describes the late-time asymptotic spectrum with $1-\rho^2<0.001$.  The best-fit parameters and goodness of fit as defined in Eq.~\eqref{eq:goodness_of_fit} for the quadratic model are shown as a function of $\Tcm{}$ in Figure~\ref{fig:alpha_evolve_quadratic}.  To further emphasize how accurately this 3-parameter model describes the neutrino distribution, we show the fractional difference between the best fit model \edit{for $\alpha$} and the output of \burst{} in Figure~\ref{fig:perdiff} \edit{and a comparison of the distribution function, the energy density, and the entropy density in Figure~\ref{fig:n_rho_s}}.

%%%%%%%%%%%%%%%%%%%%% Alpha evolution - quadratic fit %%%%%%%%%%%%%%%%%%%%
%%%%%%%%%%%%%%%%%%%%

%%%%%%% Quadratic Alpha Fit Coefficients %%%%%%%%%%%%
%%%%%%%%%%%%%%%%%%%%%%%%

The quadratic model is particularly successful at describing the neutrino occupation probability after neutrinos have fully decoupled.  At late times, each of $\alpha^{(f)}_0$, $1-\alpha^{(f)}_1$, and $\alpha^{(f)}_2$ have magnitude $\lesssim 10^{-3}$.
We note, however, that even for small values for $\alpha_2^{(f)}$, at large $\epsilon$ the total $\alpha^{(f)}$ will become large in magnitude and negative in sign, giving unphysical occupation probabilities.  As a result, we stress that our expansion in three parameters is a good fit for the $\epsilon$ range shown in Fig.\ \ref{fig:alpha_evolution_quadratic_fit}.
This motivates a perturbative treatment of the deviation from an equilibrium Fermi-Dirac distribution that we will explore further in the next subsection.

\edit{One could also consider a higher-order polynomial fit to the evolution of $\alpha$.  We tested this possibility and found that the  $\alpha_3$ parameter in a cubic fit was significantly smaller than the other $\alpha$ coefficients, and the goodness of fit only marginally improved compared to the quadratic model.  For this reason, we focus on the quadratic model in what follows.}

%%%%%%%% Alpha Fit Differential %%%%%%%%%%%%%
%%%%%%%%%%%%%%%%%%%%%%%%%%

%%%%%%%% n rho s fit %%%%%%%%%%%%%
%%%%%%%%%%%%%%%%%%%%%%%%%%

%%%%%%%%%%%%%%%%%%%%%%%%%%%%%%%%%%%%%%%
\subsection{Perturbed Densities}
\label{subsec:Neff_Calc}
%%%%%%%%%%%%%%%%%%%%%%%%%%%%%%%%%%%%%%%%

%%%%%%%%%%%%%%%%

The main cosmological observables resulting from the weak decoupling epoch are the asymptotic BBN products (to be discussed below) and the energy density in neutrinos which is often parameterized in terms of a number $N_\mathrm{eff}$, defined in Eq.~\eqref{eq:Neff_def}. The radiation density as parameterized by $N_\mathrm{eff}$ affects the expansion rate, and hence affects BBN and the evolution of cosmological perturbations. For the Standard Model, $\Nf=3.044(1)$~\cite{EscuderoAbenza:2020cmq,Akita:2020szl,Froustey:2020mcq,Bennett:2020zkv} with 3 neutrino species, the excess over 3 a reflection of extra entropy transferred  from leptonic weak interactions in the late stages of weak decoupling. Its strongest constraints come from analysis of the CMB power spectrum, and current bounds from Planck give $\Nf=2.99\pm0.17~(68\%~\mathrm{CL})$~\cite{Planck:2018vyg}.
In general, the value of $\Nf$ (as well as quantities such as number density, pressure, and entropy density) depends on $\alpha$ after weak decoupling and may differ from the value predicted in the standard scenario.

At a given order of the power series expansion of $\alpha$ in energy, $\alpha_\nu \approx \sum_{m=0}^M \alpha_m E^m$, we can relate the $\alpha_m(t)$ to the mean $m$-th powers of the energy.
Since we have shown $M=2$ is all that is needed for an excellent description of  $\alpha_\nu (E,t) $, the mean number density, the mean energy density, 
and the density of the fluctuations in the energy $\langle \delta E^2 \rangle$
are the quantities of interest.  Significant insight can be gained by calculating the perturbations to these quantities that are linear in $\Delta \alpha_{\nu f} = \alpha_{\nu f} - \epsilon$.

For small values of the parameters  defining the model for $\alpha(\epsilon,\Tcm)$, it is straightforward to calculate the perturbations to quantities like the neutrino energy density, pressure, and entropy.  To first order in $\alpha_0$, $\alpha_1-1$, and $\alpha_2$, the distribution function of neutrinos is given by
\begin{equation}
    \left\langle \widehat{n}_{\nu_f} \right\rangle = \frac{1}{e^{\alpha(\epsilon,\Tcm{})}+1} \simeq
    \frac{1}{e^{\epsilon}+1}-\frac{\left(\alpha_0^{(f)}(\Tcm{}) + (\alpha_1^{(f)}(\Tcm{}) -1)\epsilon + \alpha_2^{(f)}(\Tcm{})\epsilon^2\right)e^\epsilon}{\left(e^{\epsilon}+1\right)^2} + \ldots
    \label{eq:Perturbed_Distribution}
\end{equation}
The number density for such a distribution is
\begin{align}
    N_{\nu_f} &=  \frac{g_f}{2\pi^2}\Tcm{}^3 \int d \epsilon \, \epsilon^2 \frac{\partial \mathcal{F}}{\partial \alpha_0} = 
    \frac{g_f}{2\pi^2}\Tcm{}^3 \int d \epsilon \, \epsilon^2 \left\langle \widehat{n}_{\nu_f} \right\rangle  \nonumber \\
    &\simeq \frac{g_f\Tcm{}^3}{2\pi^2} 
    \left(\frac{3\zeta(3)}{2} - \frac{\pi^2}{6}\alpha_0^{(f)}(\Tcm{}) - \frac{9\zeta(3)}{2}(\alpha_1^{(f)}(\Tcm{})-1) - \frac{7\pi^4}{30}\alpha_2^{(f)}(\Tcm{}) \right) + \ldots \, ,
    \label{eq:Perturbed_Number_Density}
\end{align}
the energy density is
\begin{align}
    \rho_{\nu_f} &= \frac{g_f}{2\pi^2}\Tcm{}^3 \int d \epsilon \, \epsilon^2 \frac{\partial \mathcal{F}}{\partial \alpha_1} = 
    \frac{g_f}{2\pi^2}\Tcm{}^4 \int d \epsilon \, \epsilon^3 \left\langle \widehat{n}_{\nu_f} \right\rangle  \nonumber \\
    &\simeq \frac{g_f\Tcm{}^4}{2\pi^2} \left(\frac{7\pi^4}{120} 
    - \frac{9\zeta(3)}{2}\alpha_0^{(f)}(\Tcm{}) 
    - \frac{7\pi^4}{30}(\alpha_1^{(f)}(\Tcm{})-1) 
    - \frac{225\zeta(5)}{2}\alpha_2^{(f)}(\Tcm{}) \right) + \ldots \, ,
    \label{eq:Perturbed_Energy_Density}
\end{align}
while the pressure is $P_{\nu_f} = \rho_{\nu_f}/3$, and the entropy density is
\begin{align}
    s_{\nu_f} &= -\frac{g_f}{2\pi^2}\Tcm{}^3 \int d \epsilon \, \epsilon^2 \left[ \left\langle \widehat{n}_{\nu_f} \right\rangle \ln \left\langle \widehat{n}_{\nu_f} \right\rangle + \left( 1 - \left\langle \widehat{n}_{\nu_f} \right\rangle \right) \ln \left( 1 -  \left\langle \widehat{n}_{\nu_f} \right\rangle \right) \right] \nonumber \\
    &\simeq \frac{g_f\Tcm{}^3}{2\pi^2} \left(\frac{7\pi^4}{90} 
    - \frac{9\zeta(3)}{2}\alpha_0^{(f)}(\Tcm{}) 
    - \frac{7\pi^4}{30}(\alpha_1^{(f)}(\Tcm{})-1) 
    - \frac{225\zeta(5)}{2}\alpha_2^{(f)}(\Tcm{}) \right) + \ldots \, .
    \label{eq:Perturbed_Entropy_Density}
\end{align}

%We can also calculate the perturbation to $\neff \equiv \frac{8}{7} \left(\frac{11}{4}\right)^{4/3} \frac{\rho_\nu}{\rho_\gamma}$ in terms of the coefficients appearing in $\alpha(\epsilon,\Tcm{})$.
%However, there is one additional ingredient we need for $\neff$, which is the ratio $\Tcm{}/T_\gamma$.  This can be computed from the output of \burst, where it is found that $\Tcm{}/T_\gamma = 0.71477$ for times well after neutrino decoupling, which is slightly larger than the value $(4/11)^{1/3}$ predicted in the simplified case where neutrinos instantaneously decouple prior to electron-positron annihilation.
\edit{
The perturbative expansions of the above quantities only rely on a single energy scale, namely, \tcm.  This energy scale is inherent in our dual-$\alpha$ description of the distribution functions, as the neutrino energy $E_\nu=\epsilon\,\tcm$ and $\alpha$ is a function of $\epsilon$.  We can also determine perturbative expansions of other neutrino-related quantities, although doing so may require additional energy scales indirectly related to \tcm.  For example, we adopt the standard cosmological parameter $\neff \equiv \frac{8}{7} \left(\frac{11}{4}\right)^{4/3} \frac{\rho_\nu}{\rho_\gamma}$ to characterize the radiation energy density at epochs after electron-positron annihilation.
The photon energy density is explicit in the definition of \neff, and therefore the photon temperature $T_\gamma$ is a required ingredient to determine \neff perturbatively with $\alpha$.  More accurately, we need the ratio $\tcm/T_\gamma$ to determine \neff.  This ratio depends on the transfer of energy from the electromagnetic plasma to the neutrino seas.  As we have stressed and explicitly shown in this work, that transfer of energy distorts the neutrino distributions and precipitates non-zero values for $\alpha_0,\,\alpha_1-1,$ and $\alpha_2$.  In practice, $\tcm/T_{\gamma}$ cannot be extracted from examining the non-zero $\alpha$ coefficients in isolation of the other components in the system. To obtain a perturbative expansion of \neff with the $\alpha$ coefficients, we must adopt the output ratio $\tcm/T_{\gamma}$ from the \burst Boltzmann-transport simulations and use that value to calculate \neff perturbatively.
We find $\Tcm{}/T_\gamma = 0.71477$ in the \burst output
}
for times well after neutrino decoupling, which is slightly larger than the value $(4/11)^{1/3}$ predicted in the simplified case where neutrinos instantaneously decouple prior to electron-positron annihilation.
Treating the change to the fourth power of the temperature ratio and the perturbations to the distribution function as being of the same order and working to first order in these quantities, we find that the change to $\neff$ is
\begin{align}
    \Delta \neff \simeq \sum_f \frac{g_f}{2} \left[  \left(\frac{11}{4}\right)^{4/3} \left( \frac{\Tcm}{T_\gamma}\right)^4 -1 - \frac{540 \zeta(3)}{7\pi^4}\alpha_0^{(f)}(\Tcm{}) - 4(\alpha_1^{(f)}(\Tcm{})-1) - \frac{13500 \zeta(5)}{7\pi^4} \alpha_2^{(f)}(\Tcm{})  \right] \, ,
\end{align}
which for the best-fit values of the 3-parameter model for $\alpha(\epsilon,\Tcm{})$ defined in Sec.~\ref{ssec:3ParamAlpha} gives $\neff=3.034$ for $\Tcm{} <10~\mathrm{keV}$.
%This agrees very well with the value of $\neff{}$ computed directly from \burst as described in detail in Ref.~\cite{Grohs:2015tfy}, giving further support to the notion that our compact description provides an accurate model of the distribution calculated with a full transport treatment.
\edit{This agrees very well with the value of $\neff{}$ computed directly from \burst as described in detail in Ref.~\cite{Grohs:2015tfy}, giving further support to the notion that our compact description provides an accurate model of the out-of-equilibrium distributions.  If we were to do a full calculation of neutrino decoupling using the $\alpha$ coefficients --- and coupling those distributions to the other components of the system --- we would follow the ratio $\tcm/T_{\gamma}$ to late times and subsequently deduce the value of \neff, essentially coupling the $\alpha$ evolution to the thermodynamics of the early universe.  We leave this calculation to future work.}

We can also consider a more general class of quantities that can be computed from our framework.
For example the perturbed heat capacity of the cosmic neutrino background is given by
\begin{align}
    C_{\nu_f} &= \frac{g_f}{2\pi^2}\Tcm{}^3 \int d \epsilon \, \epsilon^2 \left(-\alpha_1^2 \frac{\partial^2 \mathcal{F}}{\partial \alpha_1^2} \right) = 
    \frac{g_f}{2\pi^2}\Tcm{}^3 \int d \epsilon \, \epsilon^2 \frac{\alpha_1^2 \epsilon^2 e^{\alpha}}{\left(1+e^{\alpha}\right)^2} \nonumber \\
    &\simeq \frac{g_f\Tcm{}^3}{2\pi^2} \left(\frac{7\pi^4}{30} 
    - 18\zeta(3)\alpha_0^{(f)}(\Tcm{}) 
    - \frac{7\pi^4}{10}(\alpha_1^{(f)}(\Tcm{})-1) 
    - 675\zeta(5)\alpha_2^{(f)}(\Tcm{}) \right) + \ldots \, .
    \label{eq:Perturbed_Heat_Capacity}
\end{align}

We have shown in this section that we can achieve a very precise description of the \cnub for a standard thermal history using just three parameters.  In the subsequent sections, we further apply our formalism to gain some intuition for how physics beyond the Standard Model may alter the properties of \cnub and its impact on observables.

%%%%%%%%%%%%%%%%%%%%%%%%%%%%%%%
\section{Differential Neutrino Productivity}
\label{sec:NuDiffVis}
%%%%%%%%%%%%%%%%%%%%%%%%%%%%%%%

\subsection{Determining the Differential Visibility}

Processes beyond the Standard Model, such as dark matter decay or flavor oscillation involving sterile neutrino states, may alter the standard neutrino decoupling process and the resulting neutrino occupation probabilities described above. These processes may add or remove neutrinos from the system, ultimately leading to changes in late-time cosmological observables.  Disturbances to the neutrino distribution introduced at sufficiently high temperatures would be expected to be smoothed away due to the high rate of weak interactions compared to the expansion rate which would quickly relax the neutrinos to their equilibrium distribution. If, however, a disturbance was introduced at low temperatures, the scattering rates may be small compared to the Hubble rate such that the disturbance is not able to relax, resulting in a deviation from the neutrino occupation probability that may be imprinted in late-time observables. We can achieve a more quantitative description of the conditions on disturbances that leave a lasting imprint on the neutrino occupation probability by drawing an analogy with CMB physics.  In photon decoupling, it is convenient to define the optical depth and differential visibility to describe the probability that CMB photons experienced their last scattering at a given redshift. Here we will use a similar treatment to describe the last scattering of neutrinos.

First, we can express the Boltzmann equation for the neutrinos [Eq.~\eqref{eq:boltzmann}] in a slightly more convenient form
\begin{equation} 
    \frac{d}{dt}n_{\nu_f} = -\Gamma_{\nu_f} n_{\nu_f} + P_{\nu_f}(1-n_{\nu_f}) \, ,
    \label{eq:boltzmann_mod}
\end{equation}
where $\Gamma_{\nu_f}$ is the rate of neutrinos scattering out of a state, and $P_{\nu_f}$ is the production rate of neutrinos flowing into a state. As in Sec.~\ref{sec:CollisionRate}, we will refer to the expectation value of the distribution function $\left \langle \widehat{n}_{\nu_f} \right\rangle $ as simply $n_{\nu_f}$ in this section in order to streamline our notation. Fermi blocking is responsible for the factor $(1-n_{\nu_f})$ multiplying the production rate, which suppresses the flow of neutrinos into a given state. We introduce the variable $\Gamma_{\nu_f}' = \Gamma_{\nu_f} + P_{\nu_f}$, which is the effective scattering rate out of states, taking account of Fermi-blocking. The Boltzmann equation can then be rewritten as
\begin{equation} 
    \frac{d}{dt} n_{\nu_f} = -\Gamma'_{\nu_f} n_{\nu_f} + P_{\nu_f} \, .
    \label{eq:boltzmann_mod_2}
\end{equation}
In this notation, we can relate the probability of neutrino last scattering to the combination $\frac{\Gamma_i'}{H}$, which gives the effective scattering rate relative to the expansion rate.

The optical depth for a given neutrino species $\nu_f$ is a summation of the scattering rates along the line of sight, and is given as
\begin{equation}
    \tau_{\nu_f} = \int_{a}^{a_{0}}\frac{\Gamma'_{\nu_f}}{H}  d \, \ln a' 
    = \int_{T_{0}}^{\Tcm{}}\frac{\Gamma'_{\nu_f}}{H} \frac{1}{\Tcm{}'} d\Tcm{}' \, ,
    \label{eq:optical_depth_def}
\end{equation}
where $a$ is the scale factor, and a subscript $0$ refers to a quantity evaluated today. The neutrino differential visibility for neutrino species $\nu_f$ is given by
\begin{equation}
    \diffvis_{\nu_f} \equiv \frac{d}{d \, \ln a}e^{-\tau_{\nu_f}} = \frac{\Gamma'_{\nu_f}}{H}e^{-\tau_{\nu_f}} \, .
\end{equation}
Figure~\ref{fig:diff_vis} shows the differential visibility of electron-type neutrinos as a function of $\Tcm{}$ and $\epsilon$.  The magnitude of the differential visibility provides a measure of probability that a neutrino of comoving energy $\epsilon$ last scattered at a particular value of $\Tcm{}$.

%%%%%%%%%%%%%%%%%%%%
\begin{figure}[!htbp]
    \centering
    \includegraphics[width=0.9\textwidth]{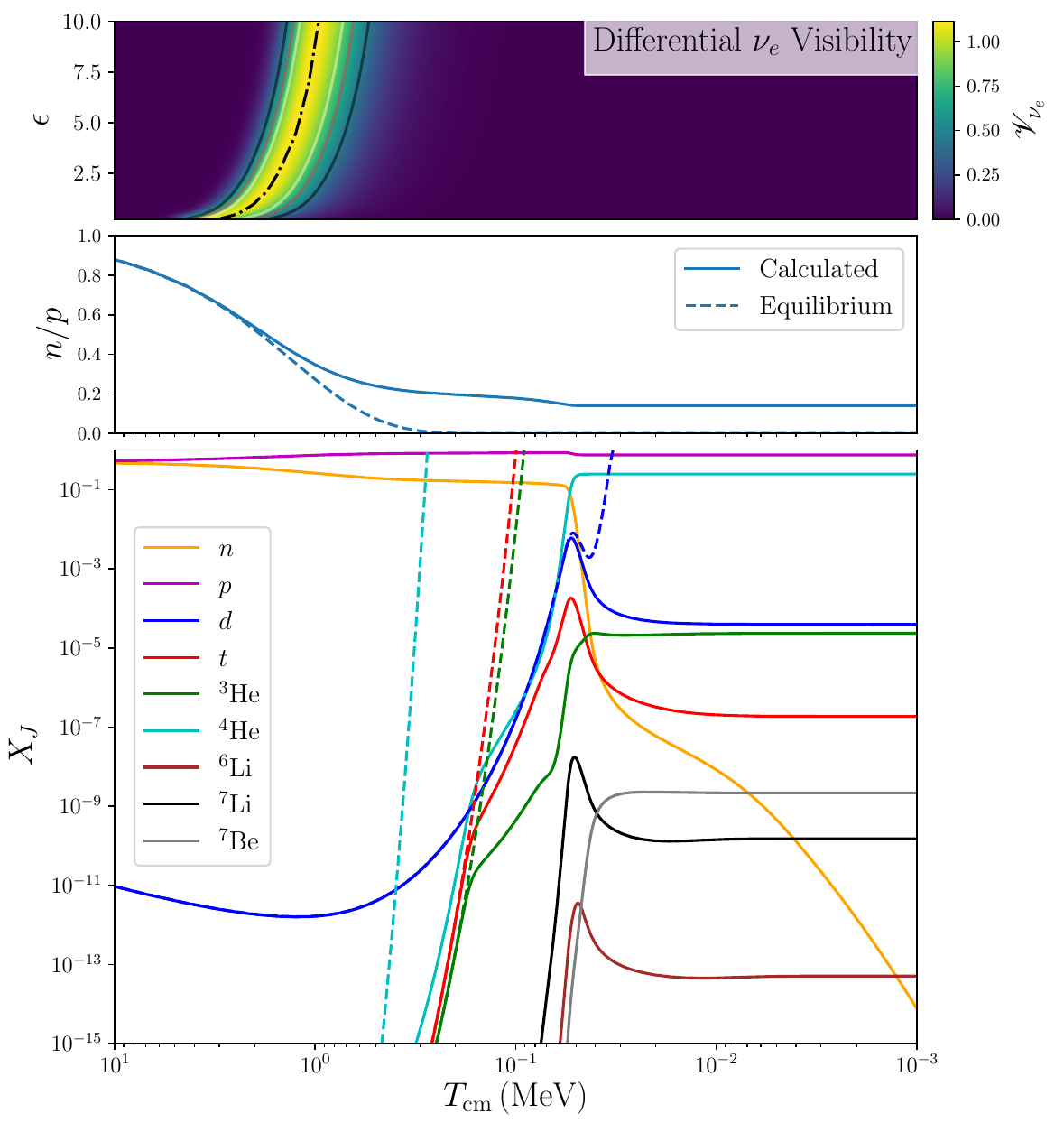}
    \caption[Neutrino visibility and nucleosynthesis]{
    (\textit{Top:}) Differential $\nu_e$ visibility $\diffvis_{\nu_e}$ 
    %$\frac{d}{d \ln a}e^{-\tau_{\nu_{e}}}$
    as a function of $T_\mathrm{cm}$ and $\epsilon$.  Peak visibility is shown in a black dash-dot line along with contours denoting 50\%, 75\%, and 95\% of peak visibility in solid colored lines. 
    (\textit{Middle:}) Total (free and bound) neutron to proton ratio as a function of $\Tcm{}$ for the full BBN calculation compared to the equilibrium expectation given by $\left(n/p\right)_{\text{eq}} = \exp(-Q/T)$. 
    (\textit{Bottom:}) Evolution of nuclide abundances during nucleosynthesis.  Solid lines show the evolution of the mass fractions by integrating the nuclear reaction network.  Dashed lines show NSE mass fractions, as given by Eq.~\eqref{nse}.}
    \label{fig:x_nse}
\end{figure}
%%%%%%%%%%%%%%%%%%%%

Figure~\ref{fig:diff_vis} showcases a novel way of visualizing the weak decoupling of neutrinos from the plasma and provides useful insight about the neutrino freeze-out process. The differential visibility peak defines a border that can be referred to as the neutrino-sphere. Inside the neutrino-sphere (to the left of the peak), scattering rates are large compared to the Hubble expansion rate, so energy injections/non-equilibrium perturbations are easily dispersed across the distribution functions. Outside the neutrino-sphere, scattering rates are small compared to the Hubble rate, so any perturbations or injection of energy cannot be easily dissipated. The latter situation is what causes the late-time neutrino distribution function to deviate from equilibrium, and which can have impacts on late-time observables.

The process of primordial nucleosynthesis provides one example of how neutrino spectral distortions may impact late-time observables.
The neutrino differential visibility is shown compared to the evolution of the neutron-to-proton ratio and the light nuclide abundances in Figure~\ref{fig:x_nse}, to be discussed further in Section~\ref{sec:WeakInteraction}.  One can see from Figure~\ref{fig:x_nse} that the broad peak of neutrino last scattering overlaps with the period when the neutron-to-proton ratio begins to deviate from the equilibrium expectation; this deviation plays an essential role in determining the primordial light element abundance yields.

%%%%%%%%%%%%%%% \nu_\mu Differential Visibility v2 %%%%%%%%%%
\begin{figure}[!tbp]
\centering
\includegraphics[width=0.85\textwidth]{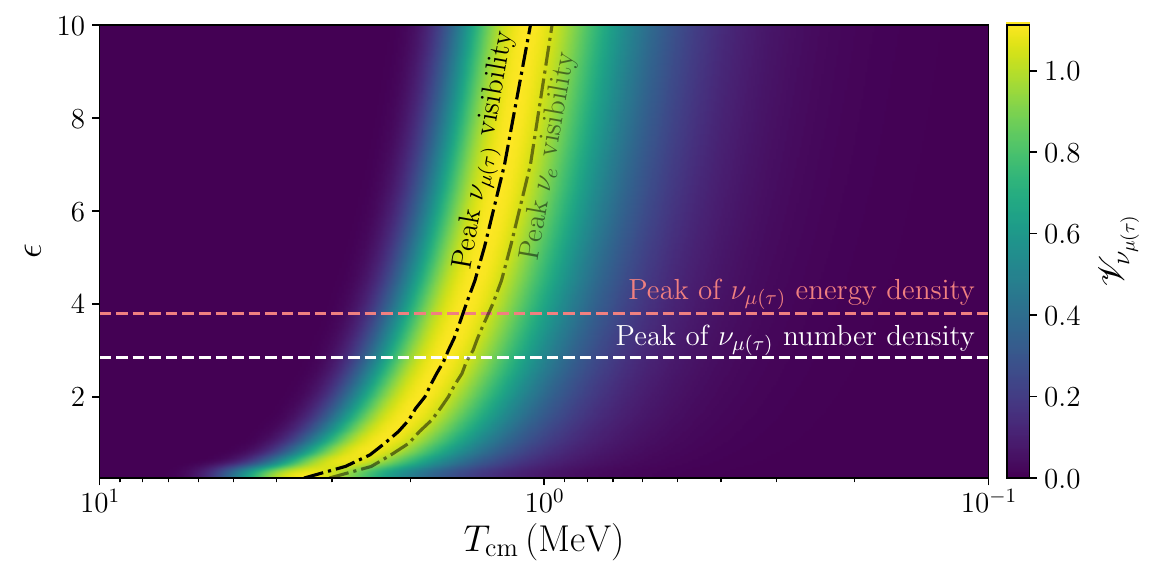}
    \caption[Differential visibility of $\nu_{\mu(\tau)}$.]{Differential visibility for $\mu$- and $\tau$-type neutrinos.    As can be seen from the difference in the peak visibility curves, the neutrino-sphere for $\nu_{\mu(\tau)}$ appears at a slightly higher $\Tcm{}$ compared to $\nu_{e}$, as expected from the larger interaction rate of electron-type neutrinos.}
\label{fig:diff_vis_num_v2}
\end{figure}
%%%%%%%%%%%%%%%%%%%%%%%%

Due to the fact that there are charged-current interaction channels for electron-type neutrinos with the plasma that are not present for the $\nu_\mu$ and $\nu_\tau$, we expect that the differential visibility will differ for the $\mu$ and $\tau$ neutrinos compared to the electron neutrinos.
We compare the differential visibility and neutrino-sphere of $\nu_{e}$ and $\nu_{\mu(\tau)}$ in Figure~\ref{fig:diff_vis_num_v2}. Recall that since we neglect flavor oscillations and $\nu_{\mu}$ undergo the same interactions as $\nu_{\tau}$, the scattering rates are identical, and thus their differential visibilities are the same as well.
The neutrino-sphere for $\nu_{\mu(\tau)}$ appears at slightly larger values of $\Tcm{}$ than that of $\nu_{e}$. This is due to the higher scattering rate of the latter as can be seen in Table \ref{colltab}. Only electron-neutrinos have a direct effect on BBN production processes via charged-current interactions, while all neutrino species have an indirect effect through their impact on the Hubble expansion rate.

\subsection{Differential productivity from neutrino decoupling}

Out of equilibrium processes involving neutrinos including, for example, $e^- + e^+ \leftrightarrow \nu + \bar{\nu}$, cause a transfer of entropy from the photon-$e^\pm$-baryon plasma into the decoupling neutrino seas.
This redistribution of energy and degrees of freedom results in a change in the phasing of the temperature, scale factor, and time that, in turn, results in altered light-element abundance yields.
We define the differential productivity by rewriting Eq.~\eqref{eq:boltzmann_mod_2} in the form
\begin{equation}
    \frac{d}{d \, \ln a}\left( e^{-\tau_{\nu_f}} n_{\nu_f} \right) = \frac{\Gamma_{\nu_f}'}{H}e^{-\tau_{\nu_f}} \sum_{r} \frac{P_{r}}{\Gamma_{\nu_f}'} = \frac{\Gamma_{\nu_f}'}{H}e^{-\tau_{\nu_f}} \sum_{r} n_{\mathrm{eq}} \frac{P_{r}}{P_{\mathrm{tot}}}.
    \label{eq:boltzmann_mod_3}
\end{equation}
Here, $P_{r}$ is the production from process $r$, $P_{\mathrm{tot}}$ is the total production rate from all neutrino-scattering processes, and $n_{\mathrm{eq}} = P_{\mathrm{tot}}/\Gamma'_{\nu_f}$ is the equilibrium distribution function found by setting Eq.~\eqref{eq:boltzmann_mod_2} equal to zero. Using this formalism, we are able to observe how a particular production process, or set of production processes, will impact the distribution functions over time. In other words, we can visualize when a given process will have the most impact on the occupation probabilities.

\begin{figure}[!tbp]
\centering
\includegraphics[width=0.95\textwidth]{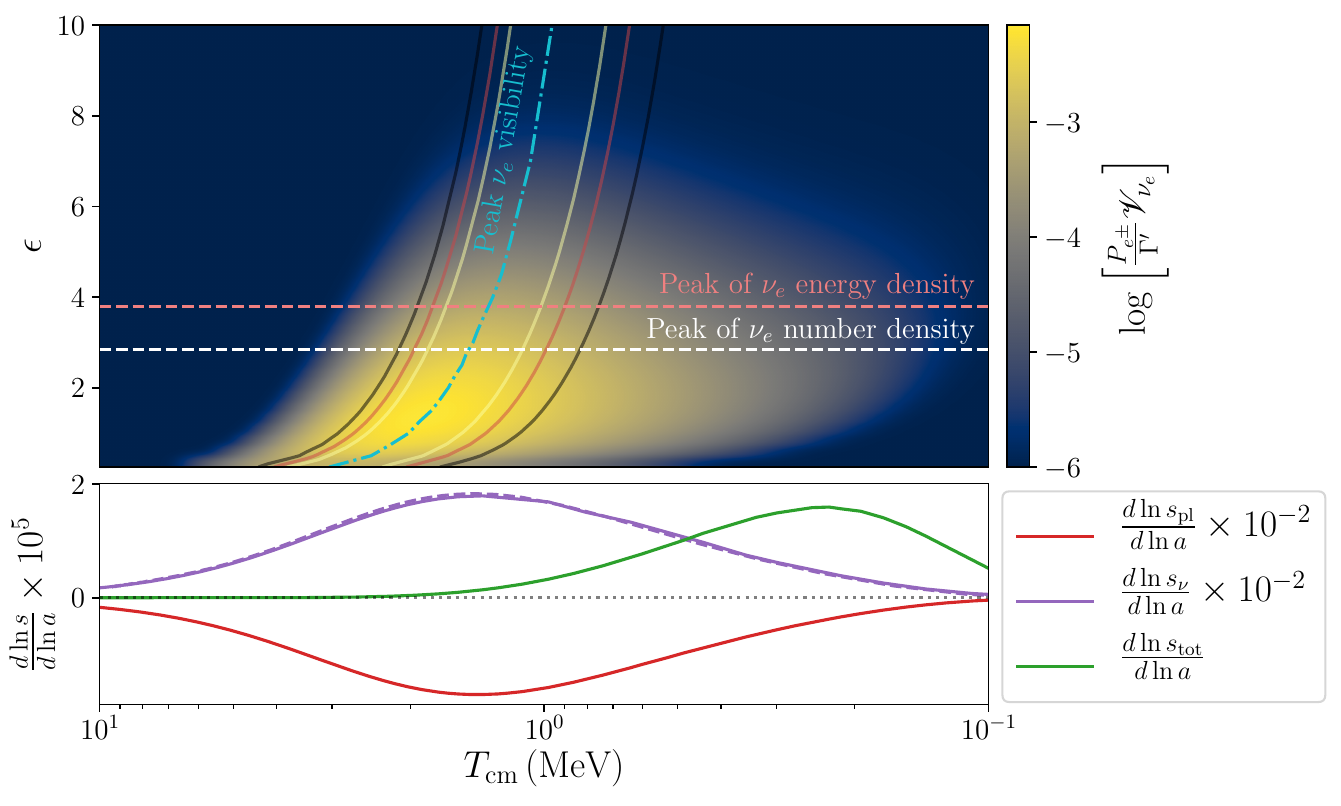}
    \caption[Visible production of $\nu_{e}$ from $e^{-}$ $e^{+}$ annihilation.]
    {(\textit{Top:}) Logarithm of the density-weighted visible production of $\nu_{e}$ from $e^{-}$ $e^{+}$ annihilation ($P_{e^{-}e^{+}}$), found by multiplying the differential visibility $\diffvis_{\nu_e}$ by $\frac{P_{e^{-}e^{+}}}{\Gamma_{\nu_{e}}'}$ ($\Gamma'$ is the effective scattering rate for $\nu_{e}$) with a weighting of $\epsilon^{2}\cdot n_{\nu_{e}}(\epsilon)$. We also show contours for 50\%, 75\%, and 95\% of peak visibility for $\nu_{e}$ for the sake of comparison.
    (\textit{Bottom:}) Logarithmic change to the neutrino, plasma, and total entropy as a function of $\Tcm{}$.  Note that the derivative of the total entropy is scaled up by a factor of $10^2$ relative to the neutrino and plasma entropy derivatives.  The purple dashed line shows the logarithmic change to the neutrino entropy as computed from Eq.~\eqref{eq:Perturbed_Entropy_Density} with the 3-parameter fit to $\alpha(\epsilon,\tcm)$ shown in Figure~\ref{fig:alpha_evolution_quadratic_fit}, \edit{while the solid purple line shows the same quantity calculated directly from \burst{}}.  The excellent agreement between the dashed line and the solid line further demonstrates that our 3-parameter model provides a precise description of the cosmic neutrino background in the Standard Model.
    }
\label{fig:differential_productivity}
\end{figure}

Figure~\ref{fig:differential_productivity} shows the logarithm of the visible production of $\nu_{e}$ from the process $e^- + e^+ \rightarrow \nu_e + \bar{\nu}_e$, given by the product of the $\nu_e$ differential visibility $\diffvis_{\nu_e}$ with $\frac{P_{e^{-}e^{+}}}{\Gamma_{\nu_{e}}'}$. There is also a weighting of $\epsilon^{2}\cdot n_{\nu_{e}}(\epsilon)$ in order to highlight the region with the highest neutrino number density.  The change to the total entropy as a function of $\Tcm{}$ is also shown.

%%%%%%% Differential Productivity Plot %%%%%%%%%%%
%%%%%%%%%%%%%%%%%%%%%%%%%%%%%%%%%

Figure~\ref{fig:differential_productivity} provides a powerful example of a novel way to visualize processes during the weak decoupling epoch. We can clearly observe the temperatures and energies where the process $e^{-}$ + $e^{+}\rightarrow\nu_e + \bar{\nu}_e$ has the largest impact on the neutrino occupation probability.
The differential productivity provides a new tool that can be applied to processes both within the Standard Model and beyond, allowing new insights into how various mechanisms may leave imprints on the late-time neutrino distribution function.

%%%%%%%%%%%%%%%%%%%%
\section{Charged-current weak interactions on neutrons and protons and BBN}
\label{sec:WeakInteraction}
%%%%%%%%%%%%%%%%%%%%

Next, we turn our focus to how neutrinos impact the formation of light nuclei in the early universe. As discussed in Sec.~\ref{sec:WeakDecoupling}, cosmic neutrinos play a significant role in the charged-current, isospin-changing weak interactions that interconvert protons and neutrons.  This in turn impacts the abundances of all nuclides in the nuclear network that is ultimately responsible for the light element abundance yields produced through primordial nucleosynthesis.

\edit{
We begin by giving a qualitative description of the nuclear reaction network based upon an $\alpha$ formalism dual to nuclide abundances; the later method being the standard treatment which the \burst code employs.
}
The light element abundance of species $J$ is given by $Y_J = n_J /n_b$, and similarly $Y_e$ for electrons, $Y_p$ for protons, and $Y_n$ for neutrons.
In all cases the equilibrating reactions are fast enough to ensure the respective conjugates to the distribution functions are linear in energy.  For species $J$ with momentum $p$, spin $s$, and energy $E_{psJ}$ we have $\alpha_{{\bf X},{\bf p},t}  =  \alpha_{0J} + \alpha_{1J}E_{psJ}$. 
The conjugate to the abundance of species $J$ is $\alpha_0 = -\beta_J \mu_J$ 
and the conjugate to the energy is of the form $\alpha_{1J} \equiv \beta_J = T_J^{-1}$ where $\mu_J$ is the chemical potential and $T_J$ is the temperature of species $J$. The interactions among all species apart from neutrinos are rapid enough during the BBN epoch to ensure that the temperatures of all nuclides, electrons, and photons are the same. This is the first large reduction in control parameters needed for the reaction networks for abundances and energy.

The neutrino energy-distribution occupation probabilities are almost completely determined by purely leptonic processes, since the number density of nucleons relative to the number density of neutrinos is only  $\sim 10^{-10}$.   The leptonically-computed neutrino distributions determine the neutron-to-proton ratio, which in turn influences primordial abundances.  There are small perturbative changes in $\alpha_{\nu_e}$ and $\alpha_{\bar{\nu}_e}$ that accompany these processes and therefore could give insight into the weak interaction history of weak decoupling and BBN.

The neutron-proton interconversion channels relevant to BBN are given in Eqs.\ \eqref{eq:nuc_reac1} -- \eqref{eq:nuc_reac3}. In \burst{}, the neutron-to-proton ratio along with the nuclear reaction network for the nuclear abundances and the neutrino Boltzmann transport network are all solved together and self-consistently at each timestep throughout the weak decoupling and BBN regime. The details of the BBN nuclear reaction network are explored in \cite{Smith2009BigFunctions}, \cite{Weinberg2008Cosmology}, and \cite{Grohs2015ProbingEpochs}.
The neutron-to-proton ratio at any time $t$, follows from the integration of
\begin{equation}\label{eq:dnp_dt}
  \frac{d(n/p)}{dt} = Y_p\left(\lambda_{pe^-} + \lambda_{p\bar{\nu}_e} + \lambda_{pe^-\bar{\nu}_e}\right) - Y_n\left(\lambda_{n\nu_e} + \lambda_{ne^+} + \lambda_{n\,{\rm decay}}\right),
\end{equation}
where the reaction rates $\lambda_{\nu_{e} n}$, $\lambda_{e^{+} n}$, $\lambda_{n\, \mathrm{decay}}$ (forward), and $\lambda_{p e^{-}}$, $\lambda_{p \bar{\nu}_{e}}$, $\lambda_{p e^{-} \bar{\nu}_{e}}$ (reverse) \cite{Smith2009BigFunctions} correspond to the reactions in Eqs.\ \eqref{eq:nuc_reac1} -- \eqref{eq:nuc_reac3}.
When the neutron-to-proton interconversion rates are fast compared to the Hubble expansion rate, the instantaneous $n/p$ ratio at time $t$ is near the steady-state solution, namely
\begin{equation}\label{eq:np_lambda}
\frac{n}{p}\simeq \frac{\lambda_{p e^{-}} + \lambda_{p \bar{\nu}_{e}} + \lambda_{p e^{-} \bar{\nu}_{e}}}{\lambda_{\nu_{e} n} + \lambda_{e^{+} n} + \lambda_{n\, \mathrm{decay}}}.
\end{equation}
At high temperatures, where neutrino and charged-lepton distributions have comparable temperatures, Eq.\ \eqref{eq:np_lambda} becomes coincident with the result from chemical equilibrium \cite{Smith1993ExperimentalNucleosynthesis}
\begin{equation}\label{eq:we}
\bigg(\frac{n}{p}\bigg)_{\text{eq}} = e^{-(m_n - m_p)/T},
\end{equation}
where the neutron-proton mass difference is  $m_n - m_p \simeq1.293\, \text{MeV}$.  The correction in the neutrino-energy-distribution functions stemming from Boltzmann neutrino transport and out-of-equilibrium neutrino scattering tends to increase the $n/p$ ratio \cite{Smith2009BigFunctions}.

\burst{} actually includes the possibility of any nuclear element of atomic number $A_J = Z_J +N_J$ being present, where $Z_J$ is the proton number and $N_J$ is the neutron number. At high temperature the nuclear system is in Nuclear Statistical Equilibrium (NSE), since the time scale of the strong nuclear force is much shorter than the characteristic expansion time, so that each reaction is governed by chemical potential balance. In the NSE regime,  the many $\alpha_{0J}$ describing the abundances of the nuclides may all be related to just the $n/p$ ratio and the total baryon density $n_b$ in terms of the binding energy of each nuclide $B_X$, thus representing another major network compression.

\burst yields the light-element abundances as a function of time.
The abundance relative to baryon of nuclear species $J$ (for example, $n$, $p$, ${}^{2}\text{H}$, $^{3}\text{He}$, etc.) is  
\begin{equation}
Y_{J}(t) = \frac{n_{J}(t)}{n_{b}(t)} \, , 
\end{equation}
and the corresponding mass fraction of element $J$ with atomic number $A_J $ is $X_J = A_J Y_J$. 
The main asymptotic BBN observables are the abundances of $^{4}$He, deuterium (D), $^{3}$He, and $^{7}$Li.
(By convention, the mass fraction of helium, $X_\alpha$, is denoted by $Y_{\rm P}$, not to be confused with the abundance of protons denoted $Y_p$.)
The NSE abundance for nuclide $X$ is
\begin{equation}\label{nse}
  Y_X = Y_p^Z Y_n^{A-Z} 2^{(A-3)/2}\pi^{3(A-1)/2}g_XA^{3/2}
  \left[\frac{n_b}{(Tm_b)^{3/2}}\right]^{A-1}e^{B_X/T},
\end{equation}
where $Z$ is the atomic number, $A$ is the atomic mass number, $Y_p$ is the abundance of free protons, $Y_n$ is the abundance of free neutrons, $g_X$ is the nuclear partition function for nuclide $X$, $n_b$ is the baryon number density, $T$ is the plasma temperature, $m_b\simeq931.5\,{\rm MeV}$ is the baryon rest mass, and $B_X$ is the binding energy of nuclide $X$.

Equipped with the knowledge of weak equilibrium and NSE in Eqs.\ \eqref{eq:we} and \eqref{nse}, we return to discussing Fig.\ \ref{fig:x_nse}. The middle panel of Fig.\ \ref{fig:x_nse} shows curves for two tracks of $n/p$.  The solid curve gives $n/p$ from an explicit calculation of the $n\leftrightarrow p$ rates, encapsulated in Eq.\ \eqref{eq:dnp_dt}, whereas the dashed line gives the weak-equilibrium expression from Eq.\ \eqref{eq:we}. In the bottom panel of Fig.\ \ref{fig:x_nse}, we show the evolution of the mass fractions $X_J=A_J Y_J$ of the nuclides.  Similar to the middle panel, solid curves are output from the reaction-network solver, and dashed lines are NSE fractions from Eq.\ \eqref{nse}. We observe that at high temperatures, $n/p$ follows the equilibrium trajectory, and each of the $X_J$ for nuclides with $A_J\le4$ follow NSE trajectories ($A_J=6,7$ are in NSE at higher temperatures $T\simeq10\,{\rm MeV}$). In this case, only one chemical potential --- and therefore only one energy-averaged $\langle\alpha_b\rangle$ for baryons --- is needed to characterize the entire set of nuclides.  Equation~\eqref{nse} simplifies as $Y_n=Y_p\,e^{-(m_n-m_p)/T}$. The temperature drops, and the weak-interaction rates no longer keep $n/p$ in weak equilibrium.  The departure from weak equilibrium occurs somewhat earlier than what one may guess \cite{2016NuPhB.911..955G}, with deviations between dashed and solid starting as early as $\tcm=2\,{\rm MeV}$, and becoming starkly apparent by $1\,{\rm MeV}$.  However, strong and electromagnetic reactions keep $d$, $t$, $^3{\rm He}$, and $^4{\rm He}$ (not visible on the scale of the vertical axis) in NSE with one another.  This system at $\tcm\sim1\,{\rm MeV}$ requires two, independent values $\{\langle\alpha_n\rangle,\langle\alpha_p\rangle\}$ to describe NSE.  In other words, Eq.\ \eqref{nse} requires unique values of $Y_n$ and $Y_p$ in this temperature range.

As the temperature drops further the network evolution is characterized by fast scrambling among sub-blocks of elements with the interaction between coarser blocks slow enough that chemical potentials are required for each. This is called quasi-nuclear statistical equilibrium (qNSE), familiar from stellar nucleosynthesis when interaction between the iron and silicon blocks is slow. As opposed to a stellar interior, the entropies are so high in the early universe that the abundances in NSE are highest for $n$ and $p$ to epochs well below $\tcm\sim100\,{\rm keV}$. As BBN develops, there is an extended freeze-out where progressively more chemical potentials are needed to characterize the network blocks.
Nuclides with larger $A_J$ depart from NSE earlier, and hence require independent chemical potentials $\alpha_{0,J}$ conjugate to their mass fraction $X_J$.
$^4{\rm He}$ departs from NSE at $\tcm\simeq700\,{\rm keV}$ and the $A_J=3$ nuclides at $\tcm\simeq200\,{\rm keV}$.
Deuterium breaks the equilibrium relation last, at a plasma temperature around 80~keV.

Out-of-equilibrium neutrino-transport effects result in a decrease in $Y_{p}$ and an increase in deuterium, $^{3}$He, and $^{7}$Li abundances \cite{Grohs:2015tfy}. There are other small corrections that affect late time observables, such as Coulomb corrections and zero-temperature radiative corrections which are fairly independent of each other and of neutrino transport \cite{Grohs:2015tfy}. Other effects, such as finite-temperature radiative corrections, and in-medium renormalization of electron and positron rest masses,  may induce nonlinear changes to the abundances. 
However, all of these conclusions are based on Standard Model physics.  BSM physics may alter a number of features of weak decoupling and BBN which can be modeled in high-fidelity with, for example, an appropriately modified \burst code.  The corresponding alterations in abundances and weak decoupling history can be captured in changes to the generalized entropy parameters, i.e., the $\alpha_f(p,t)$.

%%%%%%%%%%%%%%
\section{Abundance responses to perturbations}
\label{ssec:abunds_resps}
%%%%%%%%%%%%%%

By observing light element abundances, we can place constraints on the energy spectrum of cosmic neutrinos.  In order to achieve this, we first need to know how these abundances respond to changes to the distribution function of neutrinos.  
%%%%%%%%%%%%%%%%%%%%%
A target of this paper is the linear response of the asymptotic BBN abundances $\delta Y_J$ to distortions of the C$\nu$B from the \smnu. 
We have only a few light element abundances that we observe (deuterium, helium, and possibly lithium), yet an entire energy and time-dependent set of distorted neutrino distribution functions for the $Y_J$ to respond to. At the linear response level, we could determine the neutrino spectrum that is most likely to produce the observed abundances,  
but about this mean there are very large fluctuations.  That is, from observations of light element abundances we can learn only a few modes among the very many modes possible in the full range of time and energy-dependent perturbations to the neutrino spectrum, leaving the spectrum almost completely unconstrained.
Using observed $Y_J$ in order to make C$\nu$B inferences is a bottom-up approach.

This limitation suggests that a top-down approach is better: we specify a BSM$\nu$ model, and compute simultaneously the distorted neutrino distribution functions and the BBN abundances. This is the experimental approach, using numerical calculations like those in \burst{}. 
What we present here is a framework to understand what is needed in a BSM$\nu$ in order to have impact on BBN, but rely on the \burst{} code for the $\nu$-BBN computations which provide the experimental results for our musings. An important aspect of this work is it shows clearly where in the energy-time plane perturbations to the \smnu would be needed to get a measurable BBN  response. For example, early energy injection just equilibrates to a Fermi-Dirac distribution and BBN proceeds as in the \smnu. 
%%%%%%%%%%%%%%%%%%%%%%%%%

We therefore use the \burst{} code~\cite{Grohs:2015tfy} to calculate the linear response of light element abundances to small perturbations in the neutrino distribution function.  In order to isolate the effect of changes to the spectrum, we take a set of perturbations for which the change to the distribution function of the electron-type neutrinos is equal to the electron-type antineutrinos in order to conserve flavor and lepton numbers.  Furthermore, we perturb the distribution function of muon-type neutrinos and anti-neutrinos with the same amplitude and opposite sign in order to conserve the total radiation energy density.  With these restrictions, we then compute the perturbations to the final primordial abundances (well after BBN) numerically using \burst{}~\cite{Grohs:2015tfy}.

The perturbations to the occupation numbers, $\delta n$, are proportional to the deviations from the zero-chemical-potential FD equilibrium occupation numbers
\begin{equation}
  \delta n(\epsilon,\Tcm) \equiv \frac{\left \langle \widehat{n} \right \rangle (\epsilon,\Tcm) - \nequil(\epsilon)}{\nequil(\epsilon)}, \quad
  \nequil(\epsilon) = \frac{1}{e^\epsilon + 1}.
\end{equation}
Perturbing the $\nu_e$, $\overline{\nu}_e$, $\nu_\mu$, and $\overline{\nu}_{\mu}$ occupation numbers in the manner outlined above in Standard Model evolution with out-of-equilibrium neutrino scattering, induces a response in the abundances. Introduction of BSM physics can result in different perturbations in neutrino occupation numbers and, correspondingly, a different response in the abundances.
We characterize the abundance responses as a relative difference from a baseline
\begin{equation}
  \delta X \equiv \frac{X - X^{\rm (base)}}{X^{\rm (base)}},
\end{equation}
where the baseline abundance $X^{\rm (base)}$ is the abundance calculated when $\delta n=0$ for all $\epsilon$.

\begin{figure}[tp!]
    \begin{center}
    \includegraphics[width = 0.75\columnwidth]{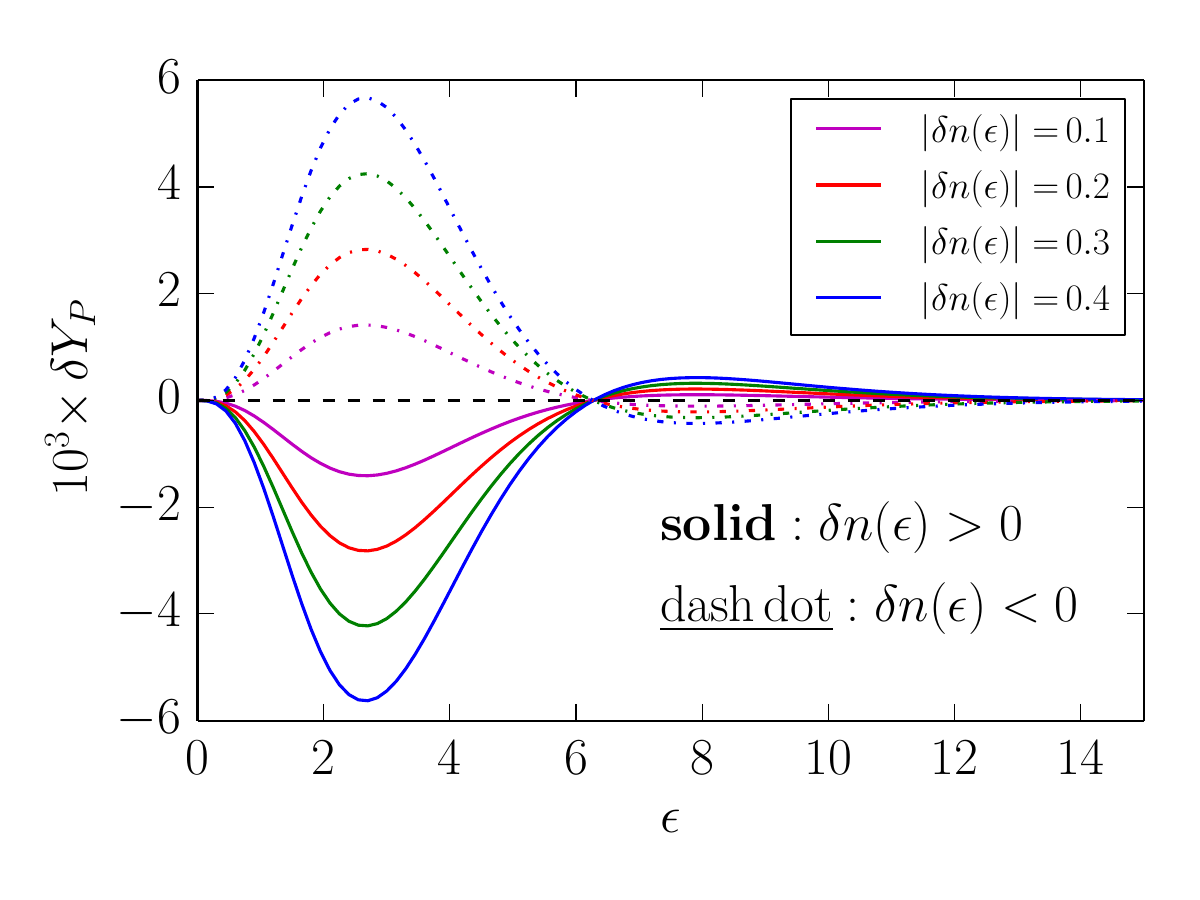}
    \caption{Relative deviations in $Y_{\rm P}$ as a function of $\epsilon$ for differing changes in occupation number $n(\epsilon)$.  Each color represents a different change to $n(\epsilon)$.  Solid lines correspond to enhancements in $n_{\nu,\overline{\nu}_e}$, while dash-dot lines correspond to \edit{suppressions} in $n_{\nu,\overline{\nu}_e}$.  The changes to $n(\epsilon)$ persist for the duration of the calculation.}
    \label{fig:yp_vs_eps_change}
    \end{center}
\end{figure}

\begin{figure}[tp!]
    \begin{center}
    \includegraphics[width = 0.75\columnwidth]{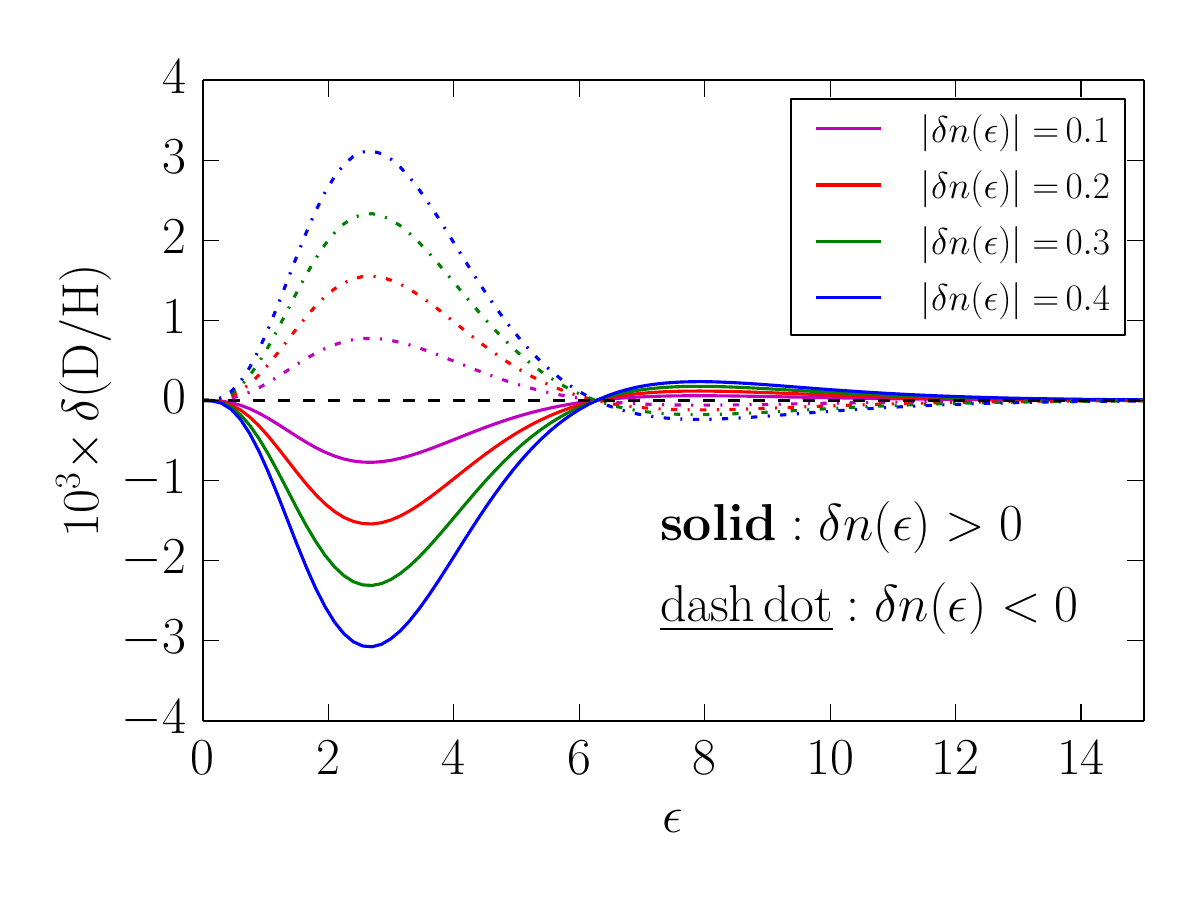}
    \caption{Same as Fig.\ \ref{fig:yp_vs_eps_change} except for relative deviations in ${\rm D/H}$}
    \label{fig:dh_vs_eps_change}
    \end{center}
\end{figure}

 We demonstrate the impact of two types of neutrino spectral distortions on the primordial abundances.  First, we calculate the impact of perturbations to the neutrino spectra that are persistent and independent of $\Tcm{}$.  In this case, the neutrino occupation numbers are perturbed by 
\begin{equation}
    \delta n(\epsilon,\Tcm) = 
    \begin{cases}
        \Delta & \text{for } \epsilon \in \epsilon_\mathrm{perturb} \\
        0 & \text{otherwise}
    \end{cases}
    \label{eq:persistent_perturbation}
\end{equation}
for all $\Tcm$.  The effects of this sort of perturbation for bins in $\epsilon$ of width $0.15$ from $0$ to $15$ to the primordial helium-4 and deuterium yields are shown in Figures~\ref{fig:yp_vs_eps_change} and \ref{fig:dh_vs_eps_change}, respectively.  Next, we calculate the impact of spectral perturbations that are isolated in both $\Tcm$ and $\epsilon$, with the form 
\begin{equation}
    \delta n(\epsilon,\Tcm) = 
    \begin{cases}
        \Delta & \text{for } \epsilon \in \epsilon_\mathrm{perturb}, \ \Tcm \in \Tcm{}_\mathrm{,perturb} \\
        0 & \text{otherwise}
    \end{cases}
\end{equation}
for bins in $\epsilon$ of width $0.15$ and bins in $\log_{10}(\Tcm/\mathrm{MeV})$ of width $0.03$.  From this latter type of perturbation, we calculate the response function of the primordial abundances to neutrino spectral distortions
\begin{equation}
    K_X(\epsilon,\Tcm) = \frac{d^2 \delta X}{\delta n(\epsilon, \Tcm) \ d\epsilon \ d\log_{10}(\Tcm/\mathrm{MeV}) } \, ,
    \label{eq:X_response_function}
\end{equation}
defined such that the fractional change to each abundance can be calculated by integrating over the perturbation to the neutrino occupation number
\begin{equation}
    \delta X = \int K_X(\epsilon,\Tcm) \ \delta n(\epsilon, \Tcm) \ d\epsilon \ d\log_{10}(\Tcm/\mathrm{MeV}) \, .
    \label{eq:resonse_integral}
\end{equation}

%%%%%%%%% Yp Change, constant perturbation ###################
%%%%%%%%%%%%%%%%%%%%%%%%%%%%%%%%%%

%%%%%%%%% D/H Change, constant perturbation %%%%%%%%%%%%%%%%%%%%%%
%%%%%%%%%%%%%%%%%%%%%%%%%%%%%%%%%%%%%%

\begin{figure}[tp!]
    \begin{center}
    \includegraphics[width = 0.95\columnwidth]{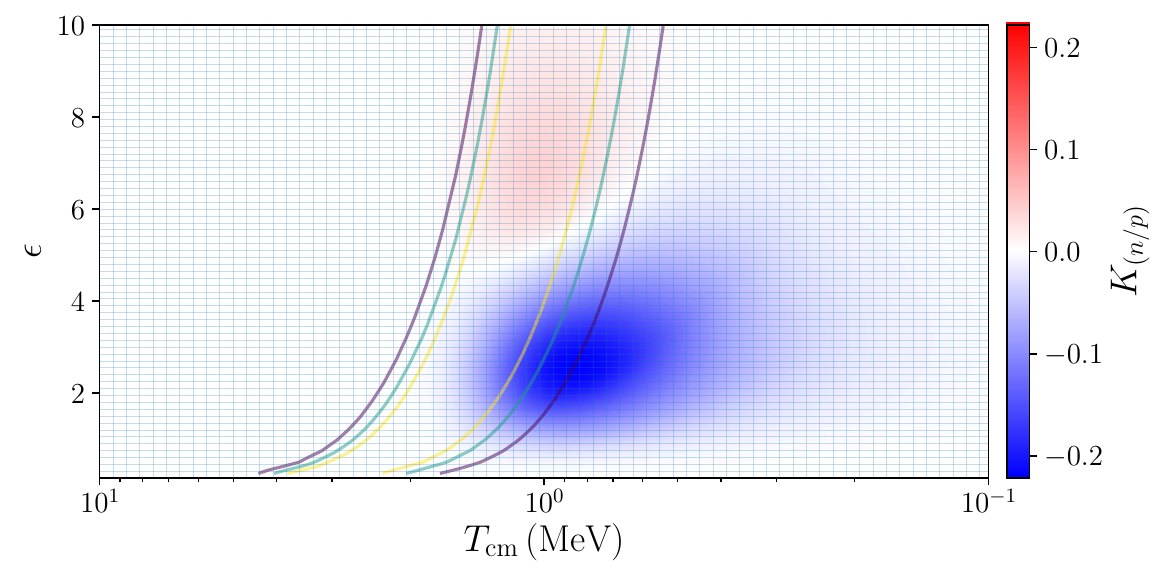}
    \caption{
    Response of the total (free and bound) neutron-to-proton ratio after the conclusion of BBN to perturbations to the $\nu_e$ energy spectrum at specified values of $\epsilon$ and $\Tcm$ in terms of the response function $K_{(n/p)}$ defined in Eq.~\eqref{eq:X_response_function}.  The perturbations involve an increase in the occupation number of electron-type neutrinos and antineutrinos, along with an equal and opposite change to the occupation number of muon-type neutrinos and antineutrinos.  The changes to the neutron-to-proton ratio shown here are therefore calculated at fixed $\neff$, and reflect how changes to the energy spectrum of electron neutrinos \edit{impact} the primordial abundance yields from BBN.  The impact on each primordial abundance can be obtained from Eq.~\eqref{eq:resonse_integral} using the scaling functions shown in Table~\ref{tab:response_scaling}.  Also shown are the contours of $50\%$, $75\%$, and $95\%$ of peak visibility for electron neutrinos (as shown in Figure~\ref{fig:diff_vis}).  The grid on which the response was calculated is overlaid.}
    \label{fig:eps_vs_tcm_n_to_p_change}
    \end{center}
\end{figure}

To an excellent approximation, the response functions for each of the primordial abundances can be calculated as a constant scaling of the response function for the neutron-to-proton ratio $K_X(\epsilon,\Tcm{}) = A_X K_{(n/p)}(\epsilon,\Tcm{})$, shown in Figure~\ref{fig:eps_vs_tcm_n_to_p_change}.  The scaling factor $A_X$ for each nuclide is given in Table~\ref{tab:response_scaling}.

It should be noted that neither of the types of perturbations to the neutrino spectra discussed in the previous paragraph represent realistic histories of the decoupling epoch, since we did not allow for evolution of the neutrino occupation numbers in the calculations of this section.  In reality, if a physical process led to an increased occupation number of neutrinos over some range of $\epsilon$ and $\Tcm$, weak interactions would cause the perturbation to be redistributed.  For example, a localized increase to the neutrino occupation number inserted at high $\Tcm$ and high $\epsilon$ would quickly be smoothed away as the system rapidly approached a thermal distribution.  Perturbations at lower $\Tcm$, particularly those introduced after the peak of the neutrino visibility function, would persist with less modification due to the low rate of weak interactions in that regime.  Despite the artificial nature of the perturbations, their effect on the primordial abundances provides insight into the regions of $n(\epsilon,\Tcm)$ to which the primordial abundances are most sensitive.

%%%%%%% n/pp Response function  %%%%%%%%%%%%%%%
%%%%%%%%%%%%%%%%%%%%%%%%%%%%%

We observe two regions different from zero in Fig.\ \ref{fig:eps_vs_tcm_n_to_p_change}.  In the blue region centered at $\tcm\simeq0.8\,{\rm MeV}$ and $\epsilon\simeq3$, $K_{n/p}$ is less than zero.  Conversely, in the light red region at $\tcm\simeq1.0\,{\rm MeV}$ and $\epsilon\simeq7$, $K_{n/p}$ is greater than zero.  The main contributors to changing $n/p$ in these regions are the four processes in Eqs.\ \eqref{eq:nuc_reac1} and \eqref{eq:nuc_reac2} which we repeat below
\begin{align}
  \nu_e + n &\leftrightarrow p + e^-,\\
  e^+ + n &\leftrightarrow p + \bar{\nu}_e,
\end{align}
where the forward rates imply neutron destruction, $d(n/p)/dt<0$, and the reverse rates imply neutron production, $d(n/p)/dt>0$.  For the first process, abbreviated $n(\nu_e,e^-)p$, there are no neutrino energy thresholds.  When we increase the distributions by adding $\delta n(\epsilon,\tcm)$, we enhance the neutron destruction rate, and likewise suppress the neutron production rate via the Pauli blocking factor.  The net result is a decrease in $n/p$ as evidenced by the blue region in Fig.\ \ref{fig:eps_vs_tcm_n_to_p_change}.  For $n(e^+,\bar{\nu}_e)p$, there is a neutrino energy threshold in the neutron creation rate, namely $m_n - m_p + m_e\sim1.8\,{\rm MeV}$.  An increase $\delta n(\epsilon,\tcm)$ only affects the rates if $\epsilon>(m_n-m_p+m_e)/\tcm$.  If $\delta n(\epsilon,\tcm)$ occurs above the $\epsilon$ threshold, then we enhance the neutron creation rate and suppress the neutron destruction rate, opposite of the effect of $\delta n(\epsilon,\tcm)$ on $n(\nu_e,e^-)p$.  However, the enhancement of the neutron production/destruction rates is not equal and does not offset one another.  Recall from Eq.\ \eqref{eq:dnp_dt}, that $d(n/p)/dt$ is a sum of the $n\leftrightarrow p$ rates weighted by the abundances of neutrons and proton, $Y_n$ and $Y_p$, respectively.  For the standard cosmology and perturbations we consider here, $Y_p\ge Y_n$ for all times.  Therefore, the neutron production rate ($\sim Y_p\lambda_{p\bar{\nu}_e}$) has a larger effect than the neutron destruction rate ($\sim Y_n\lambda_{n\nu_e}$).  The net result is that the change in $n/p$ is positive, as evidenced by the light-red region in Fig.\ \ref{fig:eps_vs_tcm_n_to_p_change}.

%%%%%%%%%%%%% Nuclide Response Table  %%%%%%%%%%%%%%%
\begin{table}[tp!]
\centering
    \begin{tabular}{|c|c|c|c|c|c|}
        \hline
        \rowcolor[HTML]{ECF4FF} 
        Nuclide  & D & ${}^3$He & ${}^4$He & ${}^6$Li & ${}^7$Li \\ \hline
        \cellcolor[HTML]{ECF4FF}$A_X$ & 0.479 & 0.167  & 0.876  & 1.63     & 0.558   \\ \hline
    \end{tabular}
    \caption{Factor relating the response function for the primordial abundance of each nuclide $K_X(\epsilon,\Tcm{})$ defined in Eq.~\eqref{eq:X_response_function} to the response function of the neutron-to-proton ratio plotted in Figure~\ref{fig:eps_vs_tcm_n_to_p_change}, such that $K_X(\epsilon,\Tcm{}) = A_X K_{(n/p)}(\epsilon,\Tcm{})$.}
    \label{tab:response_scaling}
\end{table}
%%%%%%%%%%%%%%%%%%%%%%%%%

%%%%%%%%%%%%%%%%
\section{Conclusions}
\label{sec:Conclusions}
%%%%%%%%%%%%%%%%%%%%%

We introduced a framework which allows for compact and efficient description of the cosmic neutrino background and its deviation from a relativistic Fermi-Dirac distribution.  We demonstrated that the spectrum of the cosmic neutrino background in the Standard Model is accurately captured by a three-parameter description of the thermodynamic potential throughout the epoch of weak decoupling and neutrino last scattering. \edit{While we neglected neutrino flavor oscillations in the analysis presented here, our framework could be extended to include this and other additional physics.}

Accurate description of photon decoupling has been essential in extracting detailed information from observations of the cosmic microwave background.  The precision with which we can observe the properties of the cosmic neutrino background is unfortunately quite limited compared to the rich data set provided by observations of the cosmic microwave background.  Nonetheless, indirect probes of the cosmic neutrino background can be quite incisive, and are becoming ever more so as cosmological observations improve.  It is therefore essential that the subtle aspects of neutrino decoupling are treated with sufficient precision to make the most out of the various observable imprints that are left by the influence of the cosmic neutrino background.

The  energy density of the cosmic neutrino background affects the expansion rate of the Universe, leaving observable imprints in the CMB anisotropies and the baryon acoustic oscillations observed with galaxy surveys.  This arises primarily through the impact of the radiation density, $\neff$, on the diffusion damping scale~\cite{Dvorkin:2022jyg}.  Additionally, fluctuations in the density of neutrinos, which propagate at the speed of light after decoupling, impart a characteristic phase shift of the acoustic peaks appearing in CMB and matter power spectra~\cite{Bashinsky:2003tk,Baumann:2015rya,Baumann:2017gkg}.  Both of these effects of the cosmic neutrino background depend only on the integrated energy density, and not on its energy spectrum.

The non-zero mass of neutrinos implies that at least some cosmic neutrinos are non-relativistic and contribute to the matter density of the Universe at late times.  Unlike cold dark matter and baryons, cosmic neutrinos retain a large thermal velocity through much of cosmic history, implying that cosmic neutrinos act like hot dark matter.  Their presence suppresses the growth of cosmic structure on scales smaller than the neutrino free streaming length compared to a universe containing only massless neutrinos~\cite{Dolgov:2002wy,Lesgourgues:2006nd,Wong:2011ip,Lesgourgues:2012uu,Dvorkin:2019jgs,Green:2021xzn}.  Searches for this suppression of clustering are typically characterized in terms of constraints on the sum of neutrino masses $\sum m_\nu$.  However, this parameterization assumes that cosmic neutrinos are well described by a relativistic Fermi-Dirac distribution.    A modified energy spectrum of cosmic neutrinos could manifest itself in suppression of matter clustering that differs from the standard picture, or turning things around,  what should be inferred from the measurement of suppressed matter clustering is affected by the energy spectrum of cosmic neutrinos~\cite{Alvey:2021sji}.

Impacts on the expansion rate and the suppression of matter clustering, typically framed in terms of $\neff$ and $\sum m_\nu$, depend only on the energy spectrum of cosmic neutrinos well after neutrino decoupling.  However, cosmic neutrinos play a significant role in BBN in a way that is more sensitive to the evolution of the energy spectrum.  Neutrinos participate in the interconversion of neutrons and protons through the weak interaction during the time period of weak decoupling~\cite{Cyburt:2015mya}.  As a result, the production of light elements during BBN is sensitive to the time evolution of the cosmic neutrino energy spectrum, and not just to the form that it approaches asymptotically at late times.

Within the Standard Model, the cosmic neutrino background deviates only slightly from a relativistic Fermi-Dirac spectrum.  However, even in that case, the evolution of the effective temperature of the neutrinos relative to that of the photons evolves in a non-trivial way through weak decoupling and weak freeze-out, due in large part to the fact that electron-positron annihilation overlaps with the period of weak decoupling.
Figures \ref{fig:alpha_evolution_linear_fit} and \ref{fig:alpha_evolve_linear} show that this non-trivial evolution cannot be precisely captured through two-parameter fits consistent with extending the GCE to out-of-equilibrium conditions.
Conversely, the generalized entropy approach described here provides a compact and precise description of the deviations of the Standard Model C$\nu$B from a Fermi-Dirac distribution using only one additional parameter per species.  Furthermore, our formalism provides a physical interpretation that each parameter is conjugate to macroscopic properties of the system, namely: total number of particles, total energy, and fluctuations in energy (i.e., energy-squared).

We demonstrated that the formalism we developed to describe the evolution of the cosmic neutrino background spectrum is sufficiently precise to capture the small time-dependent deviations from an equilibrium distribution that are present in the Standard Model \edit{(noting that we neglect flavor oscillations in the present work)}.  An important aspect of the framework we developed is that it provides a valuable set of tools to describe and constrain physics beyond the Standard Model.  There are many ways that new physics can impact the neutrino sector, the thermal history of the universe, and the weak decoupling epoch.  A very broad class of models of new physics can be usefully mapped onto the description presented here in terms of the evolution of the thermodynamic potential for neutrinos.  Our formalism thereby provides a bridge between the diverse phenomenology that may operate in the early universe and the constraints that can be obtained from indirect observations of the cosmic neutrino background.  For example, constraints derived from the requirement that BBN not be disrupted by new physics is often treated as a blunt instrument, requiring that the thermal history and neutrino sector are unmodified around temperatures of 0.1 to 1~MeV.  The formalism presented here provides a mechanism by which BBN constraints can be made more precise by systematically treating the observational impacts of deviations from the standard scenario.

As a concrete example, consider the out-of-equilibrium decay of a massive particle species around the time of weak decoupling.  It is relatively straightforward to compute how this type of physics impacts integrated quantities like $\neff$, but in general accounting for more detailed evolution must be treated on a case-by-case basis.  However, our formalism provides a means by which more general conclusions can be drawn.  Any physics which results in a change to the number density, energy density, or energy fluctuations of the cosmic neutrino background can be described in terms of changes to the $\alpha$ parameters we introduced.  We can derive observational constraints on these generalized modifications from measurements of primordial light element abundances using the response functions described in Section~\ref{ssec:abunds_resps}.  In this way, very general conclusions can be drawn about how BBN can be used to constrain  diverse classes of new physics, without the requirement of carrying out the full calculation of neutrino transport coupled to BBN for each scenario.  All that is required is to map the effects of any particular model onto its impact on the $\alpha$ parameters.

Furthermore, our description of the temperature and energy dependent cosmic neutrino visibility function gives us a good sense of how new physics operating in various regimes is likely to impact the thermodynamic potential.  Injection of high energy neutrinos at early times, well before weak decoupling are likely to become completely thermalized, resulting in little change to the neutrino and plasma distributions described by the standard scenario.  Production of neutrinos at late times will simply be added to the standard spectrum, since weak interactions at late times are inefficient and will not redistribute neutrinos in phase space.  Energy injection around the time of weak decoupling will result in the new contributions being partially thermalized, contributing mostly to the $\alpha_0$, $\alpha_1$, and $\alpha_2$ parameters.  Even though we demonstrated this explicitly only in the Standard Model, electron-positron annihilation provides a non-trivial example of this behavior.  One may have naively expected that electron-positron annihilation results in a bump in the neutrino spectrum at the value of $\epsilon$ associated with the electron mass at the time of annihilation.  However, we showed that partial thermalization of the resulting neutrinos allows for a precise description of the resulting neutrino spectrum in terms of just $\alpha_0$, $\alpha_1$, and $\alpha_2$.

There is a close analogy of this description with spectral distortions of the cosmic microwave background.  Energy injection during different eras in the early universe tends to result in characteristic $\mu$- and $y$-distortions of the CMB spectrum~\cite{Bond:1993qp,Chluba:2011hw,Chluba:2016bvg}.  We have shown how Standard Model processes lead to distortions of the \cnub spectrum in terms of $\alpha_0$, $\alpha_1$, and $\alpha_2$.  One can use observational constraints on these generalized spectral distortions to gain insight into a broad class of new physics, without the requirement of carrying out a detailed transport calculation for each individual model.

The types of new physics that can be constrained through limits on CMB spectral distortions can also be constrained with primordial abundance measurements through the formalism we developed here, though BBN constraints tend to apply to physics operating at an earlier time ($z\geq10^{10}$ for BBN compared to $z\geq10^3$ for CMB spectral distortions).  Examples of new physics that can be constrained by both CMB and C$\nu$B spectral distortions include annihilating or decaying particles, evaporating primordial black holes, and dissipation of small-scale acoustic oscillations~\cite{Chluba:2011hw,Jeong:2014gna,Chluba:2016bvg}.  Spectral distortions of the cosmic neutrino background may also arise from neutrino self-interactions, right-handed neutrinos, sterile neutrino oscillations, dark matter-neutrino interactions, light dark matter freeze-in scenarios, among many others.

The rich landscape of well-motivated new physics scenarios that may impact the cosmic neutrino spectrum calls out for a treatment of the constraints imposed by primordial abundance observations with more nuance than a simple rigid exclusion over some range of temperatures.  The formalism presented here enables this more precise treatment, and it does so in a convenient and compact way that allows for broad and valuable constraining power on models of new physics.

\section*{Acknowledgments}

The authors would like to thank Daniel Green, Marilena Loverde, Mark Paris, and Nashwan Sabti for helpful discussions. JRB's work was supported by Canada's National Science and Engineering Research Council's Discovery Grant and his Fellowship in the Canadian Institute for Advanced Research. 
GMF is supported in part by National Science Foundation grant PHY-2209578 at UCSD.
EG is supported by the Department of Energy Office of Nuclear Physics award DE-FG02-02ER41216.
JM is supported by the US~Department of Energy under Grant~\mbox{DE-SC0010129}.
This work was partially enabled by the National Science Foundation under grant No.\ PHY-2020275: {\it Network for Neutrinos, Nuclear Astrophysics, and Symmetries} (N3AS).  Additional support was provided by the Heising-Simons foundation under grant No.\ 2017-228.
Computational resources for this research were provided by SMU’s Center for Research Computing.
This research used resources provided by the Los Alamos National Laboratory Institutional Computing Program, which is supported by the U.S. Department of Energy National Nuclear Security Administration under Contract No. 89233218CNA000001.
GMF, EG, and JRB thank the Institute for Nuclear Theory (INT) at University of Washington, and all authors thank the Canadian Institute for Theoretical Astrophysics (CITA) at University of Toronto for their hospitality while parts of this research were being conducted.

\appendix

\section{Connected Correlation Functions}
\label{app:ConnectedComponents}

In this appendix, we provide some further details on how the free energy generating functional is related to the connected correlation functions in our generalized entropy formalism.

%%%%%%%%%%%%%%%%%%%

We start by noting that $e^{-\mathcal{F}}$ acts as the generating functional for the correlation functions of the operators $\widehat{Q}_A$.  This can be seen as follows
\begin{align}
    e^{-\mathcal{F}} &= \left\langle \exp \left[ -\sum_A \alpha_A^{(fi)} \widehat{Q}_A \right] \right\rangle_i \nonumber \\
    &= \left\langle \sum_{j=0}^\infty \frac{(-1)^j}{j!} \left[ \sum_A \alpha_A^{(fi)} \widehat{Q}_A \right]^j \right\rangle_i \nonumber \\
    &= \sum_{j=0}^\infty \frac{(-1)^j}{j!} \left\langle \left[ -\sum_A \alpha_A^{(fi)} \widehat{Q}_A \right]^j \right\rangle_i \nonumber \\
    &= 1 - \sum_A \alpha_A^{(fi)} \left\langle \widehat{Q}_A \right\rangle_i + \sum_{AB} \alpha_A^{(fi)}\alpha_B^{(fi)} \left\langle \widehat{Q}_{(A} \widehat{Q}_{B)} \right\rangle_i - \sum_{ABC} \alpha_A^{(fi)}\alpha_B^{(fi)}\alpha_C^{(fi)} \left\langle \widehat{Q}_{(A} \widehat{Q}_{B} \widehat{Q}_{C)} \right\rangle_i + \ldots \, ,
\end{align}
where parentheses indicate symmetrization on the indices 
\begin{equation}
    \widehat{Q}_{(1} \ldots \widehat{Q}_{m)} = \frac{1}{m!}\sum_\sigma \widehat{Q}_{\sigma(1)} \ldots \widehat{Q}_{\sigma(m)} \, ,
\end{equation}
and $\sigma$ labels the permutations of the indices. It is then straightforward to see that correlation functions can be obtained from functional derivatives of $e^{-\mathcal{F}}$
\begin{equation}
  \left. \frac{\delta^M e^{-\mathcal{F}(\{\alpha_A^{(fi)}\})}}{\delta\alpha_{m_1}...\delta\alpha_{m_M}} \right|_{\alpha_A = 0} = {(-1)^{M}} \left\langle \widehat{Q}_{(m_1}...\widehat{Q}_{m_M)}\right\rangle_{i} \, .
\end{equation}

We can then construct the generating functional for the connected correlation functions by taking the logarithm of the correlation function generating functional, which in this case is just the negative of the free energy generating functional
\begin{align}
    \mathcal{F} &= - \ln\left( e^{-\mathcal{F}} \right) \nonumber \\
    &= -\sum_{j=0}^\infty \frac{(-1)^j}{j!} \left\langle \left[ -\sum_A \alpha_A^{(fi)} \widehat{Q}_A \right]^j \right\rangle_{i,\mathrm{cc}} \nonumber \\
    &= -1 + \sum_A \alpha_A^{(fi)} \left\langle \widehat{Q}_A \right\rangle_i + \sum_{AB} \alpha_A^{(fi)}\alpha_B^{(fi)} \left\langle \widehat{Q}_{(A} \widehat{Q}_{B)} \right\rangle_{i,\mathrm{cc}} \nonumber \\
    &\qquad - \sum_{ABC} \alpha_A^{(fi)}\alpha_B^{(fi)}\alpha_C^{(fi)} \left\langle \widehat{Q}_{(A} \widehat{Q}_{B} \widehat{Q}_{C)} \right\rangle_{i,\mathrm{cc}} + \ldots \, .
\end{align}
Connected correlation functions can then be computed as functional derivatives of the free energy generating functional
\begin{equation}
  \left. \frac{\delta^M \mathcal{F}(\{\alpha_A^{(fi)}\})}{\delta\alpha_{m_1}...\delta\alpha_{m_M}} \right|_{\alpha_A = 0} = {(-1)^{M+1}} \left\langle \widehat{Q}_{(m_1}...\widehat{Q}_{m_M)}\right\rangle_{i,\mathrm{cc}} \, .
\end{equation}
We can verify that the connected correlation functions denoted here have the expected relationship to the ordinary correlation functions by explicit computation of the first few $M$-point functions.
\begin{align}
    \left\langle \widehat{Q}_{A}\right\rangle_{i} &= -\left. \frac{\delta e^{-\mathcal{F}}}{\delta\alpha_{A}} \right|_{\alpha_m = 0} \nonumber \\
    &= \left. e^{-\mathcal{F}} \frac{\delta \mathcal{F}}{\delta\alpha_{A}} \right|_{\alpha_m = 0} \nonumber \\
    &= \left\langle \widehat{Q}_{A}\right\rangle_{i,\mathrm{cc}}
\end{align}
\begin{align}
    \left\langle \widehat{Q}_{(A}\widehat{Q}_{B)}\right\rangle_{i} &= \left. \frac{\delta^2 e^{-\mathcal{F}}}{\delta\alpha_{A}\delta\alpha_{B}} \right|_{\alpha_m = 0} \nonumber \\
    &= -\left. e^{-\mathcal{F}} \frac{\delta^2 \mathcal{F}}{\delta\alpha_{A}\delta\alpha_{B}} \right|_{\alpha_m = 0}
     + \left. e^{-\mathcal{F}} \frac{\delta \mathcal{F}}{\delta\alpha_{A}} \frac{\delta \mathcal{F}}{\delta\alpha_{B}} \right|_{\alpha_m = 0}\nonumber \\
    &= \left\langle \widehat{Q}_{(A} \widehat{Q}_{B)}\right\rangle_{i,\mathrm{cc}} + \left\langle \widehat{Q}_{A}\right\rangle_{i,\mathrm{cc}} \left\langle \widehat{Q}_{B}\right\rangle_{i,\mathrm{cc}}
\end{align}
\begin{align}
    \left\langle \widehat{Q}_{(A}\widehat{Q}_{B}\widehat{Q}_{C)}\right\rangle_{i} &= -\left. \frac{\delta^3 e^{-\mathcal{F}}}{\delta\alpha_{A}\delta\alpha_{B}\delta\alpha_{C}} \right|_{\alpha_m = 0} \nonumber \\
    &= \left. e^{-\mathcal{F}} \frac{\delta^3 \mathcal{F}}{\delta\alpha_{A}\delta\alpha_{B}\delta\alpha_{C}} \right|_{\alpha_m = 0} \nonumber \\
     & \qquad - \left. e^{-\mathcal{F}} \frac{\delta \mathcal{F}}{\delta\alpha_{A}} \frac{\delta^2 \mathcal{F}}{\delta\alpha_{B}\delta\alpha_{C}} \right|_{\alpha_m = 0}
     - \left. e^{-\mathcal{F}} \frac{\delta \mathcal{F}}{\delta\alpha_{B}} \frac{\delta^2 \mathcal{F}}{\delta\alpha_{C}\delta\alpha_{A}} \right|_{\alpha_m = 0}
     - \left. e^{-\mathcal{F}} \frac{\delta \mathcal{F}}{\delta\alpha_{C}} \frac{\delta^2 \mathcal{F}}{\delta\alpha_{A}\delta\alpha_{B}} \right|_{\alpha_m = 0} \nonumber \\ & \qquad     + \left. e^{-\mathcal{F}} \frac{\delta \mathcal{F}}{\delta\alpha_{A}} \frac{\delta \mathcal{F}}{\delta\alpha_{B}} \frac{\delta \mathcal{F}}{\delta\alpha_{C}} \right|_{\alpha_m = 0}
     \nonumber \\
    &= \left\langle \widehat{Q}_{(A} \widehat{Q}_{B}  \widehat{Q}_{C)}\right\rangle_{i,\mathrm{cc}} \nonumber \\
    & \qquad + \left\langle \widehat{Q}_{A}\right\rangle_{i,\mathrm{cc}} \left\langle \widehat{Q}_{(B} \widehat{Q}_{C)}\right\rangle_{i,\mathrm{cc}} 
    + \left\langle \widehat{Q}_{B}\right\rangle_{i,\mathrm{cc}} \left\langle \widehat{Q}_{(C} \widehat{Q}_{A)}\right\rangle_{i,\mathrm{cc}}
    + \left\langle \widehat{Q}_{C}\right\rangle_{i,\mathrm{cc}} \left\langle \widehat{Q}_{(A} \widehat{Q}_{B)}\right\rangle_{i,\mathrm{cc}}
    \nonumber \\
    & \qquad + \left\langle \widehat{Q}_{A}\right\rangle_{i,\mathrm{cc}} \left\langle \widehat{Q}_{B}\right\rangle_{i,\mathrm{cc}} \left\langle \widehat{Q}_{C}\right\rangle_{i,\mathrm{cc}} 
\end{align}
We can then see that
\begin{align}
    \left\langle \widehat{Q}_{A}\right\rangle_{i,\mathrm{cc}} &= \left\langle \widehat{Q}_{A}\right\rangle_{i} \nonumber \\
    \left\langle \widehat{Q}_{(A}\widehat{Q}_{B)}     \right\rangle_{i,\mathrm{cc}} &= \left\langle \widehat{Q}_{(A}\widehat{Q}_{B)} \right\rangle_{i} - \left\langle \widehat{Q}_{A}\right\rangle_{i} \left\langle \widehat{Q}_{B}\right\rangle_{i} \nonumber \\
    \left\langle \widehat{Q}_{(A} \widehat{Q}_{B} \widehat{Q}_{C)}     \right\rangle_{i,\mathrm{cc}} &= \left\langle \widehat{Q}_{(A} \widehat{Q}_{B} \widehat{Q}_{C)}     \right\rangle_{i} \nonumber \\
    & \qquad - \left\langle \widehat{Q}_{A}\right\rangle_{i} \left\langle \widehat{Q}_{(B} \widehat{Q}_{C)}\right\rangle_{i,\mathrm{cc}} 
    - \left\langle \widehat{Q}_{B}\right\rangle_{i} \left\langle \widehat{Q}_{(C} \widehat{Q}_{A)}\right\rangle_{i,\mathrm{cc}}
    - \left\langle \widehat{Q}_{C}\right\rangle_{i} \left\langle \widehat{Q}_{(A} \widehat{Q}_{B)}\right\rangle_{i,\mathrm{cc}}
    \nonumber \\
    & \qquad - \left\langle \widehat{Q}_{A}\right\rangle_{i} \left\langle \widehat{Q}_{B}\right\rangle_{i} \left\langle \widehat{Q}_{C}\right\rangle_{i} \nonumber \\
    &= \left\langle \widehat{Q}_{(A} \widehat{Q}_{B} \widehat{Q}_{C)}     \right\rangle_{i} \nonumber \\
    & \qquad - \left\langle \widehat{Q}_{A}\right\rangle_{i} \left\langle \widehat{Q}_{(B} \widehat{Q}_{C)}\right\rangle_{i} 
    - \left\langle \widehat{Q}_{B}\right\rangle_{i} \left\langle \widehat{Q}_{(C} \widehat{Q}_{A)}\right\rangle_{i}
    - \left\langle \widehat{Q}_{C}\right\rangle_{i} \left\langle \widehat{Q}_{(A} \widehat{Q}_{B)}\right\rangle_{i}
    \nonumber \\
    & \qquad +2 \left\langle \widehat{Q}_{A}\right\rangle_{i} \left\langle \widehat{Q}_{B}\right\rangle_{i} \left\langle \widehat{Q}_{C}\right\rangle_{i} \, .
\end{align}
It is easy to verify for $A=B=C$, these expressions give the usual relations between moments and cumulants.  We note in passing that one could construct a generating functional for the fluctuations in the operators as
\begin{equation}
    \left\langle \exp \left[ -\sum_A \alpha_A^{(fi)} \delta\widehat{Q}_A \right] \right\rangle_i = \left\langle \exp \left[ -\sum_A \alpha_A^{(fi)} \left( \widehat{Q}_A - \left\langle \widehat{Q}_A \right\rangle \right) \right] \right\rangle_i = \left\langle \exp \left[ -\delta\widehat{s}_{fi} \right] \right\rangle_i \, ,
\end{equation}
and the corresponding generating functional for the connected correlation functions of the fluctuations is then $-\ln \left\langle \exp \left[ -\delta\widehat{s}_{fi} \right] \right\rangle_i $.

\section{Evolution of $\alpha$ coefficients}
\label{app:EvolutionAlpha}

\edit{
We start with our covariant form of the Boltzmann equation
\begin{equation}
  \frac{Dn}{dt}=\left[\frac{\partial}{\partial t} - Hp\frac{\partial}{\partial p}\right]n(t,p) = C[n(t,p)],
\end{equation}
where the symbol $D/dt$ is a covariant derivative along a worldline defined by the time coordinate.
Our first step is to define $\epsilon\equiv p/\tcm$, where $p$ is an independent variable, and $\tcm$ is a function of time as given in Eq.\ \eqref{eq:Tcm_def}
\begin{equation}
  \tcm = T_{\rm in}\frac{a_{\rm in}}{a(t)},
\end{equation}
for constants $T_{\rm in}$ and $a_{\rm in}$.  We will change coordinates to the following $t^{\prime}$ and $\epsilon$ by the transformation
\begin{align}
  t^{\prime} &= t;\\
  \epsilon &= \frac{p}{\tcm}
\end{align}
Therefore, the time derivative transforms to
\begin{align}
  \frac{\partial n}{\partial t} &= \frac{\partial n}{\partial t^{\prime}}\frac{\partial t^{\prime}}{\partial t} + \frac{\partial n}{\partial \epsilon}\frac{\partial \epsilon}{\partial t}\nonumber\\
  &= \frac{\partial n}{\partial t^{\prime}} + \frac{\partial n}{\partial \epsilon}\left[\frac{-\epsilon}{\tcm}(-H\tcm)\right]\nonumber\\
  &= \frac{\partial n}{\partial t^{\prime}} + \epsilon H\frac{\partial n}{\partial \epsilon},
\end{align}
and the momentum derivative transforms to
\begin{align}
  \frac{\partial n}{\partial p} &= \frac{\partial n}{\partial t^{\prime}}\frac{\partial t^{\prime}}{\partial p} + \frac{\partial n}{\partial \epsilon}\frac{\partial \epsilon}{\partial p}\nonumber\\
  &= \frac{1}{\tcm}\frac{\partial n}{\partial \epsilon}.
\end{align}
In this change of coordinates, the lhs of the Boltzmann equation becomes
\begin{align}
  \left[\frac{\partial}{\partial t} - Hp\frac{\partial}{\partial p}\right]n(t,p) &= \frac{\partial n}{\partial t^{\prime}} +\epsilon H\frac{\partial n}{\partial\epsilon} - H\epsilon\tcm\frac{1}{\tcm}\frac{\partial n}{\partial \epsilon}\nonumber\\
  &= \frac{\partial n}{\partial t^{\prime}} 
\end{align}
For ease in notation, assign $t=t^{\prime}$ so that the Boltzmann equation becomes
\begin{equation}
  \frac{Dn}{dt} = \frac{\partial n}{\partial t} = C[n(t,\epsilon)].
\end{equation}

With the simplified form of the Boltzmann equation, we will make a transformation into a new dependent variable $\alpha=\alpha(t,\epsilon)$
\begin{equation}
  n = \frac{1}{e^{\alpha} + 1} \implies \alpha = \ln\left(\frac{1-n}{n}\right).
\end{equation}
The covariant derivative of $\alpha$ with respect to time is
\begin{align}
  \frac{D\alpha}{dt} &= \frac{Dn}{dt}\frac{d\alpha}{dn}\nonumber\\
  &= -\frac{Dn}{dt}\frac{1}{n(1-n)}\nonumber\\
  &= -\frac{Dn}{dt}e^{-\alpha}(e^\alpha + 1)^2.\label{eq:dadt1}
\end{align}
To solve Eq.\ \eqref{eq:dadt1}, we write $\alpha(t,\epsilon)$ in terms of orthogonal polynomials
\begin{equation}
  \alpha(t,\epsilon) = \sum\limits_i a_i(t)P_i(\epsilon),
\end{equation}
where the coefficients $a_i(t)$ are time dependent [and bear no relation to scale factor $a(t)$], and the polynomials $P_i(\epsilon)$ are time-independent.  We have a choice for how we pick the $P_i$.  We could use Laguerre polynomials which form an orthogonal basis over the range $0<\epsilon<\infty$.  This would be a natural choice given that the range of $\epsilon$ is positive real numbers.  However, there is very little density at large values of $\epsilon$ so capturing the tail of the distribution would not be important for the precision at which we use the quadratic expansion in $\alpha$.  An alternative choice is to use Legendre polynomials for the orthogonal basis.  This requires a finite range, i.e., $0<\epsilon<\epsmax$ and is the protocol \burst uses where $\epsmax\sim20$.  Depending on the problem at hand, it may behoove us to use one basis over the other.  For the problem of weak decoupling in the standard cosmology, we surmise that the Legendre polynomials are a better choice.

Once we pick a set of orthogonal polynomials, we can integrate Eq.\ \eqref{eq:dadt1} over $\epsilon$ with a weight of a polynomial to find the time evolution of the coefficients
\begin{align}
  -\int d\epsilon\,P_j(\epsilon)\frac{Dn}{dt}e^{-\alpha}(e^\alpha+1)^2 &= \int d\epsilon\,P_j(\epsilon)\frac{D\alpha}{dt}\nonumber\\
   &=\int d\epsilon\,P_j(\epsilon)\frac{D}{dt}\left[\sum\limits_i a_i(t)P_i(\epsilon)\right]\nonumber\\
  &=\sum\limits_i\frac{Da_i}{dt} \int d\epsilon\,P_j(\epsilon)P_i(\epsilon)\nonumber\\
  &=\sum\limits_i\frac{Da_i}{dt} c_j\delta_{ij},
\end{align}
where $c_j\ne0$ is the coefficient for the particular orthogonal polynomial algebra.
For example: $c_j = 2/(2j+1)$ for Legendre polynomials (modulo a normalization if the domain has been rescaled).
We can solve for the time derivative of the coefficient in the expansion of $\alpha$
\begin{equation}
  \frac{Da_i}{dt} = -\frac{1}{c_i}\int d\epsilon\,P_i(\epsilon)\frac{Dn}{dt}e^{-\alpha}(e^\alpha+1)^2.
\end{equation}
We have used the symbol $Dn/dt$ for the collision term implying that we still use the $n$-distributions to write that term.  There is no need to transform $Dn/dt$ into an expression for $\alpha$: it is the same expression, but wherever we see the symbol $n(\epsilon_j)$, we use $1/(\exp[\alpha(\epsilon_j)] + 1)$.  As a result, we will set $A[\alpha]\equiv Dn/dt$ to delineate the difference in notation.  $A[\alpha]$ is a functional of $\alpha$ and a function of $t$ and $\epsilon$.
For clarity, we write the form of $A[\alpha]$ explicitly for $2\times2$ neutrino scattering ($r=1$ in Table \ref{colltab}) using the nomenclature of Sec.\ \ref{sec:CollisionRate}
\begin{equation} 
    \begin{split}
    A_{\nu_{1}}^{(r=1)}[\alpha_{f}] = \frac{1}{2E_{1}}\int& \frac{d^{3}p_{2}}{(2 \pi)^{3}2E_{2}}\frac{d^{3}p_{3}}{(2 \pi)^{3}2E_{3}}\frac{d^{3}p_{4}}{(2 \pi)^{3}2E_{4}} (2 \pi)^{4}\delta^{(4)}(p_{1}+p_{2}-p_{3}-p_{4}) \\
    & \times S_{r=1} \langle |\mathcal{M}_{r=1}|^{2} \rangle F_{r=1}(p_{1},p_{2},p_{3},p_{4}) \, ,
    \end{split}
\end{equation}
where in isotropic conditions, the statistical factor becomes
\begin{equation}
\begin{split}
F_{r=1}(\epsilon_{1},\epsilon_{2},\epsilon_{3},\epsilon_{4}) &= 
\dfrac{e^{\alpha_f(\epsilon_1)}e^{\alpha_f(\epsilon_2)} - e^{\alpha_f(\epsilon_3)}e^{\alpha_f(\epsilon_4)}}
{[e^{\alpha_f(\epsilon_1)} + 1][e^{\alpha_f(\epsilon_2)} + 1][e^{\alpha_f(\epsilon_3)} + 1][e^{\alpha_f(\epsilon_4)} + 1]},
\end{split}
\end{equation}
where the $f$ subscript on $\alpha$ gives the flavor index, and all of the neutrino lines for the $r=1$ process in Table \ref{colltab} have the same flavor.

Our final results are time derivatives of the coefficients of the orthogonal polynomials
\begin{equation}
  \frac{Da_i}{dt} = -\frac{1}{c_i}\int d\epsilon\,P_i(\epsilon)A[\alpha]e^{-\alpha}(e^\alpha+1)^2.
\end{equation}
If we expand $\alpha(t,\epsilon)$ into a power series, namely
\begin{equation}
  \alpha(t,\epsilon) = \sum_i\alpha_i(t)\epsilon^i,
\end{equation}
we can solve for the power-series coefficients in terms of the orthogonal polynomial coefficients to find the total time derivative for $\alpha_i$
\begin{align}
  \sum_i\frac{D\alpha_i}{dt}\epsilon^i &= \sum_j \frac{Da_j}{dt}P_j(\epsilon),\\
  \implies\sum_{i,k}\frac{D\alpha_i}{dt}\int d\epsilon\,P_k(\epsilon)\epsilon^i &= \sum_{j,k} \frac{Da_j}{dt}\int d\epsilon\,P_k(\epsilon)P_j(\epsilon),\nonumber\\
  &= \sum_{j,k} \frac{Da_j}{dt}c_k\delta_{jk}\nonumber\\
  &= \sum_{k} \frac{Da_k}{dt}c_k\\
  \implies \mathbf{M}\cdot\frac{D\boldsymbol{\alpha}}{dt} &= \frac{D\mathbf{a}}{dt}\\
  \implies \frac{D\boldsymbol{\alpha}}{dt} &= \mathbf{M}^{-1}\cdot\frac{D\mathbf{a}}{dt},
\end{align}
where the matrix $\mathbf{M}$ is given as
\begin{equation}
  M_{ki} = \frac{1}{c_k}\int d\epsilon\,\epsilon^iP_k(\epsilon).
\end{equation}
}

\clearpage 
\phantomsection
\addcontentsline{toc}{section}{References}
\bibliographystyle{utphys}
\bibliography{CNBRefs,Mendeley}

\end{document}